\documentclass[twocolumn,showpacs,preprintnumbers,amsmath,amssymb,aps,pre]{revtex4}

\usepackage{graphicx,color}

\begin{document}

\preprint{}

\title{Statistical mechanics of Fofonoff flows in an oceanic basin}

\author{A. Naso, P.H. Chavanis and B. Dubrulle }

\affiliation{
$^1$ Laboratoire de Physique, Ecole Normale Sup\'erieure de Lyon and CNRS (UMR
5672), 46 all\'ee d'Italie, 69007 Lyon, France\\
$^2$ LMFA, Universit\'e de Lyon, Ecole Centrale de Lyon, and CNRS (UMR 5509), 69134 Ecully Cedex, France\\ 
$^3$ Laboratoire de Physique Th\'eorique (IRSAMC), CNRS and UPS, Universit\'e de
Toulouse, F-31062 Toulouse, France\\
$^4$ SPEC/IRAMIS/CEA Saclay, and CNRS (URA 2464), 91191 Gif-sur-Yvette Cedex,
France\\
\email{aurore.naso@ec-lyon.fr, chavanis@irsamc.ups-tlse.fr,
berengere.dubrulle@cea.fr}
}

   \date{To be included later }

   \begin{abstract}We study the minimization of potential enstrophy at fixed
circulation and energy in an oceanic basin with arbitrary topography. For
illustration, we consider a rectangular basin and a  linear topography $h=by$
which represents either a real bottom topography or the  $\beta$-effect
appropriate to oceanic situations. Our minimum enstrophy principle is motivated
by different arguments of statistical mechanics reviewed in the article. It
leads to steady states of the quasigeostrophic (QG) equations characterized by a
linear relationship between potential vorticity $q$ and stream function $\psi$.
For low values of the energy, we recover Fofonoff flows [J. Mar. Res. {\bf 13},
254 (1954)] that display a strong westward jet. For large values of the energy,
we obtain geometry induced phase transitions between monopoles and dipoles
similar to those found by Chavanis \& Sommeria [J. Fluid Mech. {\bf 314}, 267
(1996)] in the absence of topography. In the presence of topography, we recover
and confirm the results obtained by Venaille \& Bouchet [Phys. Rev. Lett. {\bf
102}, 104501 (2009)] using a different formalism. In addition, we introduce
relaxation equations towards minimum potential enstrophy states and perform
numerical simulations to illustrate the phase transitions in a rectangular
oceanic basin with linear topography (or $\beta$-effect).
\end{abstract}
\pacs{05.20.-y Classical statistical mechanics - 05.45.-a Nonlinear dynamics and
chaos - 05.90.+m Other topics in statistical physics, thermodynamics, and
nonlinear dynamical systems - 47.10.-g General theory in fluid dynamics -
47.15.ki Inviscid flows with vorticity - 47.20.-k Flow instabilities - 47.32.-y
Vortex dynamics; rotating fluids}

   \maketitle
%

\section{\label{sec_intro}Introduction}

The dynamics of the oceans is extremely complex due to nonlinear coupling across
many scales of motion
and the interplay between mean and fluctuating fields \cite{holloway}. Although
the oceans can be considered as ``turbulent" in a classical sense, their
dynamics also involves wavelike phenomena and coherent  structures (vortices)
like monopoles, dipoles or modons, tripoles... Furthermore, despite the
permanent action of forcing (e.g. induced by the wind) and dissipation, the
oceans present a form of global organization. This is revealed in the existence
of strong jets like the Gulf Stream or the Kuroshio Current and in the
observation  of a large-scale oceanic circulation. In order to understand the
dynamics of the oceans, one possibility is to develop numerical codes with
increasing complexity. However, the results can be affected by the method used
to parameterize the small scales. Furthermore, numerical  simulations alone do
not explain the phenomena observed. Therefore, in order to understand the
physical output of such numerical simulations it can be useful to consider in
parallel simple mathematical models that can be studied in great detail. These
academic models can serve as a basis  to develop general methods (e.g.
statistical mechanics, kinetic theories) that can be relevant in more
complicated situations.

Early models of wind-induced oceanic circulation have been developed by Stommel
\cite{stommel} and Munk  \cite{munk} but they are based on linearized equations
and on an artificial concept of eddy viscosity. Alternatively, in a seminal
paper, Fofonoff \cite{fofonoff} neglects forcing and dissipation and studies the
case of a steady free circulation in a closed ocean. He considers quasi
geostrophic (QG) flows on the $\beta$-plane, a common starting point for many
dynamical studies in meteorology and oceanography. He furthermore assumes that
the ocean has reached a steady state characterized by a linear relationship
between potential vorticity $q=\omega+h$ and stream function $\psi$ (here, $h$
denotes the topography and the ordinary $\beta$-effect corresponds to $h=by$).
Finally, he considers  an asymptotic regime of low energy and provides a simple
analytical solution representing a westward jet with a recirculation at the
boundary. This solution is now called {\it Fofonoff flow} \cite{pedlosky}.
Numerical simulations starting from a random initial condition, in forced and
unforced situations, show that the system can spontaneously generate a Fofonoff
flow characterized by a linear $q-\psi$ relationship
\cite{veronis,griffa,cummins,wang,kazantsev}. However, such a linear
relationship is not expected to be general and more complex flows with nonlinear
$q-\psi$ relationships can also be observed. We must keep in
mind that Fofonoff flows provide just an academic model of oceanic circulation
with limited applications.

On the theoretical side, several researchers have tried to justify the relevance
of Fofonoff flows. In real oceans, the flows are forced by the wind and
dissipated at small scales. Niiler  \cite{niiler} and Marshall \& Nurser
\cite{marshall}  have argued that forcing and dissipation could equilibrate each
other in average and determine a quasi stationary state (QSS) that is a steady
state of the ideal QG equations (with forcing and dissipation switched off). In
these approaches, the $q-\psi$ relationship is selected by the properties of
forcing and dissipation and the conditions to have a linear relationship are
sought. In the case of unforced oceans, the justification of a linear $q-\psi$
relationship has been first sought in a phenomenological minimum enstrophy
principle. Bretherton \& Haidvogel \cite{bretherton} argue that potential
enstrophy decays due to viscous effects (see, however, Appendix \ref{sec_visc})
while the energy and the circulation remain approximately conserved (in the
limit of small viscosity). They propose therefore that the system should reach a
state that minimizes potential enstrophy at fixed energy and circulation. This
leads to a linear $q-\psi$ relationship like for Fofonoff flows.

A linear $q-\psi$ relationship can also be justified from statistical mechanics.
A statistical theory of 2D turbulence was first developed by Kraichnan \cite{k2}
in spectral space. It is based on the truncated 2D Euler equations which 
conserve only energy and enstrophy (quadratic constraints). In the presence of a
topography, Salmon,  Holloway \& Hendershott  \cite{salmon}  show that this
approach predicts a mean flow characterized by a linear relationship between the
averaged potential vorticity $\langle q\rangle$ and the averaged stream function
$\langle \psi\rangle$. Another statistical approach has been developed by Miller
\cite{miller} and Robert \& Sommeria  \cite{rs} in real space. This theory takes
into account all the conservation laws (energy and Casimirs) of the 2D Euler
equation and predicts various  $q-\psi$ relationships depending on the initial
conditions. However, in real situations where the system undergoes forcing and
dissipation, the conservation of all the Casimirs is abusive and has been
criticized by Ellis, Haven \& Turkington \cite{eht} and by Chavanis
\cite{physicaD,aussois}. In a recent paper  \cite{ncd1}, we have proposed to
conserve only a few microscopic moments of the vorticity among the infinite
class of Casimirs. These relevant constraints could be selected by the
properties of forcing and dissipation. For example, if we maximize the
Miller-Robert-Sommeria (MRS) entropy at fixed energy $E$, circulation $\Gamma$
and microscopic potential enstrophy $\Gamma_2^{f.g.}$, we get a mean flow
characterized by a linear $\overline{q}-\psi$ relationship leading to Fofonoff
flows. This statistical approach also predicts Gaussian fluctuations around this
mean flow. Furthermore, we have shown that the maximization of MRS entropy at
fixed energy, circulation and microscopic potential enstrophy $\Gamma_2^{f.g.}$
is equivalent to the minimization of macroscopic potential enstrophy
$\Gamma_2^{c.g.}$ at fixed energy and circulation.  This  justifies an inviscid
minimum potential enstrophy principle, and Fofonoff flows,  when only the
microscopic enstrophy (quadratic) is conserved among the infinite class of
Casimirs.

The asymptotic limit treated by Fofonoff \cite{fofonoff} corresponds
to small energies $E$, a limit relevant to oceanic situations. In the
statistical theory, this corresponds to a regime of large positive
inverse temperatures $\beta\gg 1$. In that case, the Fofonoff solution
is the unique (global) entropy maximum at fixed circulation and
energy. On the other hand, Chavanis \& Sommeria \cite{jfm} studied the
case of a linear $q-\psi$ relationship in a rectangular domain without
topography. For large energies $E$, corresponding to sufficiently
negative $\beta$, they report the existence of multiple
solutions. This leads to interesting phase transitions between
monopoles and dipoles depending on the value of a single control
parameter $\Gamma/\sqrt{E}$ and on the geometry of the domain (for
example, the aspect ratio $\tau$ of a rectangular domain). For
$\Gamma=0$, the maximum entropy state is a monopole if
$\tau<\tau_c=1.12$ and a dipole if $\tau>\tau_c$. For $\Gamma\neq 0$,
when $\tau<\tau_c$ the maximum entropy state is always a monopole and
when $\tau>\tau_c$, the maximum entropy state is a dipole for small
values of $\Gamma^2/E$ and a monopole for large values of
$\Gamma^2/E$. The approach of Chavanis \& Sommeria \cite{jfm} has been
completed recently by Venaille \& Bouchet \cite{vb,vbf} who used a
different theoretical formalism and provided a detailed discussion of
phase transitions and ensemble inequivalence in QG flows with and
without topography. For low values of the energy, they recover
Fofonoff flows \cite{fofonoff} and for large values of the energy,
they obtain geometry induced phase transitions between monopoles and
dipoles similar to those found by Chavanis \& Sommeria
\cite{jfm}. They also emphasize the notions of bicritical point and
azeotropy. Their approach is well-suited for statistical mechanics but
it may appear a bit abstract to fluid mechanicians. By contrast, the
approach of Chavanis \& Sommeria \cite{jfm} is simpler. In the present
paper, we shall extend the formalism of \cite{jfm} to the case of
flows with a topography and study how the series of equilibria is
modified in this more general context. We first consider an
antisymmetric linear topography $h=by$ in a rectangular domain (like in Fofonoff's
classical study) and then generalize the results to the case of an
arbitrary topography in an arbitrary domain. We recover and confirm
the main results of Venaille \& Bouchet \cite{vb,vbf} and illustrate
them with explicit calculations and with synthetic phase
diagrams. We compute the full series of equilibria
(containing all the critical points of entropy at fixed energy and
circulations) while Venaille \& Bouchet \cite{vb,vbf} focus on global
entropy maxima. The full series of equilibria is useful to show the
relative position of the stable and unstable branches. Furthermore,
for systems with long-range interactions, metastable states (local
entropy maxima), and even saddle points of entropy, can be long-lived
and therefore relevant in the dynamics \cite{ncd1}. It is therefore
important to take them into account. We also give a special attention
to the existence of a second order phase transition that exists only
for a particular value of the circulation $\Gamma_*$ (with
$\Gamma_*=0$ for an antisymmetric topography) and study how
this phase
transition takes birth as $\Gamma\rightarrow \Gamma_*$ through the
formation of a ``spike''. Finally, we introduce simple relaxation
equations that converge towards the minimum potential enstrophy state
at fixed circulation and energy and solve these equations numerically
to illustrate the phase transitions in a rectangular oceanic basin
with linear topography (or $\beta$-effect). Numerical integration of
these equations with a nonlinear topography have been previously
performed in \cite{cnd} and the present paper develops the theory
required to interpret the results.

\section{The quasigeostrophic equations}

\subsection{A maximization problem}

We consider a 2D incompressible flow over a topography described by the
quasigeostrophic (QG) equations
\begin{eqnarray}
\label{ss1}
\frac{\partial q}{\partial t}+{\bf u}\cdot\nabla q=0, \qquad
q=-\Delta\psi+\frac{1}{R^2}\psi+h,
\end{eqnarray}
where $q=\omega+\psi/R^2+h$ is the potential vorticity, $h$  the topography,
$R$  the Rossby radius, $\omega {\bf z}=\nabla\times {\bf u}$  the
vorticity, $\psi$ the stream function and ${\bf u}=-{\bf z}\times\nabla\psi$ the
velocity field
(${\bf z}$ is a unit vector normal to the
flow). For illustration and explicit calculations, we shall consider a linear
topography of the form $h=by$. This term can equivalently be interpreted as a
``$\beta$-effect'' in the oceans due to the Earth's sphericity.

The QG equations admit an infinite number of steady states of the form
\begin{eqnarray}
\label{ss2}
q=f(\psi),
\end{eqnarray}
where $f$ is an arbitrary function. They are obtained by solving the
differential equation
\begin{eqnarray}
\label{ss3}
-\Delta\psi+\frac{1}{R^2}\psi+h=f(\psi),
\end{eqnarray}
with $\psi=0$ on the domain boundary. The QG equations conserve  the energy
\begin{eqnarray}
\label{ss4}
E=\frac{1}{2}\int \left (q-h\right )\psi \, d{\bf r}=\frac{1}{2}\int
\left\lbrack (\nabla\psi)^2+\frac{\psi^2}{R^{2}}\right\rbrack\, d{\bf r},
\end{eqnarray}
and an infinite number of integral constraints
that are the Casimirs
\begin{eqnarray}
\label{ss5}
I_g=\int g(q)\, d{\bf r},
\end{eqnarray}
where $g$ is an arbitrary function. In particular, all the moments of
the potential vorticity $\Gamma_n=\int q^n\, d{\bf r}$ are conserved. The
first moment $\Gamma=\int q\, d{\bf r}$ is the potential circulation and the
second moment $\Gamma_2=\int q^2\, d{\bf r}$ is the potential enstrophy.

Let us consider the maximization problem
\begin{eqnarray}
\label{dim}
\max_{q}\lbrace S[q] \, | \, E[q]=E, \, \Gamma[q]=\Gamma\rbrace,
\end{eqnarray}
where $E$ and $\Gamma$ are the energy and the circulation and $S$ is a
 functional of the form
\begin{eqnarray}
\label{dims}
S=-\int C(q)\, d{\bf r},
\end{eqnarray}
where $C$ is an arbitrary convex function (i.e. $C''\ge 0$). The {critical
points}
of $S$ at fixed $E$ and $\Gamma$ are given by the variational
principle $\delta S-\beta\delta E-\alpha\delta\Gamma=0$ where $\beta$
and $\alpha$ are Lagrange multipliers. This gives
$C'(q)=-\beta\psi-\alpha$. Since $C$ is convex, we can inverse this
relation to obtain $q=F(\beta\psi+\alpha)=f(\psi)$ where
$F(x)=(C')^{-1}(-x)$.  We note that
$q'(\psi)=-\beta/C''(q)$. Therefore, a critical point of $S$ at
fixed $E$ and $\Gamma$ determines a steady state of the QG equations
with a monotonic $q-\psi$ relationship that is increasing for $\beta<0$ and
decreasing for $\beta>0$ . On the other hand, this state is a (local) maximum of
$S$ at fixed $E$ and $\Gamma$ iff
\begin{eqnarray}
\label{ss10}
-\frac{1}{2}\int C''(q)(\delta q)^2\, d{\bf r}-\frac{1}{2}\beta\int\delta
q\delta\psi\, d{\bf r}<0,
\end{eqnarray}
for all perturbations $\delta q$ that conserve energy and
circulation at first order.

\subsection{Its different interpretations}
\label{sec_inter}

The {maximization} problem (\ref{dim}) can be given several interpretations (see
\cite{proc} for a more detailed discussion):

(i) It determines a steady state of the QG equations that is
nonlinearly dynamically stable according to the stability criterion of Ellis
{\it et al.}
\cite{eht}. In that case, $S$ will be
refered to as a ``pseudo entropy'' \cite{proc}.  This criterion is more refined
than the
well-known Arnol'd theorems \cite{arnold} that provide only sufficient
conditions of
nonlinear dynamical stability. We note, however, that the criterion (\ref{dim})
provides itself just a sufficient condition of nonlinear dynamical stability. An
even more refined criterion of nonlinear dynamical stability is given by the
Kelvin-Arnol'd principle \cite{kelvin,arnold2}. The connections between these
different criteria of dynamical stability are reviewed in \cite{proc}. For the
particular choice $S=-(1/2)\int q^2\, d{\bf r}$ (neg-enstrophy), the
maximization problem (\ref{dim}) determines steady states of the QG equations
with a linear $q-\psi$ relationship that are nonlinearly dynamically stable.

(ii) The maximization problem (\ref{dim}) can be viewed as a phenomenological
selective decay principle (for $-S$) due to viscosity \cite{mm,proc}. In the
presence of a small viscosity $\nu\rightarrow 0$, the fragile integrals $-S$
decay (see, however, Appendix \ref{sec_visc}) while the robust integrals $E$ and
$\Gamma$ remain approximately conserved. This suggests that the system will
reach a steady state that is a maximum of a certain functional $S$ at fixed $E$
and $\Gamma$. If we assume that this functional is $S=-(1/2)\int q^2\, d{\bf
r}$, we recover the ordinary minimum enstrophy principle introduced by
Bretherton \& Haidvogel \cite{bretherton} (however, these authors mention in
their Appendix that other functionals $-S$ of the form (\ref{dims}), that they
call generalized enstrophies, could be minimized as well). We can also justify
an {\it inviscid}
selective decay principle due to coarse-graining \cite{jfm,proc}. For an ideal
evolution (no viscosity), the integrals $S[q]=-\int C(q)\, d{\bf r}$ of the
fine-grained PV are conserved by the QG equations (they are particular Casimirs)
while the integrals $S[\overline{q}]=-\int C(\overline{q})\, d{\bf r}$ of the
coarse-grained PV increase (see Appendix A of \cite{super}). These functionals
are  called generalized $H$-functions \cite{super,tremaine}. This suggests that
the system will reach a steady state that is a maximum of a certain H-function
$S[\overline{q}]$ at fixed $E$ and $\Gamma$.  If we assume that this functional
is $S=-(1/2)\int \overline{q}^2\, d{\bf r}$, we justify an inviscid minimum
enstrophy principle due to coarse-graining \cite{jfm}. However, other
generalized $H$ functions could be maximized as well \cite{proc}. It is
important to emphasize that these principles are purely phenomenological and
that they are not based on rigorous arguments. As such, they are not always true
\cite{brands,proc}.

(iii)  The maximization problem (\ref{dim}) provides a necessary and sufficient
condition of thermodynamical stability in the Ellis-Haven-Turkington (EHT)
approach \cite{eht}   where the Casimir constraints (fragile) are treated
canonically so that they are replaced by the specification of a prior vorticity
distribution $\chi(\sigma)$ encoding small-scale turbulence.
Indeed, a vorticity distribution $\rho({\bf r},\sigma)$ is a maximum of
relative entropy $S_{\chi}[\rho]$ at fixed circulation and energy
(EHT thermodynamical stability) iff the corresponding
coarse-grained PV field $\overline{q}({\bf r})$ is a maximum of a ``generalized
entropy'' $S[\overline{q}]$ at fixed $E$ and $\Gamma$ \cite{eht,aussois,cnd}. In
that case, the generalized entropy $S[\overline{q}]$ is determined by the prior
$\chi(\sigma)$. For a Gaussian prior, we find that the generalized entropy
$S[\overline{q}]=-(1/2)\int \overline{q}^2\, d{\bf r}$ is proportional to minus
the macroscopic coarse-grained enstrophy, justifying  a minimum potential
enstrophy principle in that context.

(iv) Since the solution of a variational problem is always solution of a more
constrained dual variational problem (but not the converse) \cite{ellis}, the
maximization problem (\ref{dim}) provides a sufficient condition of
thermodynamical stability in the  Miller-Robert-Sommeria (MRS) approach
\cite{miller,rs} where all the Casimirs are conserved. Indeed, a vorticity
distribution $\rho({\bf r},\sigma)$ is a maximum of MRS entropy $S[\rho]$ at
fixed energy, circulation and Casimirs (MRS thermodynamical
stability) if it is a maximum of $S_{\chi}[\rho]$ at fixed energy and
circulation.  However, the converse is wrong since the Casimir constraints have
been treated canonically. According to (iii), we conclude that a vorticity
distribution $\rho({\bf r},\sigma)$ is a maximum of MRS entropy $S[\rho]$ at
fixed energy, circulation and Casimirs if the corresponding coarse-grained PV
field $\overline{q}({\bf r})$ is a maximum of $S[\overline{q}]$ at fixed $E$ and
$\Gamma$ (but not the converse) \cite{bouchet}.   In that case, the generalized
entropy $S[\overline{q}]$ is determined by the initial condition \cite{proc}.
For initial conditions that lead to a Gaussian PV distribution at statistical
equilibrium, we find that  $S[\overline{q}]=-(1/2)\int \overline{q}^2\, d{\bf
r}$. Therefore a minimum of coarse-grained potential enstrophy at fixed
circulation and energy is a MRS thermodynamical equilibrium but the converse is
wrong in case of ensemble inequivalence.  In the MRS approach, a minimum
enstrophy principle can be justified only if the  microcanonical and grand
microcanonical ensembles are equivalent \cite{proc}.

(v) The maximization problem (\ref{dim}) provides a sufficient condition of
thermodynamical stability in the Naso-Chavanis-Dubrulle (NCD) approach
\cite{ncd1} where only a few Casimirs are conserved (the reason is the same as
in (iv)). In that case, the generalized entropy $S[\overline{q}]$ is determined
by the set of conserved Casimirs. For example, if we only conserve the
microscopic potential enstrophy $\Gamma_{2}^{f.g.}=\int \overline{q^2}\, d{\bf
r}$, we find that $S[\overline{q}]=-(1/2)\int \overline{q}^2\, d{\bf r}$. In
that case, the generalized entropy  is proportional to minus the coarse-grained
enstrophy $\Gamma_{2}^{c.g.}=-\int \overline{q}^2\, d{\bf r}$. Furthermore, in
this specific case where the constraints are linear or
quadratic (energy-enstrophy-circulation statistical
mechanics), it can be proven that the maximization of MRS entropy at fixed
energy, circulation and microscopic potential enstrophy $\Gamma_2^{f.g.}$
(NCD thermodynamical stability) is {\it equivalent} to the
minimization of macroscopic potential enstrophy $\Gamma_2^{c.g.}$ at fixed
energy and circulation \cite{ncd1}.  This  justifies an inviscid minimum
potential enstrophy principle, and Fofonoff flows,  when only the microscopic
enstrophy is conserved among the infinite class of Casimirs.

\subsection{Relaxation equations}
\label{sec_relax}

Some relaxation equations associated with the maximization problem (\ref{dim})
have been introduced in \cite{proc}. They can serve as numerical algorithms to
solve this maximization problem. They also provide non trivial
dynamical systems whose study is interesting in its own right.

(i) The first type of equations is of the form
\begin{equation}
\label{two16}
\frac{\partial q}{\partial t}+{\bf u}\cdot \nabla q=\nabla\cdot \left\lbrack
D\left (\nabla q+\frac{\beta(t)}{C''(q)}\nabla\psi\right )\right\rbrack,
\end{equation}
\begin{equation}
\label{two17}
\beta(t)=-\frac{\int D\nabla q\cdot\nabla\psi\, d{\bf r}}{\int
D\frac{(\nabla\psi)^{2}}{C''(q)}\, d{\bf r}},
\end{equation}
where $D({\bf r},t)\ge 0$ is the diffusion coefficient. The  boundary conditions
are ${\bf J}\cdot {\bf n}=0$ where ${\bf J}=-D
(\nabla q+\frac{\beta(t)}{C''(q)}\nabla\psi)$ is the current
and ${\bf n}$ is a unit vector normal to the boundary. With these boundary
conditions, the circulation is
clearly conserved. On the other hand, the inverse
``temperature'' $\beta(t)$ evolves in time according to
Eq. (\ref{two17}) so as to conserve energy ($\dot E=0$).  Easy calculations lead
to the $H$-theorem:
\begin{equation}
\label{two18}
\dot S=\int DC''(q)\left (\nabla q+\frac{\beta(t)}{C''(q)}\nabla\psi \right
)^{2}\, d{\bf r}\ge 0.
\end{equation}
Therefore, the relaxation equations (\ref{two16})-(\ref{two17}) relax towards a
(local) maximum of $S$ at fixed $E$ and $\Gamma$ (see \cite{proc} for a more
precise statement).

{\it Example:} If we take $S$ to be
the opposite of the potential enstrophy $S=-(1/2)\int {q}^{2}\, d{\bf
r}$, we
get
\begin{equation}
\label{sdp3}
\frac{\partial{q}}{\partial t}+{\bf u}\cdot \nabla {q}=\nabla\cdot \left\lbrack
D\left (\nabla {q}+\beta(t)\nabla\psi\right )\right\rbrack,
\end{equation}
\begin{equation}
\label{sdp4}
\beta(t)=-\frac{\int D\nabla {q}\cdot\nabla\psi\, d{\bf r}}{\int D
(\nabla\psi)^{2}\, d{\bf r}}.
\end{equation}
This equation monotonically dissipates the potential enstrophy ($\dot
\Gamma_{2}=-2\dot S=-2\int D (\nabla q+\beta(t)\nabla\psi) ^{2}\, d{\bf r}\le
0$)  at fixed energy and circulation ($\dot\Gamma=\dot E=0$) until the minimum
potential enstrophy state is reached. If we take $D$ constant and $R=\infty$
(for simplicity), the foregoing equations reduce to
\begin{equation}
\label{sdp3p}
\frac{\partial{q}}{\partial t}+{\bf u}\cdot \nabla{q}=D\left \lbrack \Delta
{q}-\beta(t)\left (q-h\right )
\right \rbrack,
\end{equation}
\begin{equation}
\label{sdp4p}
\beta(t)=\frac{S(t)+\frac{1}{2} \langle qh
\rangle}{E}=\frac{-\Gamma_{2}(t)+{\langle qh \rangle}}{2E},
\end{equation}
where we have used an integration by parts in Eq. (\ref{sdp4}) to obtain Eq.
(\ref{sdp4p}). In particular, for $h=0$, we have
$\beta(t)=S(t)/E=-\Gamma_{2}(t)/2E$. As shown in Appendix \ref{sec_nept}, these
relaxation equations are compatible with the ``Neptune effect'' discovered by
Holloway \cite{hollowaynept} and playing an important role in
oceanic modeling.

(ii) The second type of relaxation equations is of the form
\begin{equation}
\label{two22}
\frac{\partial q}{\partial t}+{\bf u}\cdot \nabla
q=-D(C'(q)+\beta(t)\psi+\alpha(t)),
\end{equation}
where $\beta(t)$ and $\alpha(t)$ evolve in time according to
\begin{equation}
\label{two23}
\langle
C'(q)\psi\rangle+\beta(t)\langle\psi^{2}\rangle+\alpha(t)\langle\psi\rangle=0,
\end{equation}
\begin{equation}
\label{two24}
\langle C'(q)\rangle+\beta(t)\langle\psi\rangle+\alpha(t)A=0,
\end{equation}
in order to satisfy the conservation of energy and circulation ($A$ is
the domain area, $\langle X\rangle=\int X\, d{\bf r}$ and we have assumed $D$
constant for simplicity). We shall consider boundary conditions of the form
$C'(q)+\alpha(t)=0$ on the boundary so as to be consistent with the steady state
for which $C'(q)+\beta\psi+\alpha=0$ in the whole domain
(recall that $\psi=0$ on the boundary).  Easy calculations lead to the $H$
theorem:
\begin{equation}
\label{two25}
\dot S=D\int  (C'(q)+\beta(t)\psi+\alpha(t))^{2}\, d{\bf r}\ge 0.
\end{equation}
Therefore, the relaxation equations (\ref{two22})-(\ref{two24}) relax towards a
(local) maximum of $S$ at fixed $E$ and $\Gamma$ (see \cite{proc} for a more
precise statement).

{\it Example:} If we take $S$ to be
the opposite of the potential enstrophy $S=-(1/2)\int {q}^{2}\, d{\bf
r}$, the relaxation equations reduce to
\begin{equation}
\frac{\partial q}{\partial t}+{\bf u}\cdot \nabla q=-D\left \lbrack
q+\beta(t)\psi+\alpha(t)\right \rbrack \label{rel_1}
\end{equation}
with
\begin{eqnarray}
\beta(t)=\frac{\Gamma\langle\psi\rangle- {A(2E+ \langle h\psi
\rangle)}}{A\langle\psi^2\rangle-\langle\psi\rangle^2}, \label{rel_2}
\end{eqnarray}
\begin{eqnarray}
\alpha(t)=-\frac{\Gamma\langle\psi^2\rangle-{\langle \psi \rangle (2E+ \langle
h\psi \rangle)}}{A\langle\psi^2\rangle-\langle\psi\rangle^2}. \label{rel_3}
\end{eqnarray}
These equations monotonically dissipate the potential enstrophy ($\dot
\Gamma_{2}=-2\dot S=-2D\int (q+\beta\psi+\alpha)^{2}\, d{\bf r}\le 0$)  at fixed
energy and circulation ($\dot E=\dot\Gamma=0$) until the minimum potential
enstrophy state is reached.

\section{Thermodynamics of Fofonoff flows}
\label{sec_tf}

\subsection{The maximization problem}
\label{sec_de}

We shall study the maximization problem
\begin{eqnarray}
\label{f1}
\max_{q}\lbrace S[q] \, | \, E[q]=E, \, \Gamma[q]=\Gamma\rbrace,
\end{eqnarray}
with
\begin{eqnarray}
\label{f2}
S=-\frac{1}{2}\int q^2\, d{\bf r},
\end{eqnarray}
\begin{eqnarray}
\label{f3}
E=\frac{1}{2}\int (q-h)\psi\, d{\bf r}=\frac{1}{2}\int (\nabla\psi)^2\, d{\bf
r},
\end{eqnarray}
\begin{eqnarray}
\label{f4}
\Gamma=\int q\, d{\bf r},
\end{eqnarray}
\begin{eqnarray}
\label{f6}
q=-\Delta\psi+h.
\end{eqnarray}
For simplicity, we assume  $R\rightarrow +\infty$ but the case of finite Rossby
radius can be treated similarly and the main results are unchanged. As discussed
in Sec. \ref{sec_inter}, the maximization
problem (\ref{f1}) can be interpreted as a refined condition of
nonlinear dynamical stability (in that case $S$ is a Casimir or a pseudo
entropy), as a sufficient condition of thermodynamical stability in the MRS
approach, or as a necessary and sufficient condition of thermodynamical
stability in the EHT and NCD approaches (in these cases $S$ is a
generalized entropy). Noting that $S$ is proportional to the opposite of the
potential enstrophy $\Gamma_2=\int q^2\, d{\bf r}$, the maximization problem
(\ref{f1}) is also equivalent to the phenomenological minimum potential
enstrophy principle. In the following, to simplify the terminology, $S$ will be
called the ``entropy".

The critical points of entropy at fixed circulation and energy are
given by the variational principle
\begin{eqnarray}
\label{f5}
\delta S-\beta\delta E-\alpha\delta\Gamma=0,
\end{eqnarray}
where $\beta$ and $\alpha$ are Lagrange multipliers that will be called
``inverse temperature'' and ``chemical potential''. This yields
\begin{eqnarray}
\label{f7}
q=-\beta\psi-\alpha.
\end{eqnarray}
We consider a domain of unit area $A=1$ and we define $\langle X \rangle=\int
X\, d{\bf r}$. Then, we have $\alpha=-\Gamma-\beta\langle \psi\rangle$ and we
can rewrite the previous relation as
\begin{eqnarray}
\label{f8}
q=-\beta(\psi-\langle \psi\rangle)+\Gamma.
\end{eqnarray}
Substituting this relation in Eq. (\ref{f6}), we obtain
\begin{eqnarray}
\label{f9}
-\Delta\psi+\beta\psi=\Gamma+\beta\langle \psi\rangle-h,
\end{eqnarray}
with $\psi=0$ on the domain boundary. This is the fundamental
differential equation of the problem. It has the form of an
inhomogeneous Helmholtz equation.

Using Eq. (\ref{f8}), the energy and the entropy can be written
\begin{eqnarray}
\label{f10}
E=-\frac{1}{2}\beta\left (\langle \psi^2\rangle-\langle \psi\rangle^2\right
)+\frac{1}{2}\Gamma\langle \psi\rangle-\frac{1}{2}\langle h\psi\rangle,
\end{eqnarray}
\begin{eqnarray}
\label{f11}
S=-\frac{1}{2}\beta^2\left (\langle \psi^2\rangle-\langle \psi\rangle^2\right
)-\frac{1}{2}\Gamma^2.
\end{eqnarray}
Using Eq. (\ref{f10}), an equivalent expression of the entropy  is
\begin{eqnarray}
\label{f12}
S=\beta E-\frac{1}{2}\beta\Gamma\langle\psi\rangle+\frac{1}{2}\beta\langle
h\psi\rangle-\frac{1}{2}\Gamma^2.
\end{eqnarray}
The last term in Eqs. (\ref{f11}) and (\ref{f12}) will be ignored in the
following since it is just an unimportant additional constant (for fixed
$\Gamma$).

\subsection{The solution of the differential equation and the equation of state}
\label{sec_cc}

To study the maximization problem (\ref{f1}), we shall follow the general
methodology developed
by Chavanis \& Sommeria \cite{jfm}. The main novelty with respect to their study
is the presence of the topography $h$. We shall write the topography in the form
$h(x,y)=bH(x,y)$ where $H$ is dimensionless.

To study the differential equation
\begin{eqnarray}
\label{f13}
-\Delta\psi+\beta\psi=\Gamma+\beta\langle \psi\rangle-b H,
\end{eqnarray}
we first assume that $\Gamma+\beta\langle \psi\rangle\neq 0$ (i.e. $\alpha\neq
0$) and we define
\begin{eqnarray}
\label{f14}
\phi=\frac{\psi}{\Gamma+\beta\langle \psi\rangle},
\end{eqnarray}
and
\begin{eqnarray}
\label{f15}
c=\frac{b}{\Gamma+\beta\langle \psi\rangle}.
\end{eqnarray}
We note that $c=-b/\alpha$ so that $c$ plays the role of the inverse of the
chemical potential (see Sec. \ref{sec_chpo}). Therefore, the condition
$\alpha\neq 0$ is equivalent to $c$ finite.
With these notations, the differential equation (\ref{f13}) becomes
\begin{eqnarray}
\label{f16}
-\Delta\phi+\beta\phi=1-c H,
\end{eqnarray}
with $\phi=0$ on the domain boundary. The solution of this equation can be
written
\begin{eqnarray}
\label{f28}
\phi=\phi_1+c\phi_2,
\end{eqnarray}
where $\phi_1$ and  $\phi_2$ are the solutions of
\begin{eqnarray}
\label{f18}
-\Delta\phi_1+\beta\phi_1=1,
\end{eqnarray}
\begin{eqnarray}
\label{f19}
-\Delta\phi_2+\beta\phi_2=-H,
\end{eqnarray}
with $\phi_1=0$ and $\phi_2=0$ on the domain boundary. For given $\beta$, the
functions
$\phi_1$ and $\phi_2$ can be obtained by solving the differential
equation numerically or by decomposing the solutions on the modes of
the Laplacian operator (cf Appendix \ref{sec_md}). For the moment, we assume
that
$\beta$ is not equal to an eigenvalue $\beta_{n}$ of the Laplacian so that the
solutions of Eqs. (\ref{f18}) and
(\ref{f19}) are unique and finite. The cases $\beta= \beta_{n}$ must be
studied specifically (see following sections).

Taking the average of Eq. (\ref{f14}) and solving for
$\langle\psi\rangle$, we obtain
\begin{eqnarray}
\label{f20}
\langle \psi\rangle=\frac{\Gamma\langle
\phi\rangle}{1-\beta \langle \phi\rangle}.
\end{eqnarray}
Using Eqs. (\ref{f14}) and (\ref{f20}), the solution of Eq. (\ref{f13}) is
\begin{eqnarray}
\label{f21}
\psi=\frac{\Gamma\phi}{1-\beta \langle \phi\rangle}.
\end{eqnarray}
Using Eqs. (\ref{f15}) and (\ref{f20}), the constant $c$ is given by
\begin{eqnarray}
\label{f22}
c=\frac{b}{\Gamma}(1-\beta\langle\phi\rangle),
\end{eqnarray}
where $\langle\phi\rangle$ itself depends on $c$.  Substituting
Eq. (\ref{f28}) in Eq. (\ref{f22}), we finally obtain
\begin{eqnarray}
\label{f23}
c=\frac{1-\beta\langle\phi_1\rangle}{\frac{\Gamma}{b}+\beta\langle\phi_2\rangle}
.
\end{eqnarray}
For a given normalized circulation $\Gamma/b$, this relation
completely determines $c$ as a function of $\beta$. Therefore, the solution of
Eq. (\ref{f13}) is given
by Eq. (\ref{f21}) where $\phi$ is determined by Eqs. (\ref{f28}) and
(\ref{f23}). We have thus completely solved the differential
equation  (\ref{f13}) for $\beta\neq \beta_{n}$. We must now relate $\beta$ to
the energy. Substituting Eq. (\ref{f21}) in the energy constraint
(\ref{f10}), we get
\begin{eqnarray}
\label{f24}
(1-\beta\langle\phi\rangle)^2=\frac{\Gamma^2}{{2E}}
(\langle\phi\rangle-\beta\langle\phi^2\rangle-c\langle \phi H\rangle).
\end{eqnarray}
Similarly, the entropy (\ref{f11}) can be written
\begin{eqnarray}
\label{f25}
\frac{2S}{\Gamma^2}=-\frac{\beta^2}{(1-\beta\langle\phi\rangle)^2}
(\langle\phi^2\rangle-\langle\phi\rangle^2).
\end{eqnarray}

In the absence of topography ($b=c=0$), we recover the equations of
Chavanis \& Sommeria \cite{jfm}. In that case, there is a single
control parameter $\Lambda=\Gamma/\sqrt{2E}$. In the present case,
there are two control parameters: $\Lambda=\Gamma/\sqrt{2E}$ and
$\mu=\Gamma/b$. In order to have a well-defined limit
$\Gamma\rightarrow 0$, it is more convenient to take ${\cal
E}=2E/b^2=\mu^2/\Lambda^2$ and $\mu=\Gamma/b$ as independent control
parameters. Using Eq. (\ref{f22}), we find that the equations of the problem are
given in a very compact form by
\begin{eqnarray}
\label{f26}
\frac{2E}{b^2}=\frac{1}{c^2}
(\langle\phi\rangle-\beta\langle\phi^2\rangle-c\langle \phi H\rangle),
\end{eqnarray}
\begin{eqnarray}
\label{f27}
\frac{2S}{b^2}=-\frac{\beta^2}{c^2}(\langle\phi^2\rangle-\langle\phi\rangle^2),
\end{eqnarray}
where $\phi$ is given by Eq. (\ref{f28}) and $c$ by Eq. (\ref{f23}).

Note that the r.h.s. of Eqs. (\ref{f26}) and (\ref{f27}) are functions
of $\beta$ which can be easily computed numerically. There are two
control parameters: the energy ${\cal E}=2E/b^2$ and the circulation
$\mu=\Gamma/b$ (normalized by the $\beta$-effect parameter $b$ or by the
amplitude of the topography). For the sake of simplicity we will
denote, in the figures and in the discussion, $E$ and $\Gamma$ the
energy and the circulation thus normalized (while $b$ will be
explicitly written in the formulae). For given $\Gamma$,
Eq. (\ref{f26}) determines $\beta$ as a function of ${E}$, i.e. the
caloric curve $\beta(E)$. Of course, it is easier to proceed the other
way round.  We first fix $\Gamma$. Then, for each $\beta$ we can
determine $c$ by Eq. (\ref{f23}) and ${E}$ by Eq. (\ref{f26}) to
obtain ${E}(\beta)$. Inverting this relation we get $\beta({E})$ for
fixed $\Gamma$.

We now consider the case $\Gamma+\beta\langle \psi\rangle=0$ (i.e. $\alpha=0$).
Equation (\ref{f13}) then becomes
\begin{eqnarray}
\label{f13bg}
-\Delta\psi+\beta\psi=-b H,
\end{eqnarray}
and the solution is
\begin{eqnarray}
\label{f13bgb}
\psi=b\phi_2.
\end{eqnarray}
Substituting this relation in Eqs. (\ref{f10}) and (\ref{f11}) and using
$\Gamma+\beta b \langle \phi_2\rangle=0$, we find that the energy and the
entropy are given by
\begin{eqnarray}
\label{f26vf}
\frac{2E}{b^2}=-\beta\langle\phi_2^2\rangle-\langle \phi_2 H\rangle,
\end{eqnarray}
\begin{eqnarray}
\label{f27hj}
\frac{2S}{b^2}=-\beta^2(\langle\phi_2^2\rangle-\langle\phi_2\rangle^2).
\end{eqnarray}
We can check that these equations are limit cases of the general equations
(\ref{f26}), (\ref{f27}) and (\ref{f28}) when $c\rightarrow +\infty$. Indeed,
the condition $\alpha\rightarrow 0$ is equivalent to $c\rightarrow +\infty$.

{\it Particular limits:} It is interesting to mention the connection with
previous works. For $c\rightarrow 0$, we recover the results of Chavanis \&
Sommeria \cite{jfm}. This corresponds to a limit of large energies
$1/E\rightarrow 0$. We expect geometry induced phase transitions between
monopoles and dipoles. On the other hand, for $\beta\rightarrow +\infty$, we
recover the results of Fofonoff \cite{fofonoff}.  This corresponds to a limit of
small energies $1/E\rightarrow +\infty$. In that case, the equilibrium state is
unique and corresponds to a westward jet (in an antisymmetric
domain with $\beta$-effect $H=y$).

\subsection{The chemical potential}
\label{sec_chpo}

The chemical potential is defined by
\begin{eqnarray}
\label{chem1}
\alpha=-\beta\langle\psi\rangle-\Gamma.
\end{eqnarray}
If $\beta\langle\psi\rangle+\Gamma=0$, then
\begin{eqnarray}
\label{chem2}
\alpha=0.
\end{eqnarray}
If $\beta\langle\psi\rangle+\Gamma\neq 0$, according to Eq. (\ref{f15}), we have
\begin{eqnarray}
\label{chem3}
\frac{\alpha}{b}=-\frac{1}{c}.
\end{eqnarray}
Therefore, up to a normalization constant, the chemical potential is equal to
$-1/c$. This gives to the parameter $c$ a clear physical meaning.

For a given value of the energy $E$, we can obtain the chemical potential curve
$\alpha(\Gamma)$ in parametric form in the following manner. Fixing ${E}$, we
find from Eqs. (\ref{f26}) and (\ref{f28}) that  $c$ is related to $\beta$ by a
second degree equation
\begin{eqnarray}
\label{chem4}
\left (\frac{2E}{b^2} + \beta\langle\phi_2^2\rangle+\langle\phi_2 H\rangle\right
)c^2\qquad\qquad\nonumber\\
+(\langle\phi_1 H\rangle -
\langle\phi_2\rangle+2\beta\langle\phi_1\phi_2\rangle)c
+\beta\langle\phi_1^2\rangle-\langle\phi_1\rangle=0.
\end{eqnarray}
This determines $c=c({E},\beta)$.   On the other hand, according to Eq.
(\ref{f23}), the circulation $\Gamma/b$ is related to $\beta$ by
\begin{eqnarray}
\label{chem5}
\frac{\Gamma}{b}=\frac{1}{c(E,\beta)}
(1-\beta\langle\phi_1\rangle)-\beta\langle\phi_2\rangle.
\end{eqnarray}
This determines $\Gamma=\Gamma(E,\beta)$. Therefore, for
given ${E}$, these equations allow us to obtain $c$ as a function of $\Gamma/b$ 
in parametric form with running parameter $\beta$. This yields the chemical
potential curve $\alpha(\Gamma)$. Then, we have to account for the particular
cases where $\beta$ is an eigenvalue $\beta_{n}$ of the Laplacian (see below).

\subsection{A critical circulation}
\label{sec_ccirc}

According to Eq. (\ref{f23}), we note that the expression of $c$ (related to the
chemical potential $\alpha$), involves the important function:
\begin{eqnarray}
\label{f34}
F(\beta)\equiv \beta\langle\phi_1\rangle-1.
\end{eqnarray}
We shall call $\beta_{*}^{(k)}$ the solutions of $F(\beta_*^{(k)})=0$ and we
shall denote simply $\beta_{*}=\beta_*^{(1)}$ the largest solution. The function
$F(\beta)$ and the inverse temperature $\beta_*$ were introduced by Chavanis \&
Sommeria \cite{jfm}. We will see that the temperature $\beta_*$ plays an
important
role in the problem.  For $\beta=\beta_{*}$, we find that $c=0$ except if
$\Gamma=\Gamma_{*}$ where $\Gamma_*$ is a critical circulation given
by \footnote{In fact, there exists several critical circulations
$\Gamma_*^{(k)}$, associated to each value of $\beta_*^{(k)}$, but we will see
that only the one corresponding to the highest inverse temperature $\beta_*$
matters when we consider stable states.}:
\begin{eqnarray}
\label{f51}
\frac{\Gamma_*}{b}=-\beta_*\langle\phi_2\rangle_*.
\end{eqnarray}
We will have to distinguish the cases $\Gamma=\Gamma_*$ and $\Gamma\neq
\Gamma_*$.

\subsection{The program}
\label{sec_prog}

We shall successively consider the case of an antisymmetric and
a non-symmetric topography. It will be shown that the
mathematical
expressions simplify greatly for an antisymmetric topography
so that it is
natural to treat this case first. To be specific, we will
consider
a rectangular domain and a topography of the form $h=by$. This
corresponds to the situation studied by Fofonoff
\cite{fofonoff} in
his seminal paper.  It should therefore be given particular
attention. Then, we will
consider a non-symmetric topography of the form $h=b(y-y_0)$ in a
rectangular domain. Finally, it will be shown in Secs.
\ref{sec_nst} and
\ref{sec_sg} how the results can be generalized to an arbitrary domain
and an arbitrary topography.

The inverse temperature $\beta$ is the Lagrange multiplier associated with the
conservation of the energy $E$ and the chemical potential $\alpha$ is the
Lagrange multiplier associated with the conservation of the circulation
$\Gamma$. Thus, $\beta=(\partial S/\partial E)_\Gamma$ and $\alpha=(\partial
S/\partial \Gamma)_E$. We shall first study the caloric curve $\beta(E)$ for a
given value of the circulation $\Gamma$, then the chemical potential
$\alpha(\Gamma)$ for a given value of the energy $E$.

\section{The case of an antisymmetric linear topography
(Fofonoff case)}
\label{sec_st}

We consider a complete oceanic basin as in the study of Fofonoff
\cite{fofonoff}.
The domain is rectangular with $-\sqrt{\tau}/2\le x\le \sqrt{\tau}/2$ and
$-1/(2\sqrt{\tau})\le y\le
1/(2\sqrt{\tau})$ where $\tau=L_x/L_y$ is the aspect ratio. The topography  
$h=by$ (i.e. $H=y$) is linear and antisymmetric with respect to
$y=0$. This linear topography can also represent the $\beta$-effect.
The eigenvalues of the Laplacian in a rectangular domain will be
called $\beta_{mn}$ (see Appendix \ref{sec_md}).  Assuming
$\beta\neq\beta_{mn}$, it is easy to show from considerations of symmetry that
$\phi_2$ is odd  and $\phi_1$ is even with respect to $y$. Therefore,
\begin{eqnarray}
\label{f30}
\langle\phi_2\rangle=\langle\phi_1 y\rangle=\langle\phi_1\phi_2\rangle=0.
\end{eqnarray}
When $\Gamma+\beta\langle\psi\rangle\neq 0$ ($\alpha\neq 0$, $c$ finite), the
equations of the problem  become
\begin{eqnarray}
\label{gjtr}
\psi=\frac{b}{c}\phi,\qquad (\phi=\phi_1+c\phi_2)
\end{eqnarray}
\begin{eqnarray}
\label{f31}
c=\frac{b}{\Gamma}(1-\beta\langle\phi_1\rangle),
\end{eqnarray}
\begin{eqnarray}
\label{f32}
\frac{2E}{b^2}=\frac{1}{c^2}
(\langle\phi_1\rangle-\beta\langle\phi_1^2\rangle)-\beta
\langle\phi_2^2\rangle-\langle \phi_2 y\rangle,
\end{eqnarray}
\begin{eqnarray}
\label{f33}
\frac{2S}{b^2}=-\frac{\beta^2}{c^2}
(\langle\phi_1^2\rangle-\langle\phi_1\rangle^2)-{\beta^2}\langle\phi_2^2\rangle.
\end{eqnarray}
When $\Gamma+\beta\langle\psi\rangle= 0$ ($\alpha=0$, $c\rightarrow \infty$),
the equations of the problem become
\begin{eqnarray}
\label{gj}
\psi={b}\phi_2,
\end{eqnarray}
\begin{eqnarray}
\label{f26vfb}
\frac{2E}{b^2}=-\beta\langle\phi_2^2\rangle-\langle \phi_2 y\rangle,
\end{eqnarray}
\begin{eqnarray}
\label{f27hjb}
\frac{2S}{b^2}=-\beta^2\langle\phi_2^2\rangle.
\end{eqnarray}
Finally,  since $\langle\phi_2\rangle=0$, we find that $\Gamma_*=0$. We thus
need to distinguish two cases depending on whether $\Gamma=0$ or $\Gamma\neq 0$.

\subsection{The caloric curve $\beta(E)$ for $\Gamma=0$}
\label{sec_stz}

We shall first discuss the caloric curve $\beta(E)$ for
$\Gamma=\Gamma_*=0$. Details on the construction of this
curve  can be found in Appendix \ref{techn_S_eq}. Since the multiple solutions
occur for large values of $E$, it appears more convenient to plot $\beta$ as a
function of $1/E$ like in the study of Chavanis \& Sommeria \cite{jfm}. The
corresponding curve $\beta(E)$ can be deduced easily.

For small energies, there exists only one solution of Eq.
(\ref{f13}) and it is a global entropy maximum at fixed circulation and energy.
On the other hand, for large energies, there exists an infinite
number of solutions of  Eq. (\ref{f13}), i.e. there exists an infinite number of
critical points of entropy at fixed circulation and energy. In order to select
the most probable structure, we need to compare their entropies. For $1/E=0$ we
have $S/E=\beta$ so we just need to compare their inverse temperature $\beta$.
More generally, it can be shown that the entropy is a monotonic
function of $\beta$ (for given $E$ and $\Gamma$) so, in the following, we shall
select the solution with the largest $\beta$.
For large energies, there is a  competition between the solution with inverse
temperature $\beta_*$, the solution with inverse temperature $\beta_{21}$ (where
$\beta_{21}$ is the largest eigenvalue with zero average value
$\langle\psi_{21}\rangle=0$ and $\langle\psi_{21}y\rangle=0$) and the solution
with inverse temperature $\beta_{12}$ (where $\beta_{12}$ is the largest
eigenvalue with zero average value $\langle\psi_{12}\rangle=0$ and
$\langle\psi_{12}y\rangle\neq 0$). In the region where the solutions are in
competition, the solution with the highest entropy is the one with $\beta=\max
\lbrace \beta_*,\beta_{21},\beta_{12}\rbrace$.  The selection depends on the
aspect ratio of the domain. In a rectangular domain elongated in the horizontal
direction ($\tau>1$), $\beta_{21}>\beta_{12}$. In a rectangular domain elongated
in the vertical direction ($\tau<1$), $\beta_{12}>\beta_{21}$. On the other
hand, as shown by Chavanis \& Sommeria \cite{jfm}, there exists a critical
aspect ratio $\tau_c=1.12$ such that: $\max  \lbrace
\beta_*,\beta_{21},\beta_{12}\rbrace=\beta_*$ for $1/\tau_c<\tau<\tau_c$,  $\max
\lbrace \beta_*,\beta_{21},\beta_{12}\rbrace=\beta_{21}$ for $\tau>\tau_c$ and 
$\max \lbrace \beta_*,\beta_{21},\beta_{12}\rbrace=\beta_{12}$ for 
$\tau<1/\tau_c$. To describe the phase transitions, we must therefore consider
these three cases successively.

\begin{figure}[h]
\center
\includegraphics[width=8cm,keepaspectratio]{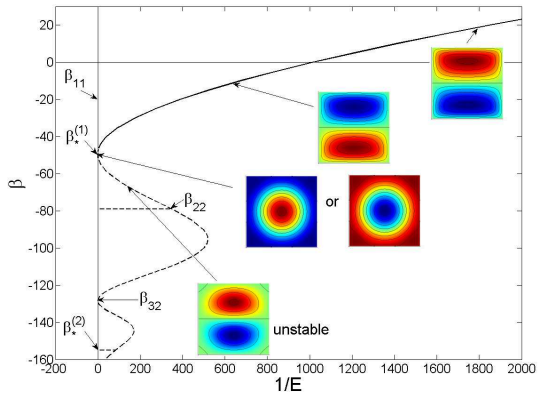}
\caption{\label{E_beta_1} Relationship between $\beta$ and $1/E$ in a square
domain $[-0.5,0.5]\times[-0.5,0.5]$ for $\Gamma=0$ ($1/\tau_c<\tau=1<\tau_c$).
The potential vorticity $q$ is shown for several values of $\beta$. For
small energies we get Fofonoff flows. For large
energies and $1/\tau_c<\tau<\tau_c$, the maximum entropy state is the monopole.
There exists a second order phase transition between Fofonoff flows and
monopoles (see zoom in Fig. \ref{E_beta_1z}). Note that this second order phase
transition exists only for $\Gamma=0$.}
\end{figure}

\begin{figure}[h]
\center
\includegraphics[width=8cm,keepaspectratio]{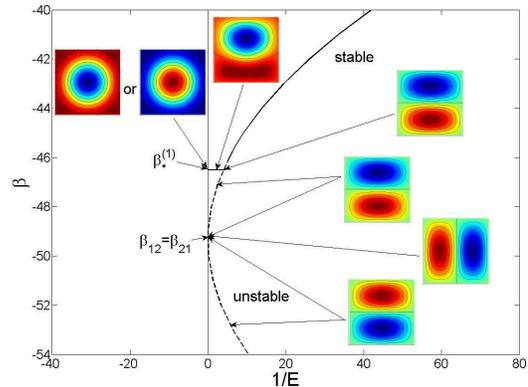}
\caption{\label{E_beta_1z} Zoom of Fig. \ref{E_beta_1} in the region of second
order phase transition.}
\end{figure}

\begin{figure}[h]
\center
\includegraphics[width=4.5cm,keepaspectratio]{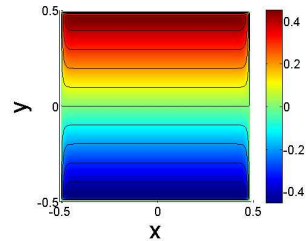}
\caption{\label{Fofonoff} Potential vorticity $q$ in a square domain
$[-0.5,0.5]\times[-0.5,0.5]$ for $\Gamma=0$ and $\beta=10^4$. This corresponds
to the so-called Fofonoff flow.}
\end{figure}

\begin{figure}[h]
\center
\includegraphics[width=4.5cm,keepaspectratio]{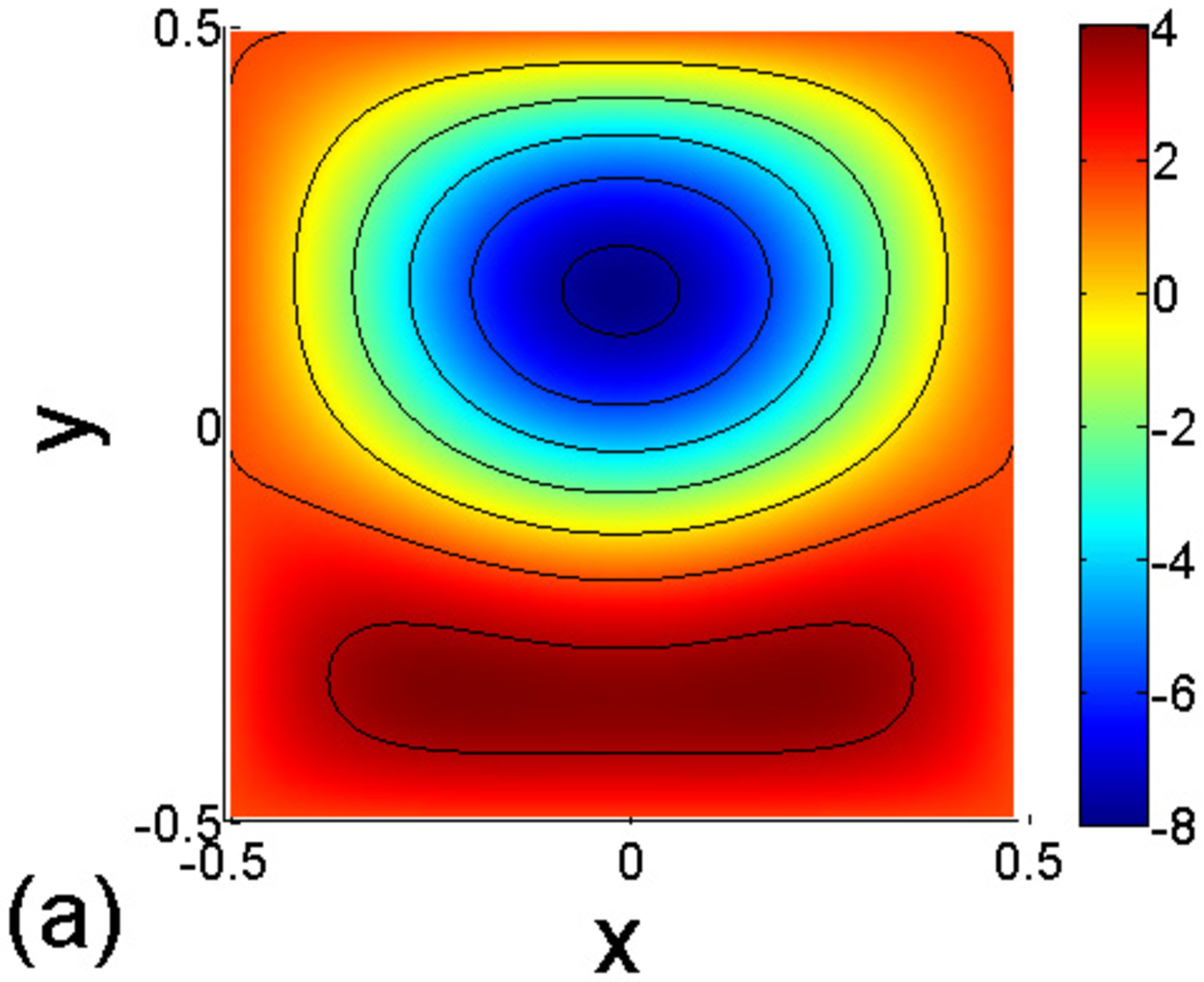}\includegraphics[
width=4.5cm,keepaspectratio]{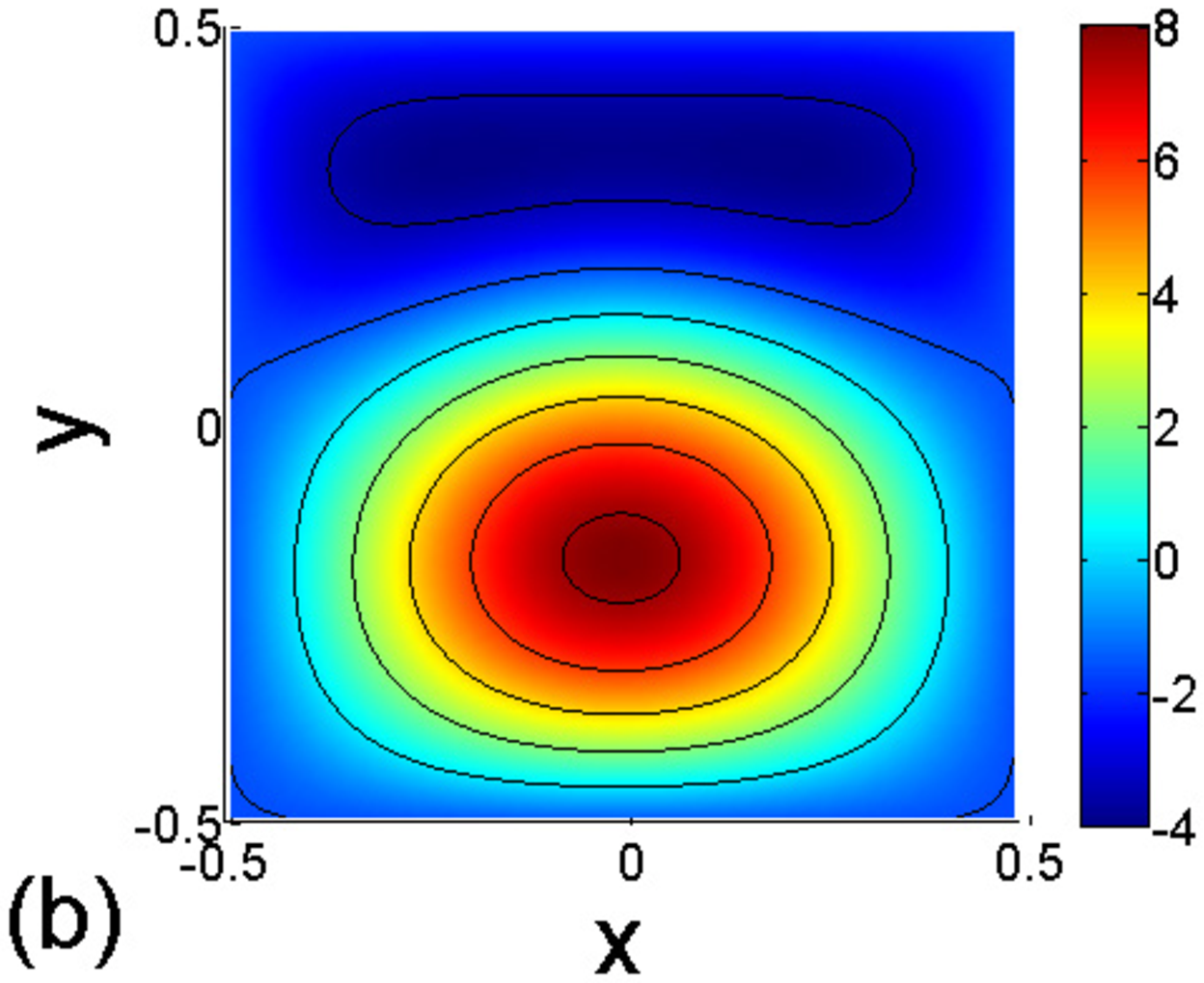}
\includegraphics[width=4.5cm,keepaspectratio]{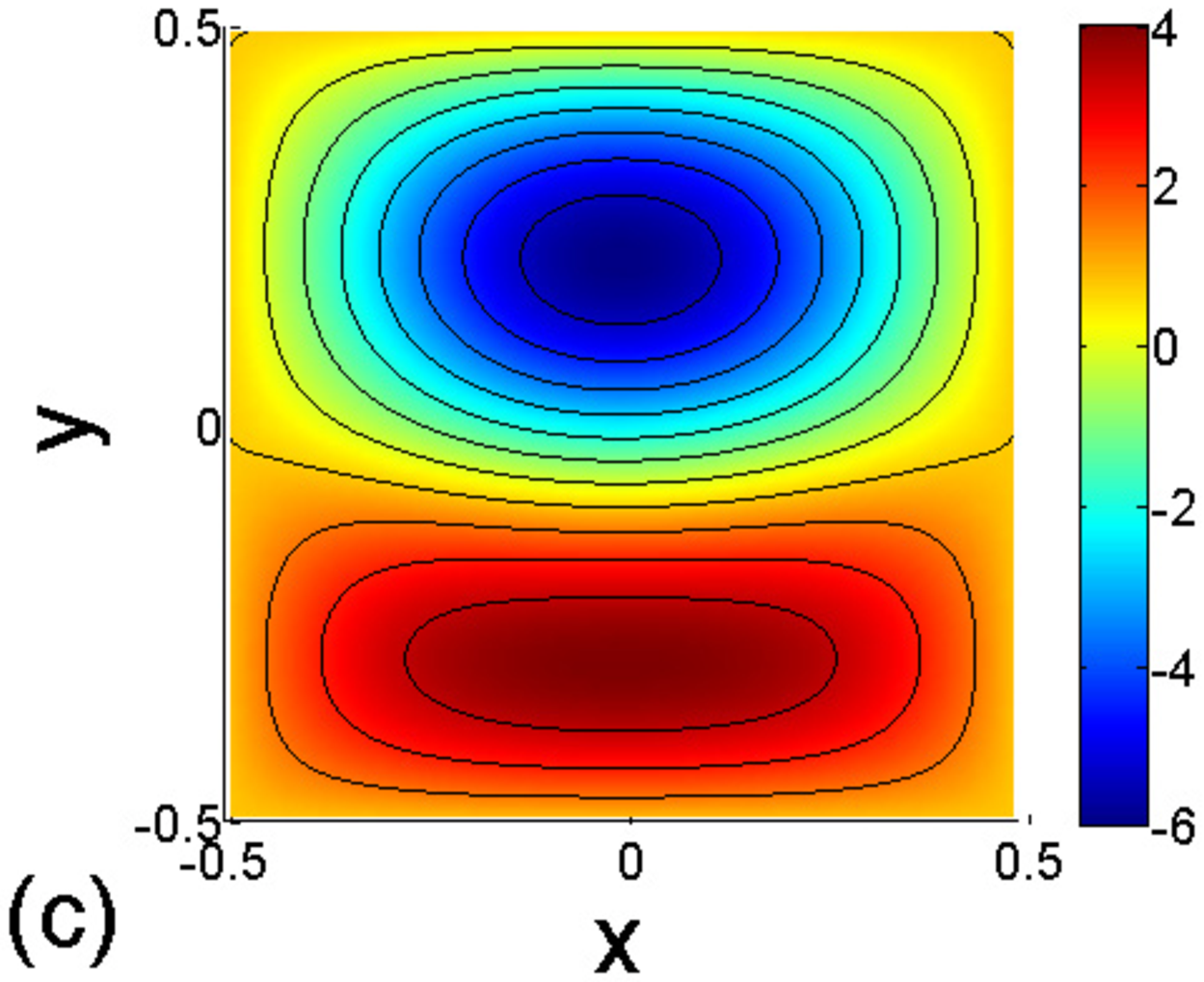}\includegraphics[
width=4.5cm,keepaspectratio]{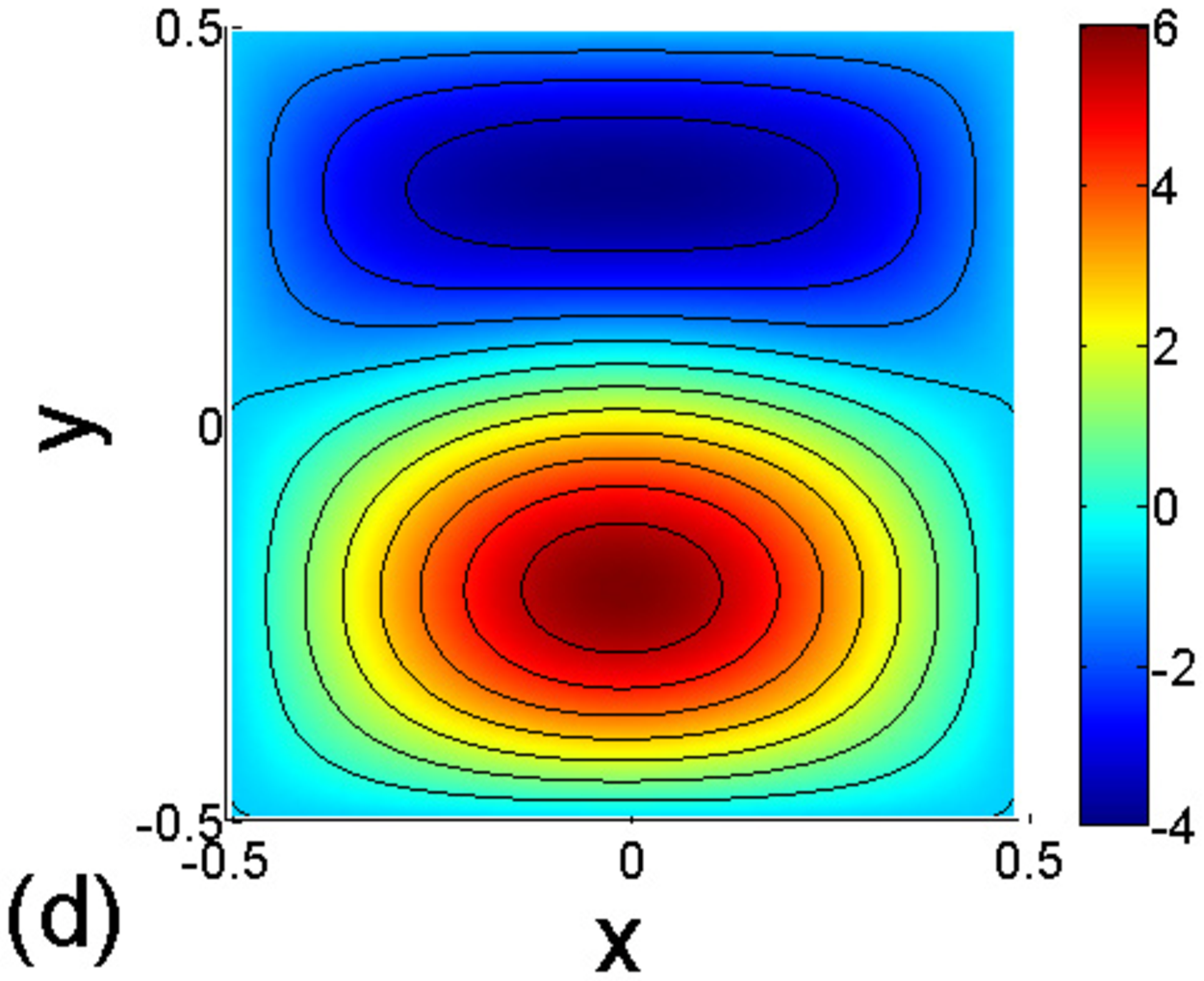}
\caption{\label{pm_a} Potential vorticity $q$ in a square domain
$[-0.5,0.5]\times[-0.5,0.5]$ for $\Gamma=0$, $\beta=\beta_*^{(1)}$ and (a)
$c=0.5$, (b) $c=-0.5$, (c) $c=1$ and (d) $c=-1$. Increasing values from blue to
red.}
\end{figure}

$\bullet$ $1/\tau_c<\tau<\tau_c$, as in Figs. \ref{E_beta_1} and
\ref{E_beta_1z}. For small energies there exists only one
solution of Eq. (\ref{f13}) and it is a global entropy maximum. For
$E\rightarrow 0$, leading to $\beta\rightarrow +\infty$, we recover the
classical Fofonoff solution (see Fig.~\ref{Fofonoff}).  Far from the boundaries,
the Laplacian term in  Eq. (\ref{f13}) can be neglected, which
leads to $\beta\psi=\Gamma+\beta\langle\psi\rangle-by$ and ${\bf
u}=-(b/\beta){\bf x}$, representing a westward jet of velocity
$u=b/\beta$.  The  eastward recirculation can be obtained from a boundary layer
approximation \cite{fofonoff}.  For intermediate energies with $\beta>0$, the
flow involves two symmetric gyres: the gyre at $y>0$ has positive PV and the
gyre at  $y<0$ has negative PV. For intermediate energies and $\beta<0$, the
situation is reversed: the gyre at $y>0$ has negative PV and the gyre at  $y<0$
has positive  PV. All these solutions, forming the upper branch of the main
curve, will be called {\it Fofonoff flows} \footnote{For the sake of simplicity,
 we will call Fofonoff flows all the states with low energy. However, it is
worth mentioning that the strict Fofonoff limit, corresponding to the case where
$\Delta\psi$ is negligible with respect to $\beta\psi$, arises only for very low
energies ($E\rightarrow 0$). For  a linear topography (or $\beta$-effect) in
an antisymmetric domain, this leads to westward jets. For
intermediate energies, and the same topography, the flow consists of two gyres
of opposite sign and  one should rather speak of ``rolls''. For other
topographies, the situation is still different but, at decreasing values of the
energy, the flow always tends to align with the topography.}. They will be
labeled (FP) and (FN) respectively. On the other hand, for large
energies, there is a competition between several solutions of Eq. (\ref{f13}).
When $1/\tau_c<\tau<\tau_c$, the maximum entropy state is the solution with
$\beta=\beta_*$. The solution with $\beta=\beta_*$ and $1/E=0$ is a {\it pure
monopole} as in the study of Chavanis \& Sommeria \cite{jfm}. It can rotate in
either direction since the monopole (MP) with positive PV at the center and the
monopole (MN) with negative PV at the center have the same entropy. For
$\beta=\beta_*$ and $1/E>0$ the monopole is mixed with a Fofonoff flow. It will
be called {\it mixed monopole/Fofonoff flow}. For a fixed value
of $E$, there exists two different solutions depending on the sign of $c$ (see
Fig.~\ref{pm_a}). The caloric curve $\beta(E)$ displays a second order phase
transition marked by the discontinuity of $\frac{\partial\beta}{\partial E}(E)$
at $E=E_*(\tau)$. In conclusion, for $1/E>1/E_*(\tau)$ we have Fofonoff flows
(with $\alpha(E)=0$), for $0<1/E<1/E_*(\tau)$ we have mixed monopole/Fofonoff
flows (with $\pm \alpha(E)\neq 0$) and for $1/E=0$ we have a pure monopole (with
$\alpha(E)\rightarrow \pm\infty$).

\begin{figure}[h]
\center
\includegraphics[width=8cm,keepaspectratio]{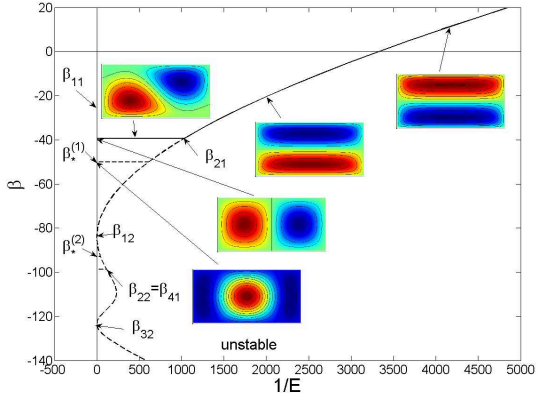}
\caption{\label{E_beta_2} Relationship between $\beta$ and $1/E$ in a
rectangular domain 
$[-1/\sqrt{2},1/\sqrt{2}]\times[-1/(2\sqrt{2}),1/(2\sqrt{2})]$ of aspect ratio
$\tau=2>\tau_c$ for $\Gamma=0$. For small energies we get
Fofonoff flows. For large energies and $\tau>\tau_c$, the
maximum entropy state is the horizontal dipole. There exists a second order
phase transition between Fofonoff flows and horizontal dipoles.}
\end{figure}

$\bullet$ $\tau>\tau_c=1.12$ (horizontally
elongated domains), as in Fig \ref{E_beta_2}. For small and
moderate energies, the situation is similar to that described previously
(Fofonoff flows). On the other hand, for large
energies, the situation is different.  When $\tau>\tau_c$, the maximum entropy
state is the solution with $\beta=\beta_{21}$. The solution with
$\beta=\beta_{21}$ and $1/E=0$ is a {\it pure horizontal dipole} as in the study
of Chavanis \& Sommeria \cite{jfm}. It can rotate in either direction since the
dipole (DP) with positive PV on the left and the dipole (DN) with negative PV on
the left have the same entropy. For $\beta=\beta_{21}$ and $1/E>0$ the dipole is
mixed with a Fofonoff flow. It will be called {\it mixed horizontal
dipole/Fofonoff flow}. The caloric curve $\beta(E)$ displays a second order
phase transition marked by the discontinuity of $\frac{\partial\beta}{\partial
E}(E)$ at $E=E_{21}(\Gamma=0,\tau)$. In conclusion, for
$1/E>1/E_{21}(\Gamma=0,\tau)$ we have Fofonoff flows (with $\alpha(E)=0$), for
$0<1/E<1/E_{21}(\Gamma=0,\tau)$ we have mixed horizontal dipole/Fofonoff flows
(with $\alpha(E)=0$ and $\pm\chi(E)$) and for $1/E=0$ we have a pure horizontal
dipole (with $\alpha(E)=0$ and $\chi\rightarrow \pm\infty$).

\begin{figure}[h]
\center
\includegraphics[width=8cm,keepaspectratio]{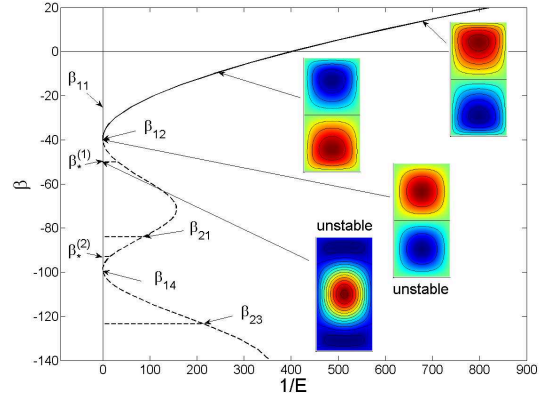}
\caption{\label{E_beta_3} Relationship between $\beta$ and $1/E$ in a
rectangular domain 
$[-1/(2\sqrt{2}),1/(2\sqrt{2})]\times[-1/\sqrt{2},1/\sqrt{2}]$ of aspect ratio
$\tau=1/2<1/\tau_c$ for $\Gamma=0$.  For  $1/E>0$ we get Fofonoff flows. For
$1/E=0$ and $\tau<1/\tau_c$, the maximum entropy state is the vertical dipole.
There is no phase transition. }
\end{figure}

$\bullet$ $\tau\le 1/\tau_c=0.893$
(vertically elongated domains), as in Fig. \ref{E_beta_3}.
For $1/E>0$ we recover Fofonoff flows as discussed previously.  On the other
hand, for $1/E=0$, when $\tau\le 1/\tau_c$, the maximum entropy state is the
{\it pure vertical dipole} with $\beta=\beta_{12}$.  In that case, there is no
phase transition: the Fofonoff flows continuously form a
vertical dipole for $1/E=0$. This can be explained by the fact
that the vertical dipole does not break the symmetry of Fofonoff flows contrary
to the monopoles and the horizontal dipoles in the previous cases. In
conclusion, for $1/E>0$ we have Fofonoff flows (with $\alpha(E)=0$) and for
$1/E\rightarrow 0$ we have a pure vertical dipole (with $\alpha(E)=0$).

\begin{figure}[h]
\center
\includegraphics[width=8cm,keepaspectratio]{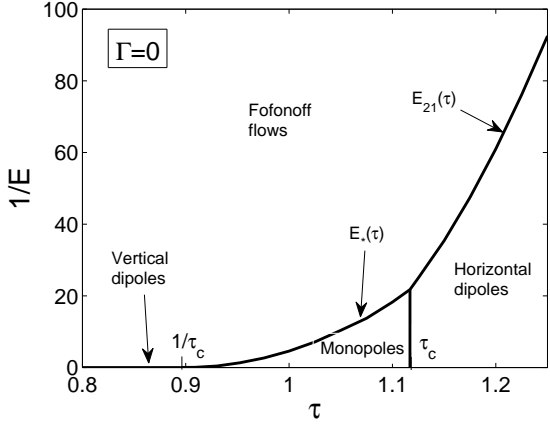}
\caption{\label{phasediag1} Phase diagram for an antisymmetric
topography in the $(\tau,E)$ plane when $\Gamma=0$. Second order phase
transitions arise, from monopoles to Fofonoff flows on the curve $E=E_*(\tau)$,
and from horizontal dipoles to Fofonoff flows on the curve $E=E_{21}(\tau)$. For
$1/E=0$, the monopoles and the dipoles can rotate in either direction (the
solution is degenerate). The line $E_0(\tau)$, corresponding to
$\beta=0$, is not plotted because it occurs for too high values of $1/E$. This
line separates the Fofonoff flows (FP) for $\beta>0$ to the Fofonoff flows (FN)
for $\beta<0$.}
\end{figure}

In Fig. \ref{phasediag1}, we plot the phase diagram in the $(\tau,E)$ plane for
$\Gamma=0$. Concerning the curve $\beta(E)$ at $\Gamma=0$, there is a second
order phase transition between Fofonoff flows and horizontal dipoles for
$\tau>\tau_c$, a second order phase transition between Fofonoff flows and
monopoles for $1/\tau_c<\tau<\tau_c$ and no phase transitions for
$\tau<1/\tau_c$ (the passage from Fofonoff flows to vertical dipoles is
regular).

\subsection{The caloric curve $\beta(E)$ for $\Gamma\neq 0$}
\label{sec_stnonz}

We describe here the caloric curve $\beta(E)$ for $\Gamma\neq
0$. Details on the construction of this curve can be found in
Appendix \ref{techn_S_neq}. Three cases must be considered.

\begin{figure}[h]
\center
\includegraphics[width=8cm,keepaspectratio]{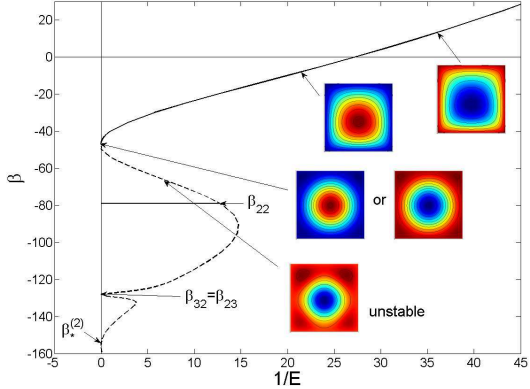}
\caption{\label{E_beta_4} Relationship between $\beta$ and $1/E$ in a square
domain $[-0.5,0.5]\times[-0.5,0.5]$ with $\Gamma=1$ ($1/\tau_c<\tau=1<\tau_c$).
For $1/E>0$, we get asymmetric Fofonoff flows. For $1/E=0$ and
$1/\tau_c<\tau<\tau_c$, the maximum entropy state is the monopole. Note that
there is no plateau at $\beta=\beta_*$ contrary to the case $\Gamma=0$. Thus,
for $\Gamma\neq 0$, there is no phase transition.}
\end{figure}

\begin{figure}[h]
\center
\includegraphics[width=8cm,keepaspectratio]{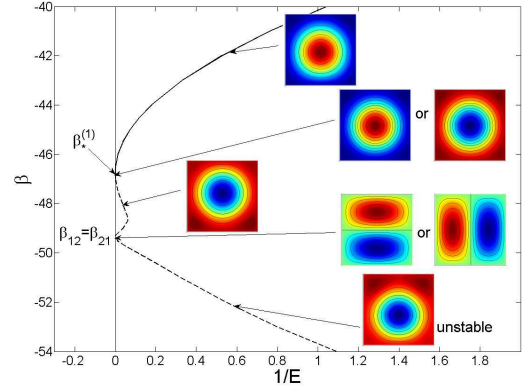}
\caption{\label{E_beta_4z} Zoom of Fig. \ref{E_beta_4}.}
\end{figure}

\begin{figure}[h]
\center
\includegraphics[width=8cm,keepaspectratio]{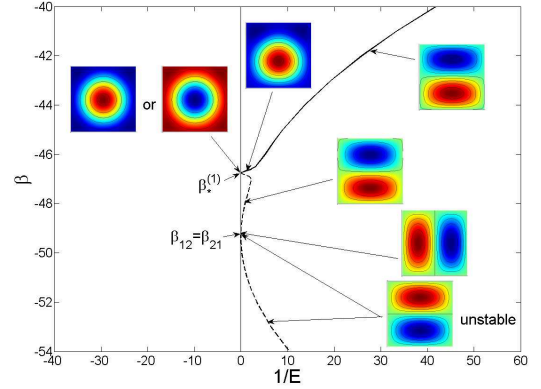}
\caption{\label{E_beta_5} Relationship between $\beta$ and $1/E$ in a square
domain $[-0.5,0.5]\times[-0.5,0.5]$ with $\Gamma=0.01$. For $\Gamma\rightarrow
0$, the curve is more and more pinched near $\beta_*$ explaining the formation
of a plateau for $\Gamma=0$.}
\end{figure}

$\bullet$ $1/\tau_c<\tau<\tau_c$, as in Figs. \ref{E_beta_4},
\ref{E_beta_4z} and \ref{E_beta_5}. For $1/E>0$, the maximum
entropy state is an {\it asymmetric Fofonoff flow} and for  $1/E=0$, the
maximum entropy state is the monopole $\beta_*$. Since there is no plateau, the
caloric curve $\beta(E)$ does not display any phase transition: for $\Gamma\neq
0$, $\beta(E)$ and $\frac{\partial\beta}{\partial E}(E)$ are continuous. This is
different from the case $\Gamma=0$.  In conclusion, for $1/E>0$ we have
asymmetric Fofonoff flows (with $\alpha(E)\neq 0$) and for $1/E=0$ we have a
pure monopole (with $\alpha(E)=\infty$) rotating in either direction.
Interestingly, when $\Gamma\rightarrow 0$ (see Fig.
\ref{E_beta_5}), the main curve is more and more ``pinched'' near the point 
($1/E=0$, $\beta=\beta_*)$. This is consistent with the formation of a plateau
(second order phase transition) when $\Gamma=0$.

\begin{figure}[h]
\center
\includegraphics[width=8cm,keepaspectratio]{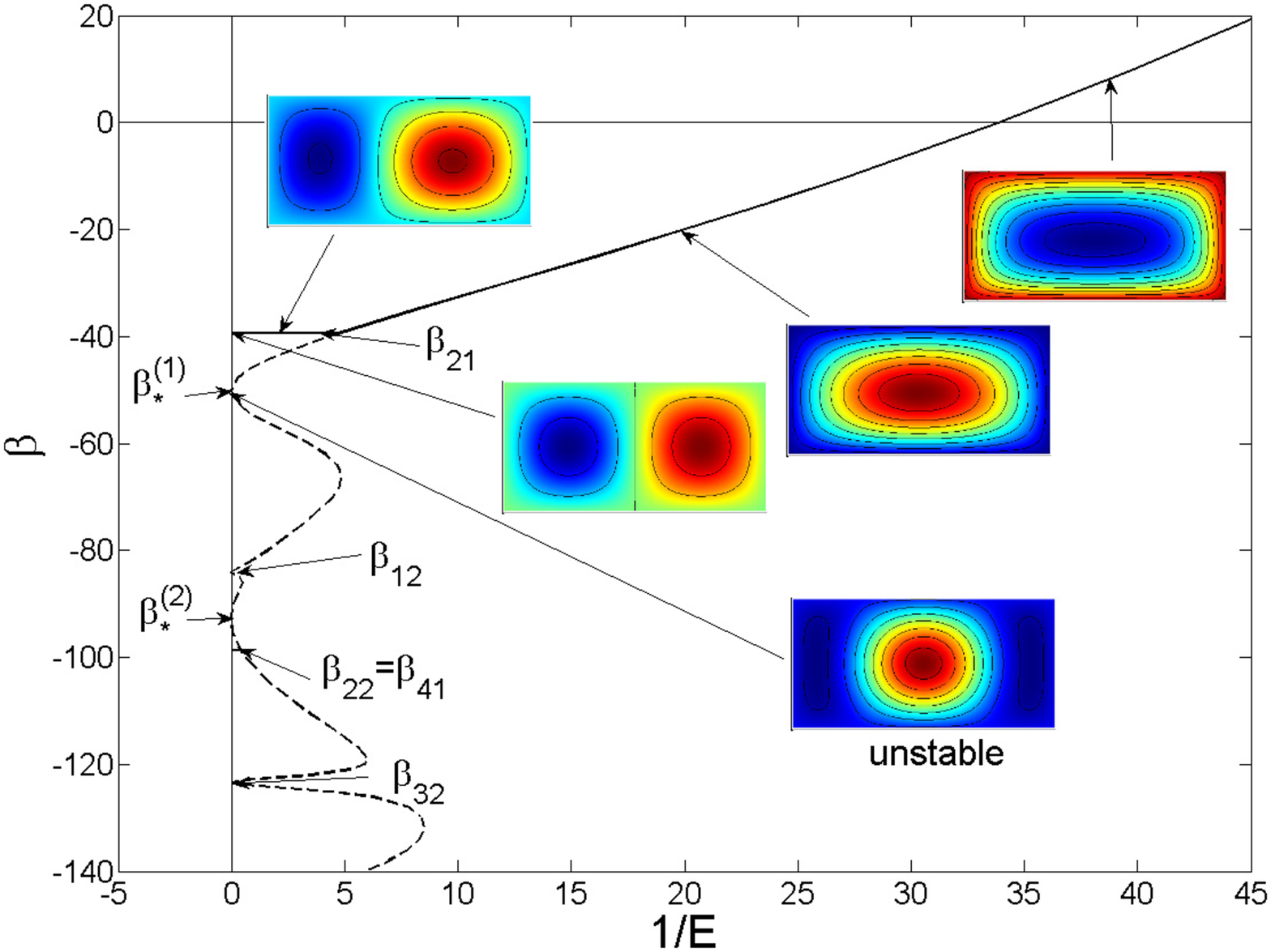}
\caption{\label{E_beta_6} Relationship between $\beta$ and $1/E$  in a
rectangular domain 
$[-1/\sqrt{2},1/\sqrt{2}]\times[-1/(2\sqrt{2}),1/(2\sqrt{2})]$ of aspect ratio
$\tau=2>\tau_c$ with $\Gamma=1$. For small energies, we get
asymmetric Fofonoff flows. For large energies and
$\tau>\tau_c$, the maximum entropy state is the horizontal dipole. There exists
a second order phase transition between Fofonoff flows and horizontal dipoles.}
\end{figure}

\begin{figure}[h]
\center
\includegraphics[width=8cm,keepaspectratio]{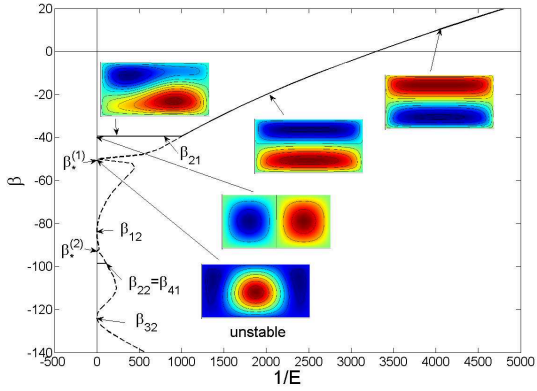}
\caption{\label{E_beta_7} Relationship between $\beta$ and $1/E$ in a
rectangular domain
$[-1/\sqrt{2},1/\sqrt{2}]\times[-1/(2\sqrt{2}),1/(2\sqrt{2})]$ of aspect ratio
$\tau=2$ with $\Gamma=0.01$.}
\end{figure}

$\bullet$ $\tau>\tau_c$ (horizontally
elongated domains), as in Figs. \ref{E_beta_6} and \ref{E_beta_7}. For
small energies, the maximum entropy state is
an asymmetric Fofonoff flow. For $1/E=0$, the maximum entropy state is the
horizontal dipole $\beta_{21}$. For intermediate energies, the maximum entropy
state is a mixed horizontal dipole/asymmetric Fofonoff solution. These solutions
 form a plateau $\beta=\beta_{21}$. In that case, the caloric curve $\beta(E)$
displays a second order phase transition marked by the discontinuity of
$\frac{\partial\beta}{\partial E}(E)$ at $E=E_{21}(\Gamma,\tau)$. In conclusion,
for $1/E>1/E_{21}(\Gamma,\tau)$ we have asymmetric Fofonoff flows (with
$\alpha(E)\neq 0$), for $0<1/E<1/E_{21}(\Gamma,\tau)$ we have mixed horizontal
dipoles/asymmetric Fofonoff flows  (with $\alpha(E)\neq 0$ and $\pm\chi(E)$) and
for $1/E=0$ we have a pure horizontal dipole (with $\alpha(E)\neq 0$ and
$\chi\rightarrow +\infty$) rotating in either direction.

\begin{figure}[h]
\center
\includegraphics[width=8cm,keepaspectratio]{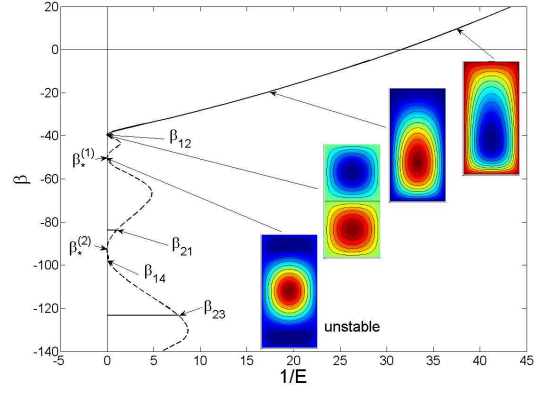}
\caption{\label{E_beta_8} Relationship between $\beta$ and $1/E$  in a
rectangular domain 
$[-1/(2\sqrt{2}),1/(2\sqrt{2})]\times[-1/\sqrt{2},1/\sqrt{2}]$ of aspect ratio
$\tau=1/2<1/\tau_c$ with $\Gamma=1$. For $1/E>0$, we get asymmetric Fofonoff
flows and for  $1/E=0$ and $\tau<1/\tau_c$, the maximum entropy state is the
vertical dipole. There is no phase transition.}
\end{figure}

\begin{figure}[h]
\center
\includegraphics[width=8cm,keepaspectratio]{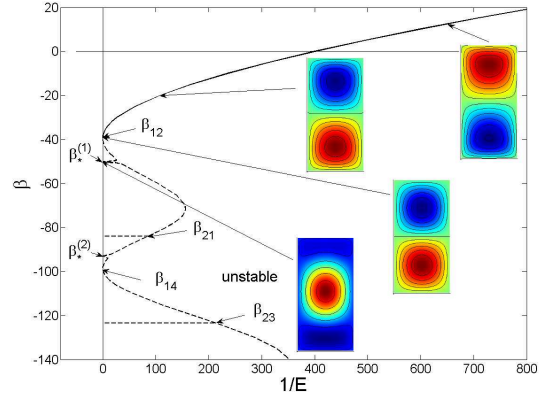}
\caption{\label{E_beta_9} Relationship between $\beta$ and $1/E$ in a
rectangular domain
$[-1/(2\sqrt{2}),1/(2\sqrt{2})]\times[-1/\sqrt{2},1/\sqrt{2}]$ of aspect ratio
$\tau=1/2$ with $\Gamma=0.01$.}
\end{figure}

$\bullet$ $\tau<1/\tau_c$ (vertically
elongated domains), as in Figs. \ref{E_beta_8} and \ref{E_beta_9}. For $1/E>0$,
the maximum entropy state is an asymmetric Fofonoff flow and
for  $1/E=0$, the maximum entropy state is the vertical dipole $\beta_{12}$.
Since there is no plateau, there is no phase transition in that case.  In
conclusion, for $1/E>0$ we have asymmetric Fofonoff flows (with $\alpha(E)\neq
0$) and for $1/E=0$ we have a vertical dipole (with $\alpha(E)\neq 0$).

\begin{figure}[h]
\center
\includegraphics[width=8cm,keepaspectratio]{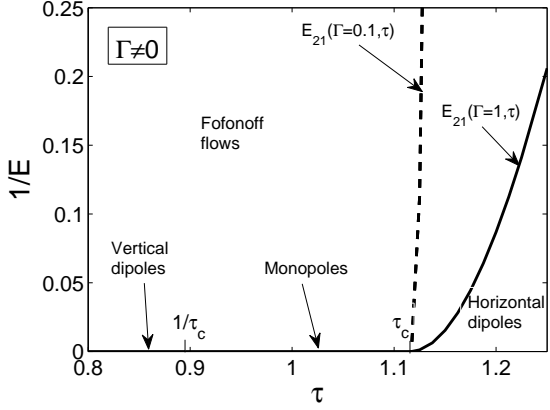}
\caption{\label{phasediag2} Phase diagram  for an antisymmetric
topography in the $(\tau,E)$ plane for different values of $\Gamma\neq 0$.
Second order phase transitions arise, from horizontal dipoles to Fofonoff flows,
on the curve $E=E_{21}(\Gamma,\tau)$.  For $1/E=0$, the monopoles and the
dipoles can rotate in either direction (the solution is degenerate).
The line $E_0(\tau)$, corresponding to $\beta=0$, is not
plotted because it occurs for too high values of $1/E$. This line separates the
Fofonoff flows (FP) for $\beta>0$ to the Fofonoff flows (FN) for $\beta<0$.}
\end{figure}

In Fig. \ref{phasediag2}, we plot the phase diagram in the $(\tau,E)$ plane for
different values of $\Gamma\neq 0$. Concerning the curve $\beta(E)$ at fixed
$\Gamma\neq 0$, there is a second order phase transition between Fofonoff flows
and horizontal dipoles for $\tau>\tau_c$ and no phase transitions for
$\tau<\tau_c$ (the passage from Fofonoff flows to monopoles and  vertical
dipoles is regular).

\subsection{The chemical potential curve $\alpha(\Gamma)$ for fixed $E$}
\label{sec_cps}

In the previous sections, we have studied the caloric curve $\beta(E)$ for a
fixed circulation $\Gamma$. We shall now study the chemical potential curve
$\alpha(\Gamma)$ for a fixed energy $E$. The general equations determining the
chemical potential are given by Eqs. (\ref{chem4}) and (\ref{chem5}). For
an antisymmetric topography, using Eq. (\ref{f30}), the term in
factor of $c$ in Eq. (\ref{chem4}) vanishes and the foregoing equations reduce
to
\begin{eqnarray}
\label{chem}
\frac{\alpha^2}{b^2}=\frac{1}{c^2}=\frac{\frac{2E}{b^2}
+\beta\langle\phi_2^2\rangle+\langle\phi_2
y\rangle}{\langle\phi_1\rangle-\beta\langle\phi_1^2\rangle},\qquad
\end{eqnarray}
\begin{eqnarray}
\label{chemb}
\frac{\Gamma}{b}=\frac{1}{c(E,\beta)}(1-\beta\langle\phi_1\rangle).
\end{eqnarray}
In that case, it is very easy to obtain the chemical potential curve
$\alpha(\Gamma)$ for fixed $E$, parameterized by $\beta$. This curve is
antisymmetric with respect to $\Gamma\rightarrow -\Gamma$. As in Secs.
\ref{sec_stz} and \ref{sec_stnonz}, three cases must be
considered.

\begin{figure}[h]
\center
\includegraphics[width=8cm,keepaspectratio]{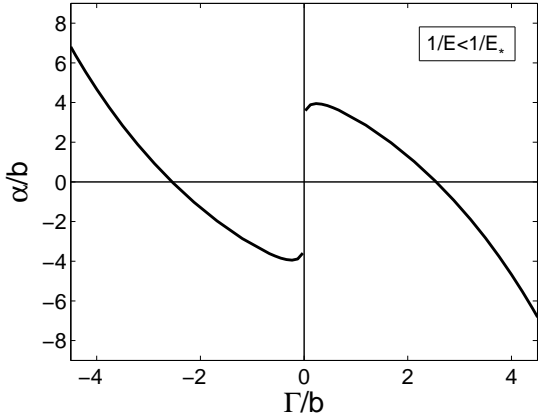}
\caption{\label{alpha1} Chemical potential as a function of the circulation in a
square domain ($1/\tau_c<\tau=1<\tau_c$). For $1/E<1/E_*(\tau)$, there is a
first order phase transition between positive and negative
monopoles (here $1/E=2$ and $1/E_*(\tau=1)\approx 4.56$).}
\end{figure}

\begin{figure}[h]
\center
\includegraphics[width=8cm,keepaspectratio]{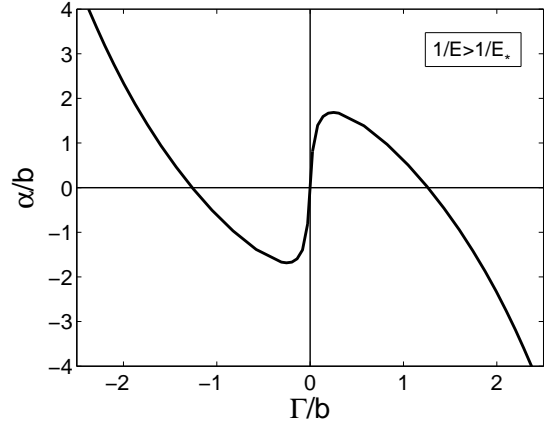}
\caption{\label{alpha2} Chemical potential as a function of the circulation in a
square domain ($1/\tau_c<\tau=1<\tau_c$). For $1/E>1/E_*(\tau)$, there is no
phase transition (here $1/E=8$ and $1/E_*(\tau=1)\approx 4.56$).}
\end{figure}

$\bullet$ $1/\tau_c<\tau<\tau_c$, as in Figs. \ref{alpha1} and
\ref{alpha2}. For $\Gamma=0$ and $1/E<1/E_*(\tau)$, we are on the plateau
$\beta=\beta_*$ of mixed monopoles/Fofonoff flows. There exists two solutions
for each energy that have two symmetric values of the chemical potential
$\pm\alpha(E)\neq 0$ (see Fig. \ref{pm_a}). Their values are given by  Eqs.
(\ref{chem3}) and (\ref{chem4}) replacing $\beta$ by $\beta_*$. At the end of
the plateau, for $1/E=1/E_*(\tau)$, we get $\pm\alpha=0$. For $\Gamma=0$ and
$1/E>1/E_*(\tau)$, we are on the main branch and the chemical potential is equal
to $\alpha(E)=0$ for each energy. For $\Gamma\neq 0$ there is a unique solution
$\alpha(E)$ for each energy. In conclusion, for $1/E<1/E_*(\tau)$,
the chemical potential curve displays a first order phase
transition at $\Gamma=0$ marked by the discontinuity of $\alpha(\Gamma)$. This
corresponds to the transition from the monopole (MP) for $\Gamma=0^+$ to the
monopole (MN) for $\Gamma=0^-$.  For  $1/E>1/E_*(\tau)$, the curve
$\alpha(\Gamma)$ is continuous and differentiable so there is no phase
transition.

\begin{figure}[h]
\center
\includegraphics[width=8cm,keepaspectratio]{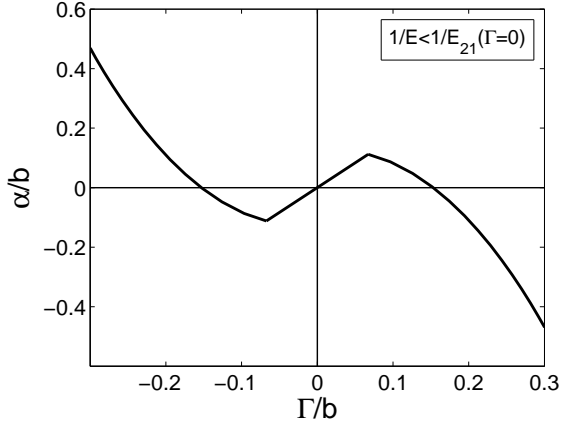}
\caption{\label{alpha3} Chemical potential as a function of the circulation in a
rectangular domain of aspect ratio $\tau=2>\tau_c$. For 
$1/E<1/E_{21}^{min}(\tau)$, there are two second order phase transitions
between horizontal dipoles and Fofonoff flows (here $1/E=500$
and $1/E_{21}^{min}(\tau=2)\approx 1045$).}
\end{figure}

\begin{figure}[h]
\center
\includegraphics[width=8cm,keepaspectratio]{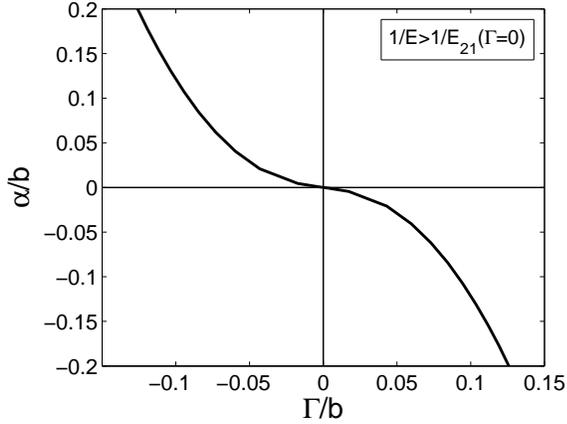}
\caption{\label{alpha4} Chemical potential as a function of the circulation in a
rectangular domain of aspect ratio $\tau=2>\tau_c$. For 
$1/E>1/E_{21}^{min}(\tau)$, there is no phase transition (here $1/E=2000$ and
$1/E_{21}^{min}(\tau=2)\approx 1045$). }
\end{figure}

$\bullet$ $\tau>\tau_c$, as in Figs. \ref{alpha3} and
\ref{alpha4}. For any $\Gamma$ and $1/E\le 1/E_{21}(\Gamma,\tau)$, we are on
the plateau $\beta=\beta_{21}$ of mixed horizontal dipole/Fofonoff flows. There
exists two solutions for each energy (with $\pm\chi(E)$),  but  they have the
same value of chemical potential $\alpha(E)$.  For  $1/E>1/E_{21}(\Gamma,\tau)$,
we are on the main branch and the chemical potential takes a unique value 
$\alpha(E)$, corresponding to a unique solution, for each energy.  For
$\Gamma=0$, we always have $\alpha(E)=0$. Since $E_{21}(\Gamma,\tau)$ is a
parabola of the form $E_{21}(\Gamma,\tau)=a(\tau)\Gamma^2+c(\tau)$ with $a\ge
0$, its minimum value is $E_{21}^{min}(\tau)=E_{21}(\Gamma=0,\tau)$ obtained for
$\Gamma=0$. Let us now assume $1/E<1/E_{21}^{min}(\tau)$. Then, as long as
$|\Gamma|<\Gamma_{21}(E)$ (so that $1/E<1/E_{21}(\Gamma)$), we are on the
plateau $\beta=\beta_{21}$ and the chemical potential $\alpha(\Gamma)$ is a
linear function of the circulation given by Eqs. (\ref{chem3}) and (\ref{chem5})
where $\beta$ is replaced by $\beta_{21}$. For $|\Gamma|>\Gamma_{21}(E)$, we are
on the main branch and $\alpha(\Gamma)$ is given by Eqs. (\ref{chem3}),
(\ref{chem4}) and (\ref{chem5}). In conclusion, if $1/E<1/E_{21}^{min}(\tau)$,
the chemical potential curve displays two second order phase
transitions between horizontal dipoles and Fofonoff flows at
$\Gamma=\pm\Gamma_{21}(E)$ marked by the discontinuity of
$\frac{\partial\alpha}{\partial \Gamma}(\Gamma)$. If $1/E>1/E_{21}^{min}(\tau)$
there is no phase transition.

\begin{figure}[h]
\center
\includegraphics[width=8cm,keepaspectratio]{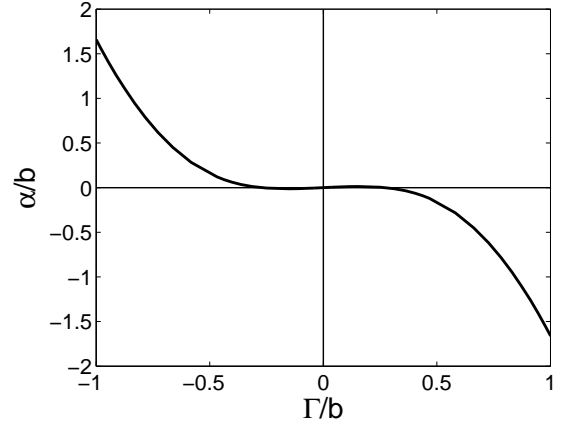}
\caption{\label{alpha5} Chemical potential as a function of the circulation in a
rectangular domain of aspect ratio $\tau=1/2<1/\tau_c$. There is no phase
transition (here $1/E=50$).}
\end{figure}

$\bullet$ $\tau<1/\tau_c$, as in Fig. \ref{alpha5}. For any
$\Gamma$, we are on the main branch and the chemical potential takes a unique
value  $\alpha(E)$ for each energy.  For $\Gamma=0$, we always have
$\alpha(E)=0$. Furthermore the curve $\alpha(\Gamma)$ is continuous and
differentiable. In conclusion, there is  no phase transition.

It is interesting to mention the existence of particular points
in the phase diagram. (i) For $1/\tau_c<\tau<\tau_c$, 
$E=E_*(\tau)$ is a {\it critical point} at which a first order phase transition
appears (see Figs.~\ref{alpha1} and \ref{alpha2}).  (ii) For energies such that
$1/E<1/E_{21}(\Gamma=0,\tau_c)=1/E_*(\tau_c)$, $(\tau=\tau_c,\Gamma=0)$ is a
{\it bicritical point}: for fixed energy, the system exhibits
a bifurcation from a first order phase transition (see Fig.~\ref{alpha1}) to two
second order phase transitions (see Fig.~\ref{alpha3}) when increasing the
aspect ratio of the domain. (iii) For $\tau >\tau_c$, a {\it second order
azeotropy} arises at $E=E_{21}(\Gamma=0,\tau)$, where two second order phase
transitions appear simultaneously from nothing (see
Fig.~\ref{alpha3} and \ref{alpha4}). The possible existence of these behaviors
in systems with long range interactions was predicted by Bouchet \& Barr\'e
\cite{bouchet_barre}. It was evidenced by Venaille \& Bouchet \cite{vb,vbf} for
Euler and geophysical flows by extending the work of Chavanis \& Sommeria
\cite{jfm}. We refer to these works for a more detailed description of these
phase transitions. We note that the critical point and the
second order azeotropy are specific to flows with topography while the 
bicritical point also exists when $h=0$.

\section{The case of a non-symmetric linear topography}
\label{sec_nst}

In the previous section, we have considered the case of a linear
topography $h=by$ that is antisymmetric with respect to the
middle axis
$y=0$ of a rectangular domain (these results remain valid for more
general antisymmetric topographies of the form $h=b H(y)$). We
now
consider the case of a linear topography $h=b(y-y_0)$, i.e. $H=y-y_0$,
with $y_0\neq 0$ that is non-symmetric (the following results remain
valid for more general non-symmetric topographies of the form $h=b
H(y)$).  We shall see that the details of calculations are a bit
different while the structure of the main curves remains finally
unchanged.

For a non-symmetric topography, there is no particular
simplification of the equations of the problem. Therefore, we must use the
general equations of Sec. \ref{sec_cc}. It
has been shown in Sec. \ref{sec_cc} that there exists a critical circulation
$\Gamma_*$. This critical circulation depends on the aspect ratio $\tau$ of the
domain \footnote{For example, for a linear topography $h=by$ in a semi-basin
$0\le y\le 1/\sqrt{\tau}$, we get $\Gamma_*(\tau)=b/(2\sqrt{\tau})$. Indeed,
this is equivalent to a complete basin $-\sqrt{\tau}\le y\le 1/\sqrt{\tau}$ with
a topography $h=by+b/(2\sqrt{\tau})$. This is in turn equivalent to a complete
basin with topography $h=by$ and a vorticity $\omega'=\omega- b/(2\sqrt{\tau})$.
The new circulation is $\Gamma'=\Gamma-b/(2\sqrt{\tau})$. Since $\Gamma'_*=0$
according to Sec. \ref{sec_st}, we get $\Gamma_*=b/(2\sqrt{\tau})$. For a
topography $h=b(y-y_0)$, we get $\Gamma_*=-b y_0$.} and on the form of the
topography. We shall consider successively the cases $\Gamma\neq \Gamma_*(\tau)$
and $\Gamma=\Gamma_*(\tau)$.

\subsection{The caloric curve $\beta(E)$ for $\Gamma\neq \Gamma_*$}
\label{sec_nsg}

The series of equilibria for a non-symmetric topography when $\Gamma\neq
\Gamma_*$ are  similar to the ones obtained in Sec.
\ref{sec_stnonz} for an antisymmetric topography when
$\Gamma\neq 0$  (recall that $\Gamma_*=0$ in an antisymmetric
domain). Details on their construction are given in Appendix
\ref{techn_NS_neq}.

$\bullet$ 1st case: $1/\tau_c<\tau<\tau_c$. For $1/E>0$ we have
asymmetric Fofonoff flows (with $\alpha(E)\neq 0$) and for $1/E=0$ we have a
pure monopole (with $\alpha(E)=\infty$). The caloric curve $\beta(E)$ does not
display any phase transition.

$\bullet$ 2nd case: $\tau>\tau_c$. For
$1/E>1/E_{21}(\Gamma,\tau)$ we have asymmetric Fofonoff flows (with
$\alpha(E)\neq 0$), for $0<1/E<1/E_{21}(\Gamma,\tau)$ we have mixed horizontal
dipoles/Fofonoff flows  (with $\alpha(E)\neq 0$ and $\chi_{1,2}(E)$) and for
$1/E=0$ we have a pure horizontal dipole (with $\alpha(E)\neq 0$ and
$\chi\rightarrow +\infty$). The caloric curve $\beta(E)$ displays a second order
phase transition at the energy $E_{21}(\Gamma,\tau)$.

$\bullet$ 3rd case: $\tau<1/\tau_c$. For $1/E>0$, we have
asymmetric Fofonoff flows (with $\alpha(E)\neq 0$) and for $1/E=0$ we have a
vertical dipole (with $\alpha(E)\neq 0$). There is no phase transition.

\subsection{The caloric curve $\beta(E)$ for $\Gamma=\Gamma_*$}
\label{sec_cchj}

The series of equilibria in a non-symmetric domain when $\Gamma= \Gamma_*$ are
similar to the ones obtained in Sec.  \ref{sec_stz} for
an antisymmetric domain when $\Gamma= 0$ (recall that
$\Gamma_*=0$ in an antisymmetric domain).
Details on their construction are given in  Appendix
\ref{techn_NS_eq}. They can also be understood from the curves  $\Gamma\neq
\Gamma_*$ of Sec. \ref{sec_nsg} by considering the limit $\Gamma\rightarrow
\Gamma_*$. When $\Gamma\rightarrow \Gamma_*$, the main curve is more and more
``pinched'' near the point  ($1/E=0$, $\beta=\beta_*)$ and for $\Gamma=\Gamma_*$
a plateau appears at temperature $\beta=\beta_*$ between $1/E=0$ and
$1/E=1/E_*(\tau)$.

$\bullet$ If $1/\tau_c<\tau<\tau_c$: for $1/E>1/E_*(\tau)$ we
have Fofonoff flows (with $\alpha(E)\neq 0$), for $0<1/E<1/E_{*}(\tau)$ we have
mixed monopoles/Fofonoff flows  (with $\alpha_{1,2}(E)\neq 0$) and for $1/E=0$
we have a pure monopole (with $\alpha_{1,2}(E)\rightarrow \pm\infty$). The
caloric curve $\beta(E)$ displays a second order phase transition at the energy
$E_{*}(\tau)$.

$\bullet$ If $\tau>\tau_c$: for $1/E>1/E_{21}(\Gamma_*,\tau)$
we have Fofonoff flows (with $\alpha(E)\neq 0$), for
$0<1/E<1/E_{21}(\Gamma_*,\tau)$ we have mixed horizontal dipoles/Fofonoff flows 
(with $\alpha(E)\neq 0$ and $\chi_{1,2}(E)$) and for $1/E=0$ we have a pure
horizontal dipole (with $\alpha(E)\neq 0$ and $\chi\rightarrow +\infty$).  The
caloric curve $\beta(E)$ displays a second order phase transition at the energy
$E_{21}(\Gamma_*,\tau)$.

$\bullet$ If $\tau<1/\tau_c$: for $1/E>0$ we have  Fofonoff
flows (with $\alpha(E)\neq 0$) and for $1/E=0$ we have a vertical dipole (with
$\alpha(E)\neq 0$). There is no phase transition.

\subsection{The chemical potential curve $\alpha(\Gamma)$ for fixed $E$}

In a non-symmetric domain, the results are similar to those obtained in Sec.
\ref{sec_cps} even if the general equations are a little more complicated and
the curve $\alpha(\Gamma)$ is non-symmetric.

$\bullet$ If $1/\tau_c<\tau<\tau_c$: for $1/E<1/E_*(\tau)$, there is a first
order phase transition at $\Gamma=\Gamma_*$ marked by the discontinuity of
$\alpha(\Gamma)$. For  $1/E>1/E_*(\tau)$, the curve $\alpha(\Gamma)$ is
continuous and differentiable so there is no phase transition.

$\bullet$ If $\tau>\tau_c$: for $1/E<1/E_{21}^{min}(\tau)$, there are two 
second order phase transitions at $\Gamma=\Gamma_{21}^{-}(E)$ and
$\Gamma=\Gamma_{21}^{+}(E)$ marked by the discontinuity of
$\frac{\partial\alpha}{\partial \Gamma}(\Gamma)$. If $1/E>1/E_{21}^{min}(\tau)$,
there is no phase transition.

$\bullet$ If $\tau<1/\tau_c$:  there is  no phase transition.\\

In conclusion: (i) for $1/\tau_c<\tau<\tau_c$, $E_*(\tau)$ is
a critical point; (ii)  $(\tau=\tau_c,\Gamma=\Gamma_*(\tau_c))$ is a bicritical
point; (iii) a second order azeotropy arises for $\tau >\tau_c$ at
$E=E_{21}^{min}(\tau)$.

\section{Summary and generalizations}
\label{sec_sg}

We shall now generalize the previous results to the case of an
arbitrary domain and an arbitrary topography $h(x,y)=bH(x,y)$. Some
illustrations of phase transitions in geophysical flows with different
topographies are given in
\cite{cnd}. We here develop the theory needed to interpret them.

In the series of equilibria containing all the critical points of
entropy at fixed circulation and energy, we must distinguish:

$\bullet$ The main curve $\beta(E)$: each point of this curve corresponds to a
unique solution with a unique value of the chemical potential $\alpha$. For
$1/E\rightarrow +\infty$ (small energies), leading to $\beta\rightarrow
+\infty$, we obtain Fofonoff flows with an arbitrary topography. Far from the
boundaries, we can neglect the Laplacian term in  Eq. (\ref{f13}) leading to a
stream function $\beta\psi=\Gamma+\beta\langle\psi\rangle-bH$ and a velocity
field ${\bf u}=\frac{b}{\beta}{\bf z}\times \nabla H$. The recirculation at the
boundary can be obtained from a boundary layer approximation.

$\bullet$ The inverse temperature $\beta=\beta_*$: for $\Gamma\neq \Gamma_*$,
there is only one solution with $\beta=\beta_*$ that exists at $1/E=0$. This is
a limit point of the main curve. For $\Gamma=\Gamma_*$,  the solutions with
$\beta=\beta_*$  form at plateau going from $1/E=0$ to $1/E_{*}$. On this
plateau, each value of the energy $1/E$ determines two solutions with the same
temperature $\beta_*$ but with different chemical potentials $\alpha_{1}(E)$ and
$\alpha_{2}(E)$.

$\bullet$ The inverse temperature $\beta=\beta''_1$ (largest eigenvalue with
$\langle\psi''_1\rangle\neq 0$): this is a particular point of the main curve
that corresponds to $1/E''_{1}(\Gamma)$ where $E''_{1}(\Gamma)$ is a parabola.
This point marks the domain of inequivalence between the grand canonical
ensemble on the one hand and the canonical and microcanonical ensembles on the
other hand (see \cite{vb,vbf} and Sec. \ref{sec_ths}).

$\bullet$ The inverse temperature $\beta=\beta'_{1a}$ (largest eigenvalue with
$\langle\psi'_{1a}\rangle= 0$ and $\langle H\psi'_{1a}\rangle= 0$; eigenmode
with zero mean orthogonal to the topography): these solutions  form a plateau
going from $1/E=0$ to $1/E'_{1a}(\Gamma)$ where $E'_{1a}(\Gamma)$ is a parabola.
On this plateau, each value of the energy $1/E$ determines two solutions with
the same temperature $\beta'_{1a}$, the same chemical potential $\alpha(E)$ but
with different  mixing coefficients $\chi_{1}(E)$ and $\chi_{2}(E)$.

$\bullet$ The inverse temperature $\beta=\beta'_{1b}$ (largest eigenvalue with
$\langle\psi'_{1b}\rangle= 0$ and $\langle H\psi'_{1b}\rangle\neq  0$; eigenmode
with zero mean non orthogonal to the topography): it exists only at $1/E=0$.
This is a limit point of the main curve.

\begin{figure}[h]
\center
\includegraphics[width=8cm,keepaspectratio]{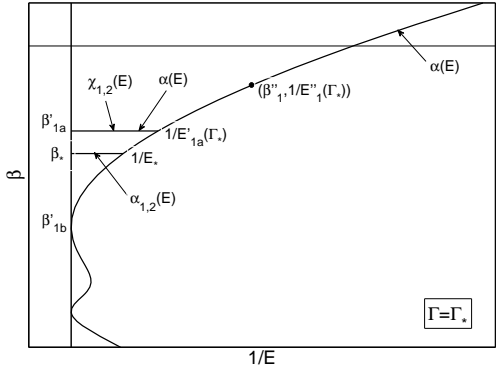}
\caption{\label{rec1} Relationship between $\beta$ and $1/E$ for
$\Gamma=\Gamma_*$. In a rectangular domain with a linear topography, $\beta''_1$
corresponds to $\beta_{11}$, $\beta'_{1a}$ corresponds to $\beta_{21}$ and 
$\beta'_{1b}$ corresponds to $\beta_{12}$. }
\end{figure}

\begin{figure}[h]
\center
\includegraphics[width=8cm,keepaspectratio]{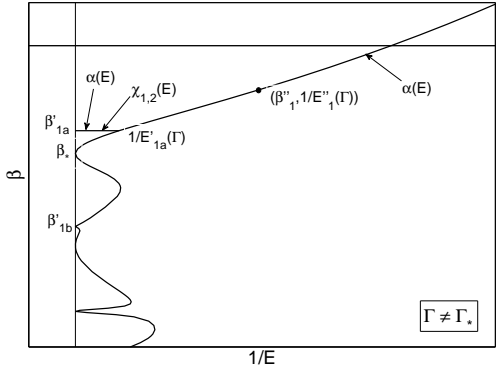}
\caption{\label{rec2} Relationship between $\beta$ and $1/E$ for
$\Gamma\neq\Gamma_*$.}
\end{figure}

The series of equilibria showing these different solutions is represented
schematically in Fig. \ref{rec1} for  $\Gamma= \Gamma_*$ and in Fig. \ref{rec2} 
for $\Gamma\neq \Gamma_*$. In the domain of competition, where there exists
several solutions for the same energy, we must select the solution with the
largest $\beta$, which is the maximum entropy state. Thus, in this range,
$\beta=\max\lbrace \beta_*,\beta'_{1a},\beta'_{1b}\rbrace$. Note that the value
of $\max\lbrace \beta_*,\beta'_{1a},\beta'_{1b}\rbrace$ only depends on the
geometry of the domain and on the form of the topography.

\subsection{The caloric curve $\beta(E)$}
\label{sec_dsa}

Let us first describe the caloric curve $\beta(E)$ for a given value
of $\Gamma$. We need to distinguish three cases:

1. If $\max\lbrace \beta_*,\beta'_{1a},\beta'_{1b}\rbrace=\beta_*$: (i) If
$\Gamma=\Gamma_*$, we have a second order phase transition since
$\frac{\partial\beta}{\partial E}(E)$ is discontinuous at $1/E=1/E_*$. (ii) If
$\Gamma\neq \Gamma_*$, there is no phase transition.

2. If $\max\lbrace \beta_*,\beta'_{1a},\beta'_{1b}\rbrace=\beta'_{1a}$:  we have
a second order phase transition since $\frac{\partial\beta}{\partial E}(E)$ is
discontinuous at $E'_{1a}(\Gamma)$. This is because $\psi'_{1a}$ is orthogonal
to the topography leading to a symmetry breaking. However, this situation is not
generic (see the remark at the end of \ref{sec_dsb}).

3. If $\max\lbrace \beta_*,\beta'_{1a},\beta'_{1b}\rbrace=\beta'_{1b}$: there is
no phase transition. This is because $\psi'_{1b}$ is not orthogonal to the
topography, so that there is no symmetry breaking.

\subsection{The chemical potential curve $\alpha(\Gamma)$}
\label{sec_dsb}

Let us now describe the chemical potential curve $\alpha(\Gamma)$ for a given
value of $E$. We need to distinguish three cases:

1. If $\max\lbrace \beta_*,\beta'_{1a},\beta'_{1b}\rbrace=\beta_*$: if
$1/E<1/E_*$, there is a first order phase transition at $\Gamma=\Gamma_*$ since
$\alpha(\Gamma)$ is discontinuous at $\Gamma=\Gamma_*$. For $1/E>1/E_*$, there
is no phase transition.

2. If $\max\lbrace \beta_*,\beta'_{1a},\beta'_{1b}\rbrace=\beta'_{1a}$: if
$1/E<1/(E'_{1a})_{min}$, there are two second order phase transitions at 
$\Gamma=\Gamma_{21}^{-}(E)$ and $\Gamma=\Gamma_{21}^{+}(E)$. For
$1/E>1/(E'_{1a})_{min}$, there is no phase transition.

3. If $\max\lbrace \beta_*,\beta'_{1a},\beta'_{1b}\rbrace=\beta'_{1b}$: there is
no phase transition.

All these results are fully consistent with those obtained by Venaille \&
Bouchet \cite{vb,vbf} using a different theoretical treatment.

{\it Remark:} in general, the topography (e.g. in the oceans) is very
complex and is generically not orthogonal to an eigenmode of the
Laplacian, so that $\psi'_{1a}$ does not exist (note also
that 
zero mean eigenmodes $\psi'_n$ exist only if the domain has specific symmetries
which is 
generally not the case in the oceans). In addition, in 
generic situations, $\Gamma\neq \Gamma_*$. Therefore, in typical
caloric curves $\beta(E)$, there are no plateaus at $\beta_*$ and
$\beta'_{1a}$ (see, e.g., \cite{cnd}). Consequently, the interesting
phase transitions (second order, bicritical points, azeotropy,...)
described in \cite{vb,vbf} and in this paper do not generically exist
in geophysical flows (with complex topographies). One notorious
exception of physical interest is the case of a 
rectangular basin with a linear topography
(treated explicitly in Secs. \ref{sec_st} and \ref{sec_nst})
corresponding to the $\beta$-effect, or the case of topographies of
the form $h=bH(x)$ or $h=bH(y)$. However, for more complex topographies
there is no phase transition in the strict sense. Nevertheless,
there always exists {\it at least} a ``smooth'' transition from a
monopole to a dipole when we stretch the domain (at fixed high
energy), and a ``smooth'' transition from a monopole or a dipole to a
Fofonoff flow when we lower the energy (at fixed domain shape). On the
other hand, $(\Gamma=\Gamma_*,\tau=\tau_c)$ is always {\it at least} a
critical point at which  a first order phase transition appears (for
sufficiently high energies).

\subsection{Thermodynamical stability}
\label{sec_ths}

In the previous sections, when several solutions were in competition
for the same values of $E$ and $\Gamma$, we have compared their
entropies to select the maximum entropy state. It turns out that the
maximum entropy state is the solution with the highest inverse
temperature \cite{jfm}. Therefore, if we consider fully stable
states (global entropy maxima at fixed energy and circulation), the strict
caloric curve corresponds to $\beta\ge \max\lbrace
\beta_*,\beta'_{1a},\beta'_{1b}\rbrace$. We can also study the thermodynamical
stability of the solutions by determining whether they are
(local) entropy
maxima or saddle points of entropy. Complementary stability results
have been obtained by Chavanis \& Sommeria \cite{jfm}, Venaille \&
Bouchet \cite{vb,vbf} and Naso {\it et al.} \cite{ncd1}.  We shall
briefly recall their results and refer to the
corresponding papers for more
details.

Chavanis \& Sommeria \cite{jfm} and Naso {\it et al.} \cite{ncd1} have obtained
sufficient conditions of microcanonical instability by considering the effect of
``dangerous'' perturbations on the equilibrium states. Using their methods, it
can be shown that the solutions with $\beta<\beta'_{1a}$ are unstable in the
microcanonical ensemble. On the other hand,  when  $\beta'_{1a}<\beta_*$, it can
be shown that the solution  $\psi'_{1a}$ corresponding to $\beta=\beta'_{1a}$
and $1/E=0$ is unstable in the microcanonical ensemble. By
continuity, all the plateau $\beta=\beta'_{1a}$ should be unstable (since the
two extremities of this plateau are unstable).

Venaille \& Bouchet \cite{vb,vbf} have shown that the solutions with
$\beta\ge \max\lbrace \beta_*,\beta'_{1a},\beta'_{1b}\rbrace$
are stable in the canonical ensemble (hence in the microcanonical ensemble)
while the other solutions are unstable in the canonical
ensemble. However, as discussed by Chavanis \& Sommeria \cite{jfm} and Naso {\it
et al.} \cite{ncd1}, some of these solutions can be metastable (local entropy
maxima) in the microcanonical ensemble.

For $\beta>\beta''_{1}$, the solutions are stable in the grand canonical
ensemble (hence in the canonical and microcanonical ensembles) \cite{vb,vbf}.
This is related to the Arnold theorem (see, e.g.,  \cite{proc}). For
$\max\lbrace \beta_*,\beta'_{1a},\beta'_{1b}\rbrace<\beta<\beta''_{1}$, the
solutions are stable in the canonical (hence microcanonical) ensemble but not in
the grand microcanonical ensemble. This corresponds to a situation of ensemble
inequivalence \cite{vb,vbf}.

\section{Numerical simulations}
\label{sec_num}

We shall now perform numerical simulations to illustrate the phase transitions
described in the previous sections. To that purpose, we use the relaxation
equations of Sec. \ref{sec_relax} that can serve as numerical algorithms to
compute maximum entropy states with relevant constraints. We shall first
integrate numerically Eq.~(\ref{rel_1}), with the constraints
(\ref{rel_2},\ref{rel_3}), in an antisymmetric square domain
with a linear topography and two different initial conditions such that
$\Gamma=0$. We use as boundary conditions:
\begin{eqnarray}
\psi|_{\partial D}=0,\\
q|_{\partial D}=-\alpha(t),
\end{eqnarray}
where $\partial D$ is the domain boundary. The first condition enforces
free-slip on the boundary, while the second one is necessary for consistency of
the steady state, characterized by Eq.~(\ref{f7}).

\begin{figure}[h]
\center
\includegraphics[width=4cm,keepaspectratio]{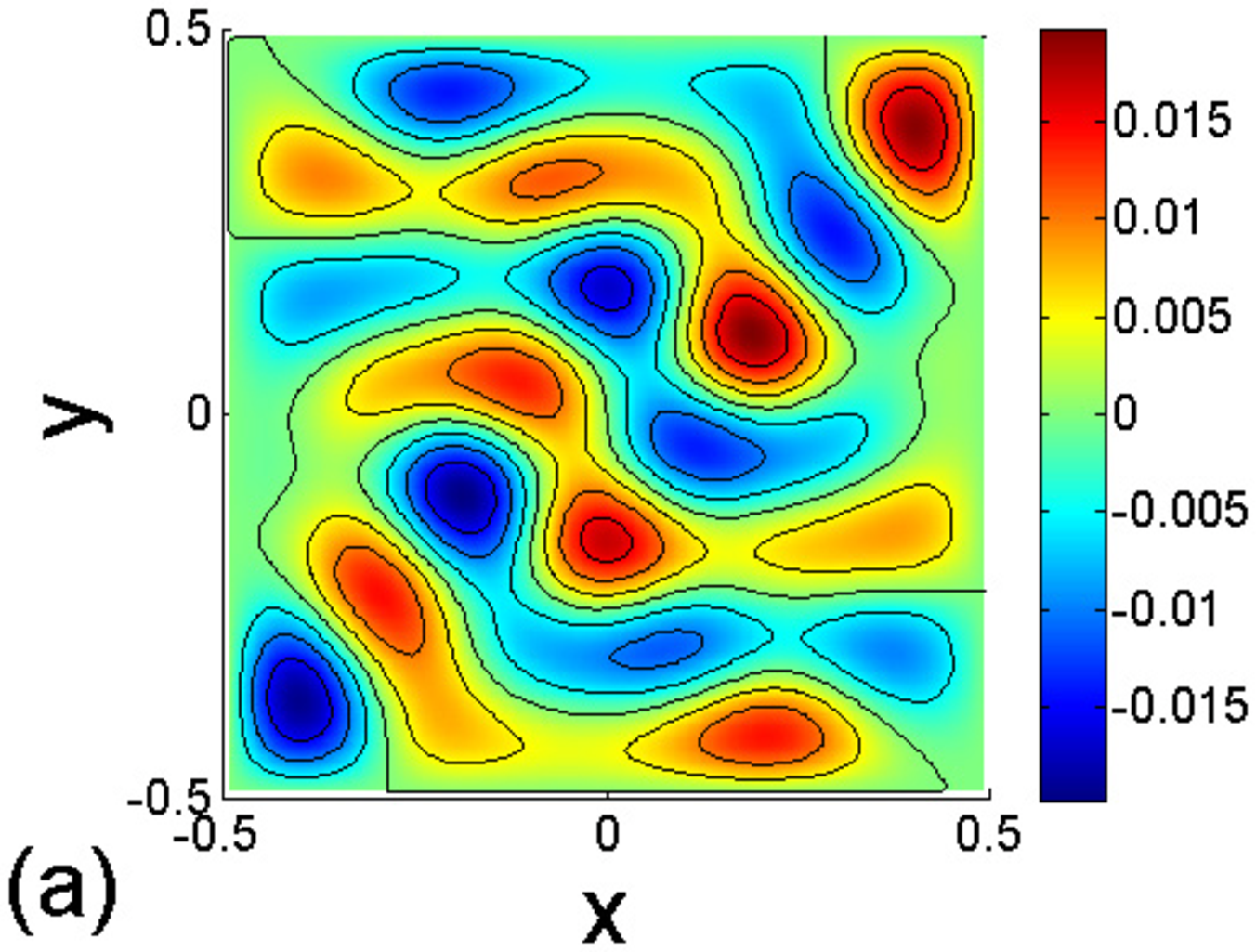}
\includegraphics[width=4cm,keepaspectratio]{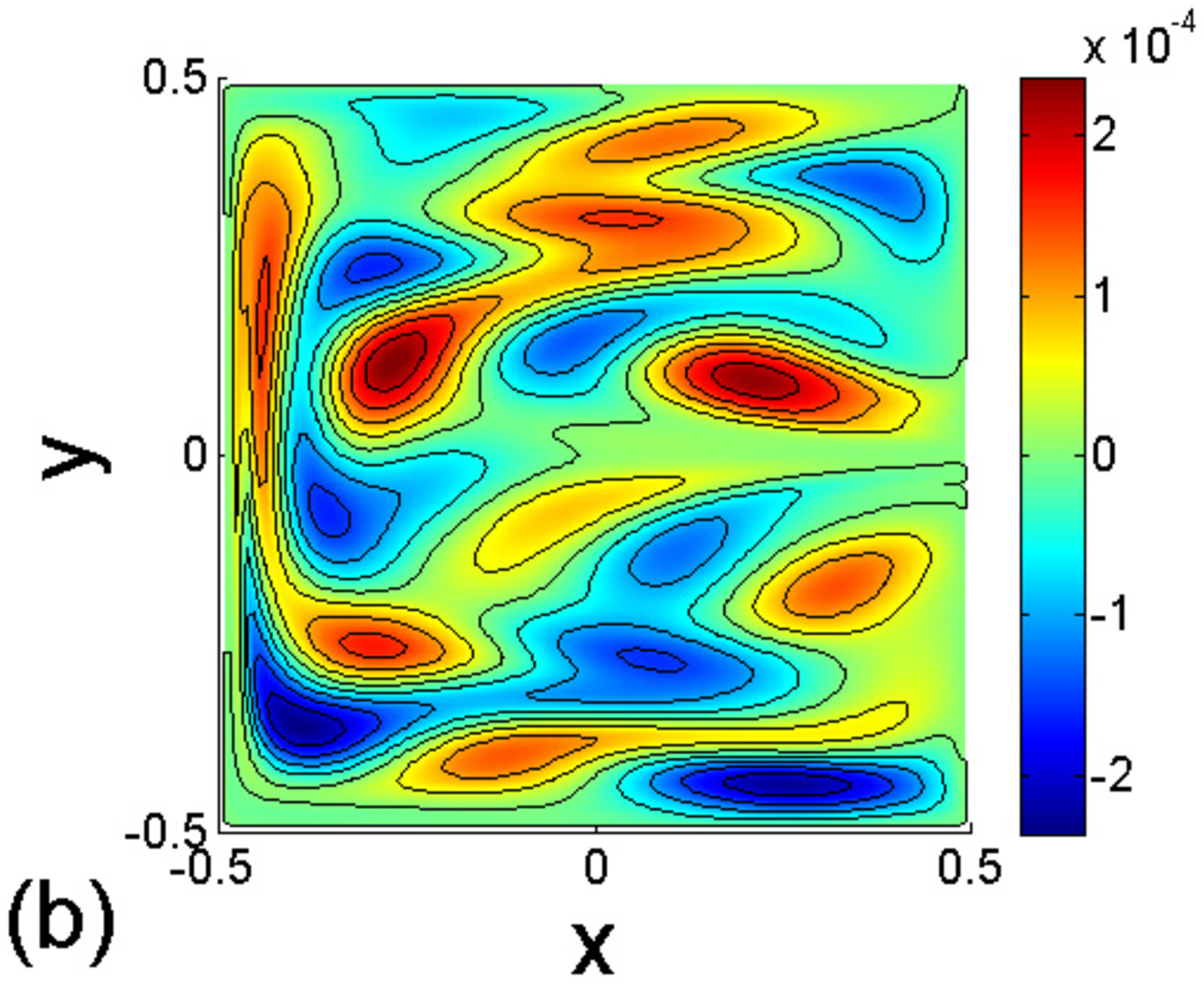}
\includegraphics[width=4cm,keepaspectratio]{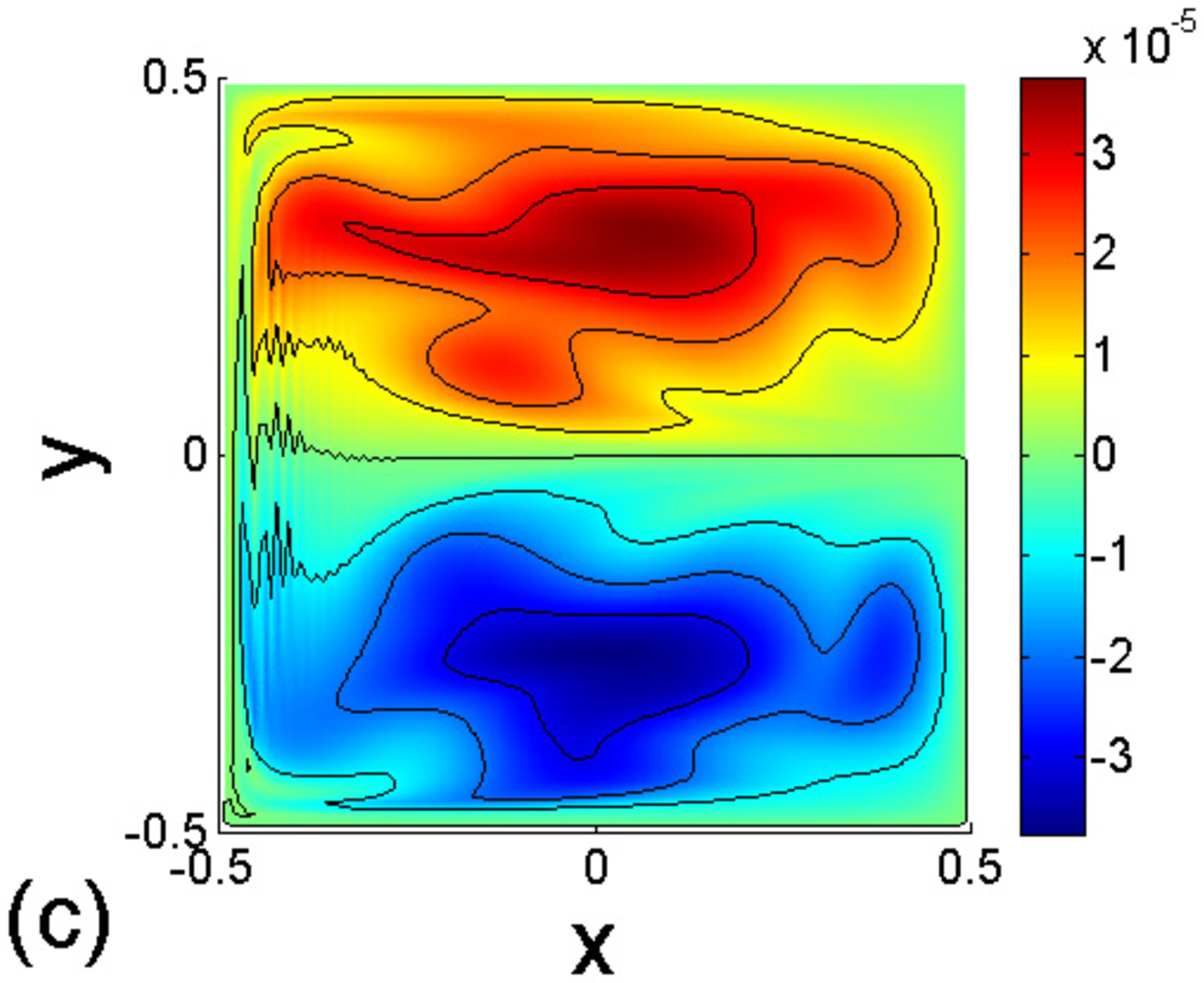}
\includegraphics[width=4cm,keepaspectratio]{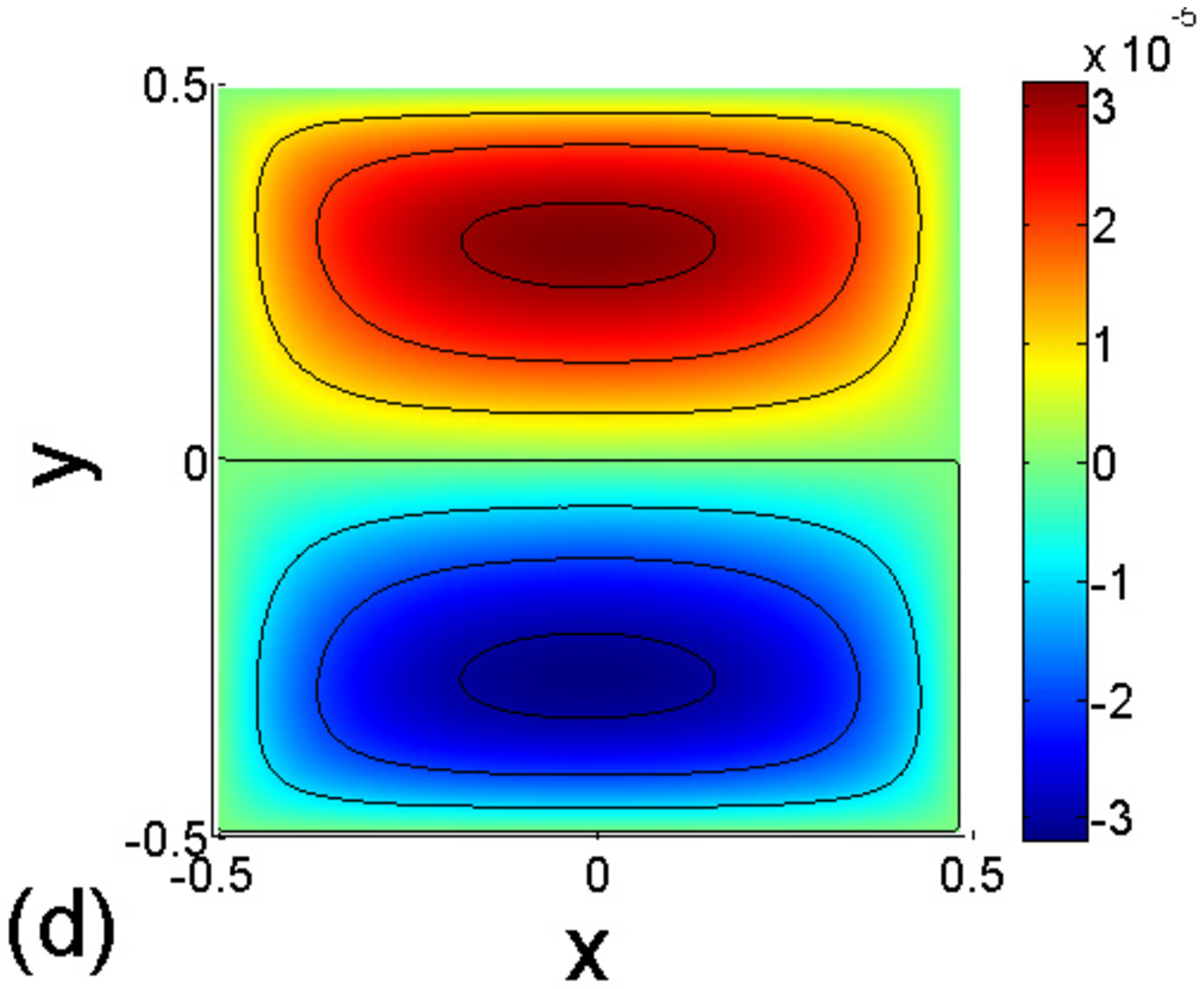}
\caption{\label{rel_f1} Potential vorticity at $t=0$, $15$, $25$, $40$, for
$\Gamma=0$ and $1/E\simeq 1050$ (low energy) in a square domain. In that case,
we obtain two rolls that are strongly influenced by the topography. For even
lower energies, we obtain strict Fofonoff flows exhibiting a westward jet.}
\end{figure}

\begin{figure}[h]
\center
\includegraphics[width=4cm,keepaspectratio]{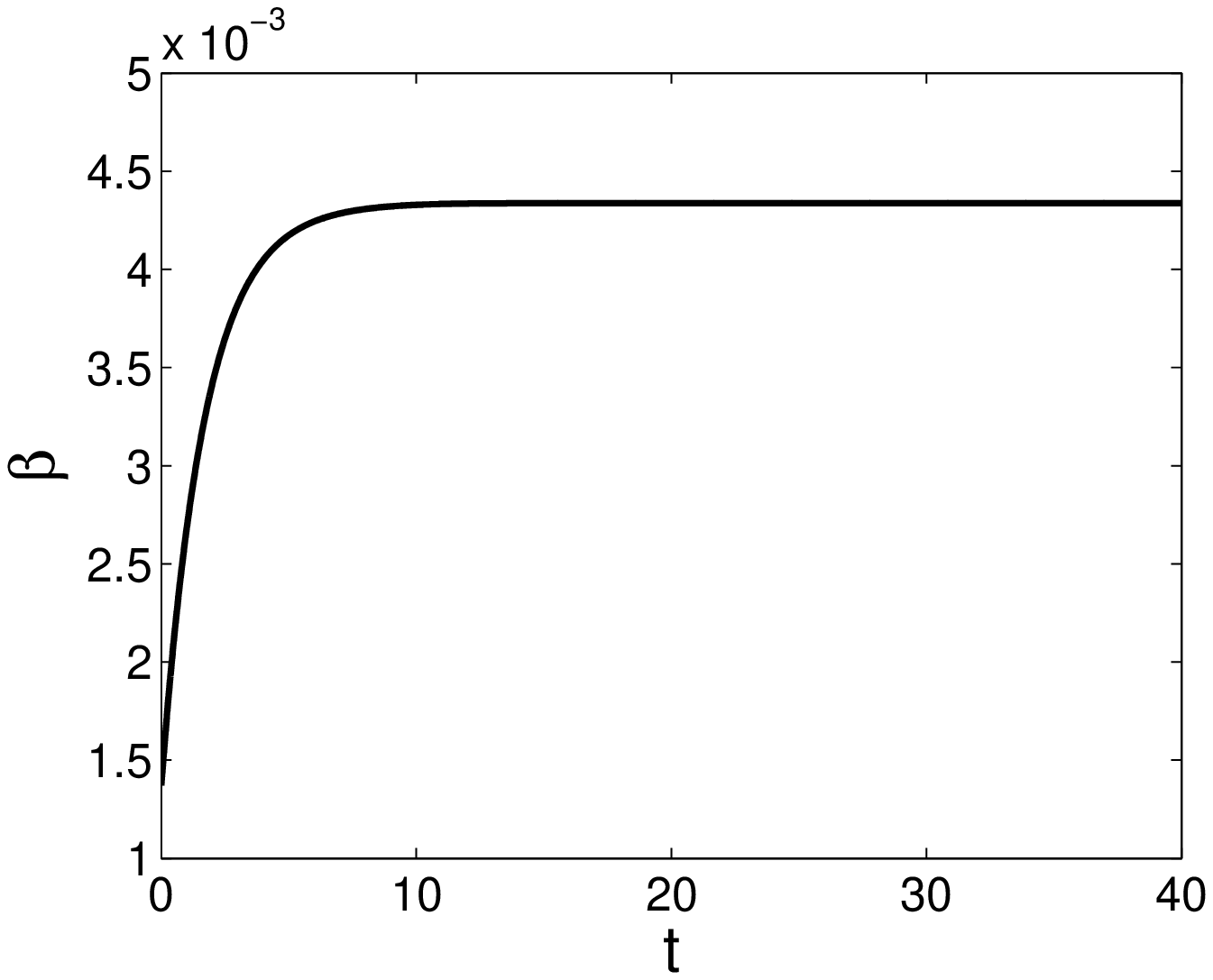}
\includegraphics[width=4cm,keepaspectratio]{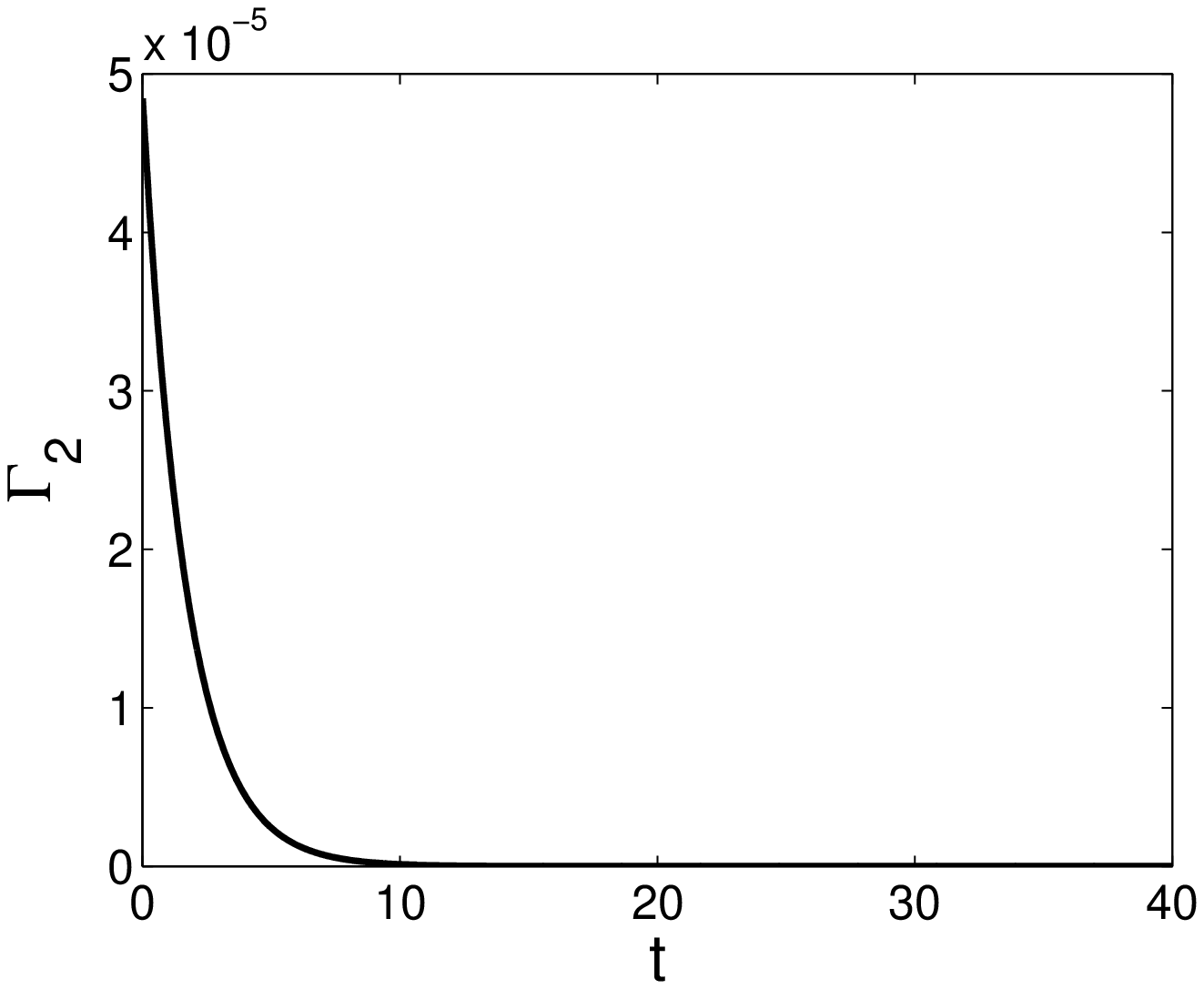}
\caption{\label{rel_f2} Time evolutions of the inverse temperature $\beta$ and
of the enstrophy $\Gamma_2$, for $\Gamma=0$ and $1/E\simeq 1050$ (low energy) in
a square domain, corresponding to Fig.~\ref{rel_f1}.}
\end{figure}

We first integrate the relaxation equations with an initial condition,
$q(x,y,t=0)$, written as the sum of sine functions with random
amplitudes and wave numbers ranging from 1 to 9 (see
Fig.~\ref{rel_f1}(a)). With such a field, $\Gamma=0$ and the energy is
rather low: $1/E\approx 1050$. The coefficient $D$ is set to $0.3$. The
resulting
density of potential vorticity at different times is plotted in
Fig. \ref{rel_f1}. As expected with such a low value of the energy
(see Fig.~\ref{E_beta_1}), the relaxation equation converges to a
Fofonoff state. The inverse temperature $\beta$ and the enstrophy are
plotted as a function of time in Fig.~\ref{rel_f2}. As expected,
$\Gamma_2$ monotonically decreases in time (equivalently, the entropy
increases). On the other hand, the inverse temperature monotonically
increases during the simulation. Both quantities remain constant once
the steady state (Fofonoff flow, maximum of entropy) has been reached.

These results can be compared with those of Wang \& Vallis \cite{wang}. In this
study, the authors integrate the quasigeostrophic equation with linear
topography in an antisymmetric square domain, with initial
conditions (random eddies) and boundary conditions (free-slip) similar to ours.
After about 10 eddy turnover times, the time averaged flow converges to a state
close to the Fofonoff solution. Comparing our Fig.~\ref{rel_f1} with Fig. 4 of
\cite{wang}, it is clear that the dynamical behavior of the coarse-grained
potential vorticity $q$ solution of the relaxation equation is reminiscent of
that of the time averaged $q$ solution of the quasigeostrophic equation. In both
cases, the eddies are first pushed to the western boundary, then two gyres form.
These structures grow and fill out the northern and southern parts of the
domain. Therefore, even if the relaxation equations are not supposed to provide
a realistic parametrization of 2D turbulence, they may however give an idea of
the true evolution of the flow towards statistical equilibrium. It is worth
noticing that, with the relaxation equations, the steady state has been reached
after about $8000$ time steps, while it took more than several ten thousands
time steps for the quasigeostrophic equation to reach the statistically steady
state. Indeed, by construction, the relaxation equations ``push'' the system in
the direction of the statistical equilibrium state.

\begin{figure}[h]
\center
\includegraphics[width=4cm,keepaspectratio]{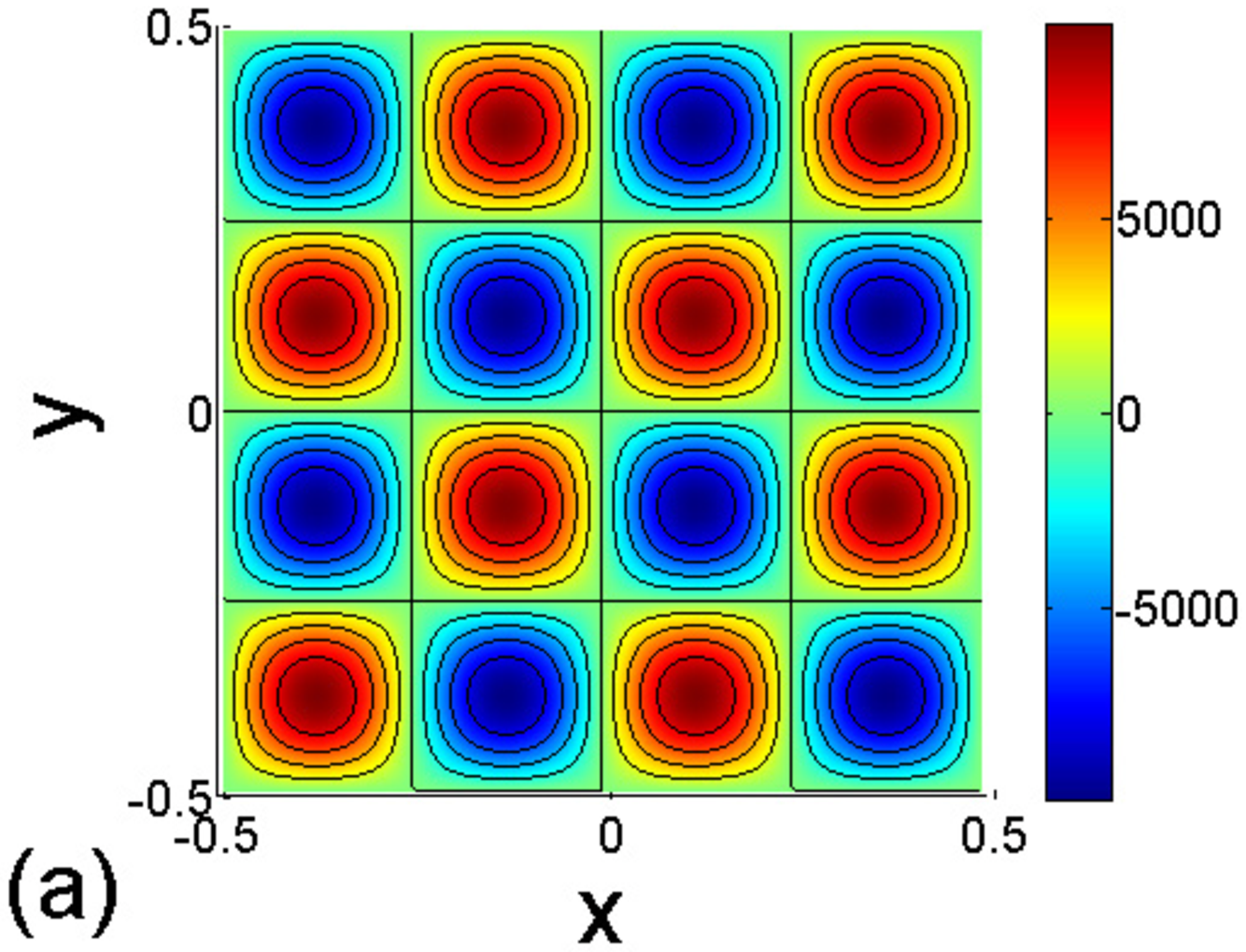}
\includegraphics[width=4cm,keepaspectratio]{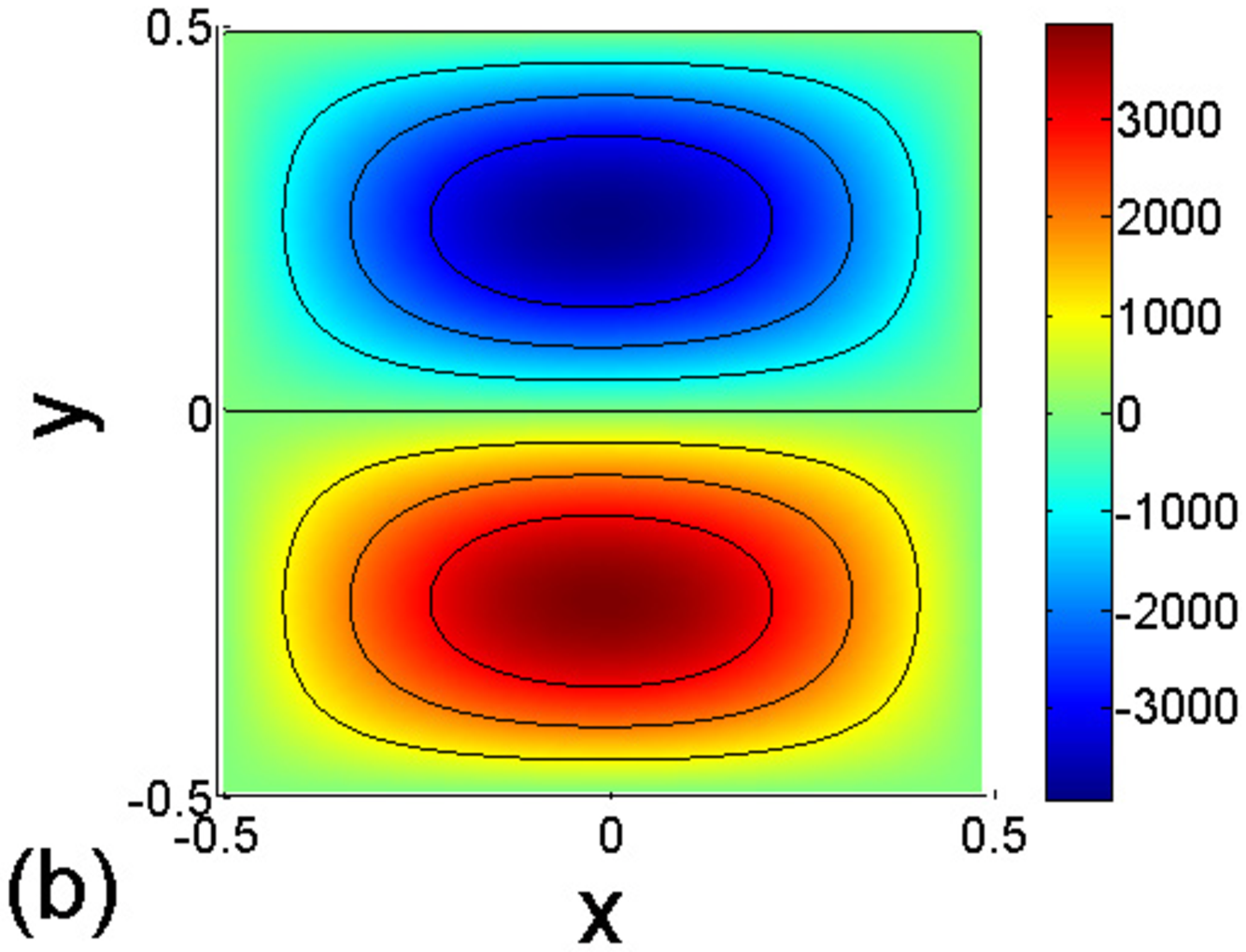}
\includegraphics[width=4cm,keepaspectratio]{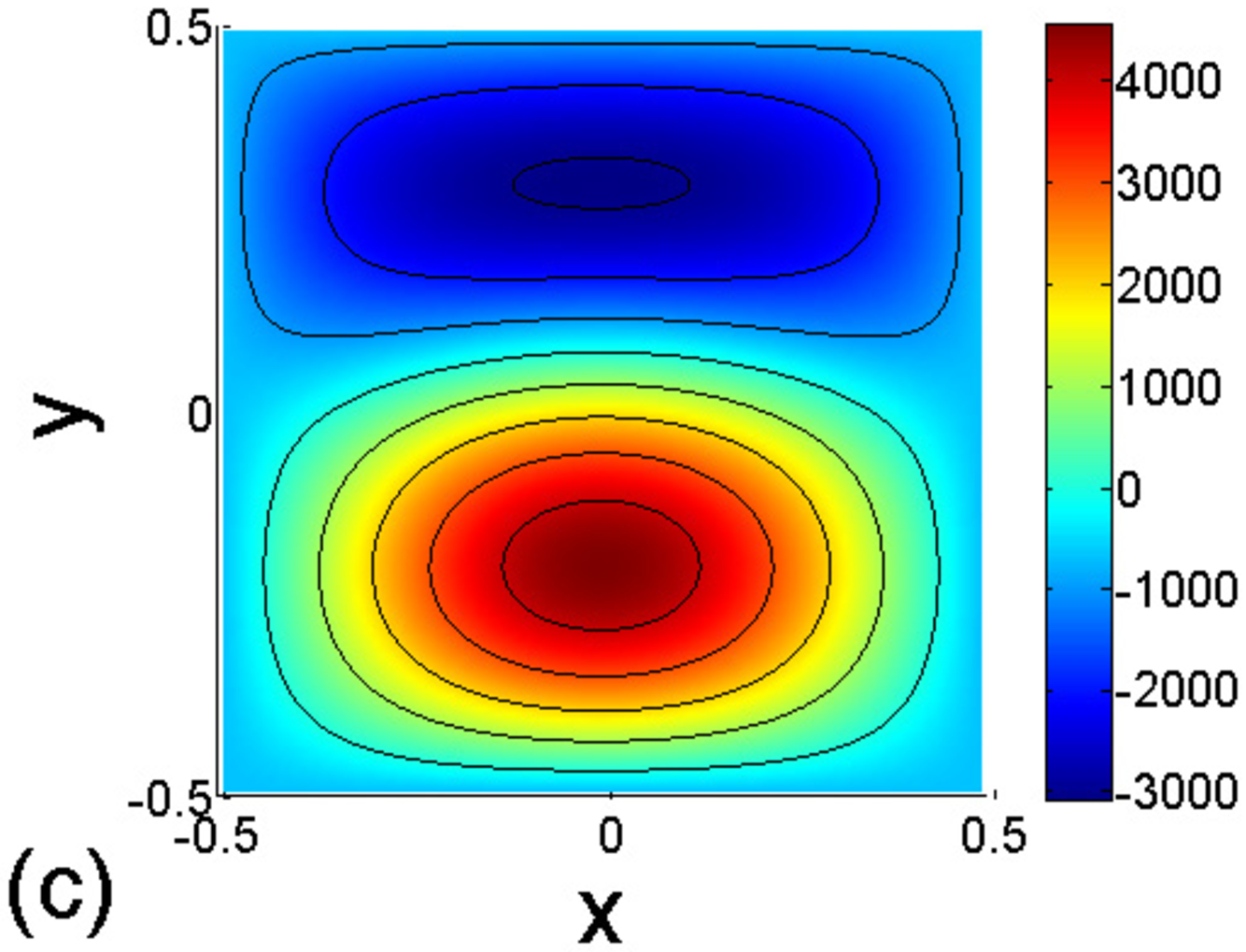}
\includegraphics[width=4cm,keepaspectratio]{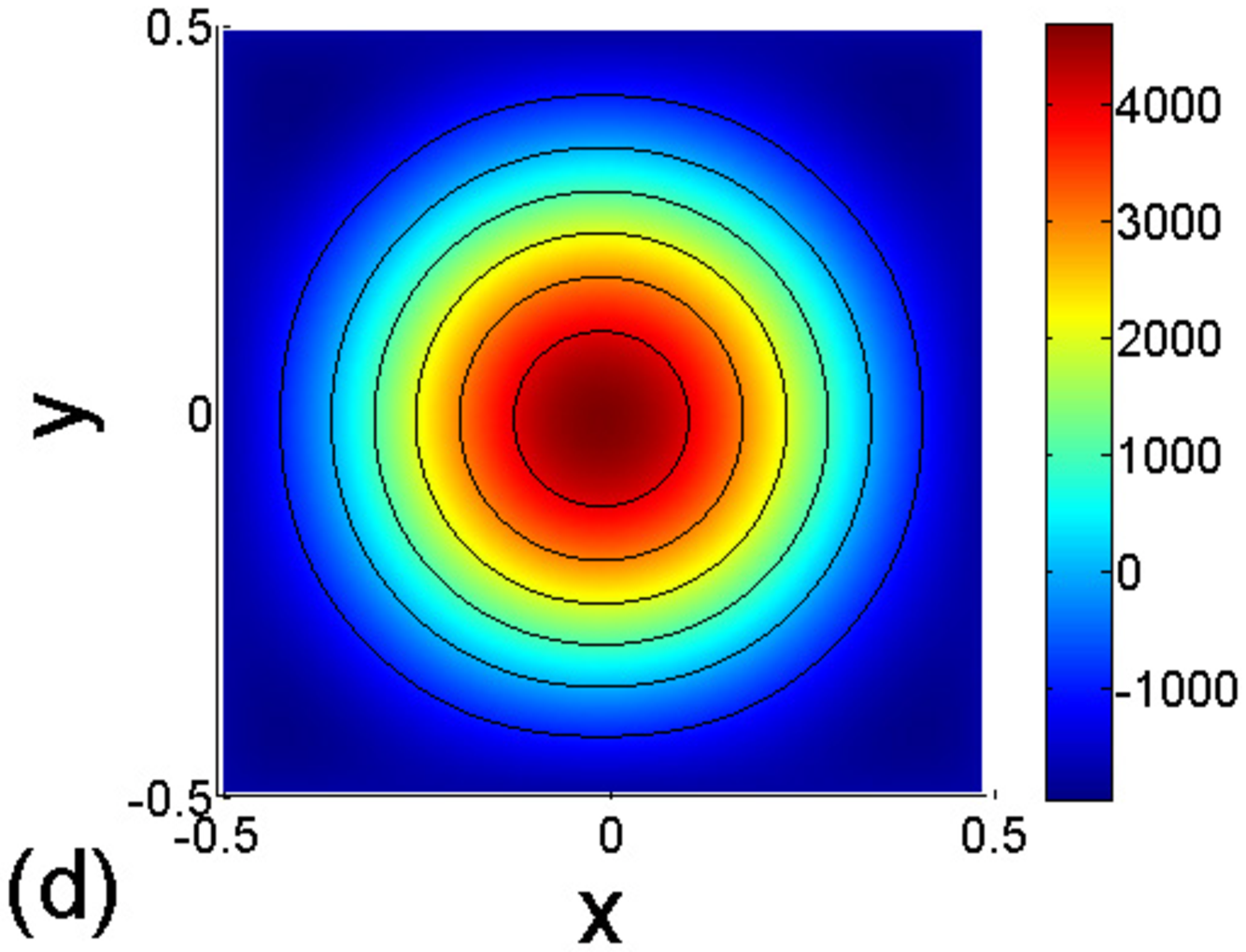}
\caption{\label{rel_f3} Potential vorticity at $t=0$, $10$, $90$, $140$, for
$\Gamma=0$ and $1/E\simeq 1.3.10^{-5}$ (high energy) in a square domain. The
system first relaxes towards the dipole (saddle point). At $t=10$, the optimal
perturbation is applied to the system. It then converges towards the monopole
(stable steady state). In that case, the stable state is not influenced by the
topography.}
\end{figure}

\begin{figure}[h]
\center
\includegraphics[width=4cm,keepaspectratio]{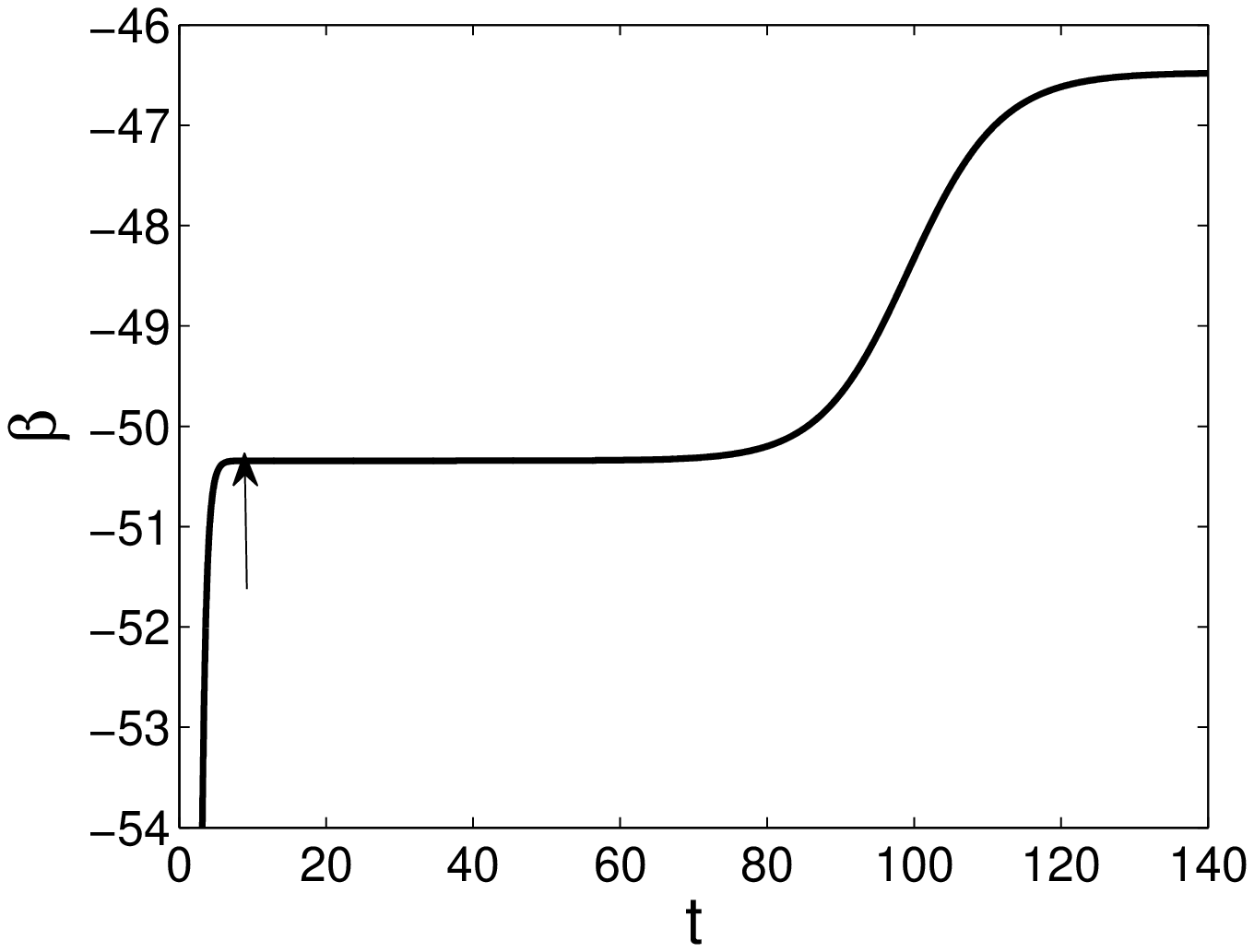}
\includegraphics[width=4cm,keepaspectratio]{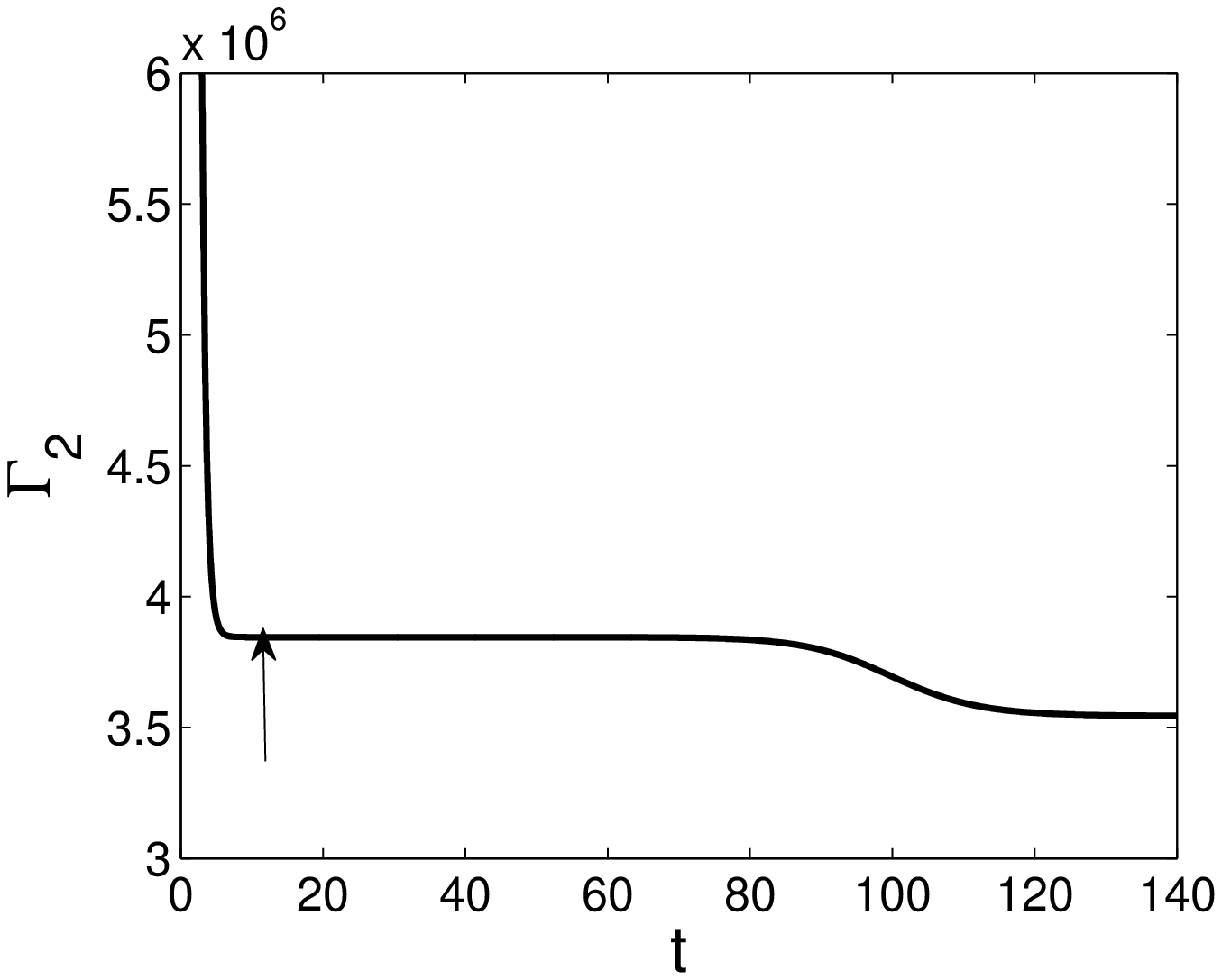}
\caption{\label{rel_f4} Time evolutions of the inverse temperature $\beta$ and
of the enstrophy $\Gamma_2$, for $\Gamma=0$ and $1/E\simeq 1.3.10^{-5}$ (high
energy) in a square domain, corresponding to Fig.~\ref{rel_f3}. At $t=10$
(indicated by the arrows), the optimal perturbation is applied to the system.}
\end{figure}

We then impose an initial condition of high energy, $1/E\simeq 1.3.10^{-5}$
(see Fig.~\ref{rel_f3}(a)) and integrate numerically
Eq.~(\ref{rel_1}), with the constraints
(\ref{rel_2},\ref{rel_3}). Since we are only interested in the final
state, and not in the way the system converges towards it, we do not
implement the advection term.  The coefficient
$D$ is set to $1$. The $q$ density is plotted at different times in
Fig.~\ref{rel_f3}, and the time evolutions of the inverse temperature
and of the enstrophy are shown in Fig.~\ref{rel_f4}. We find that the
system first relaxes spontaneously towards a horizontal dipole (see
Fig.~\ref{rel_f3}(b) and first plateaus of Fig.~\ref{rel_f4}). In the
absence of external perturbation, the system remains in this state for
a long time, even if it is predicted to be unstable (see
Fig.~\ref{E_beta_1z}). It can be noticed that the value of the inverse
temperature is slightly smaller than $\beta_{12}$, confirming that
the system is on the lower branch of Fig.~\ref{E_beta_1z}. As already
shown in the case of the 2D Euler equations \cite{ncd1}, we illustrate
here that saddle points can be very long-lived. Inspired by the
stability analysis performed in \cite{ncd1}, we add to the system, at
$t=10$, a perturbation of the form $\delta q=1-\beta_*\phi_{1,*}$,
where $\phi_{1,*}$ is the solution of Eq.~(\ref{f18}) for
$\beta=\beta_*$ (see
\cite{ncd1}).  As expected (see
Fig.~\ref{E_beta_1z}), the system is then immediately destabilized,
and relaxes towards the monopole which is the  maximum entropy state
(Fig.~\ref{rel_f3}(c-d) and second plateau of
Fig.~\ref{rel_f4}). Depending on the sign of the perturbation, the
system can relax towards the direct or towards the inverse monopole.

\begin{figure}[h]
\center
\includegraphics[width=4cm,keepaspectratio]{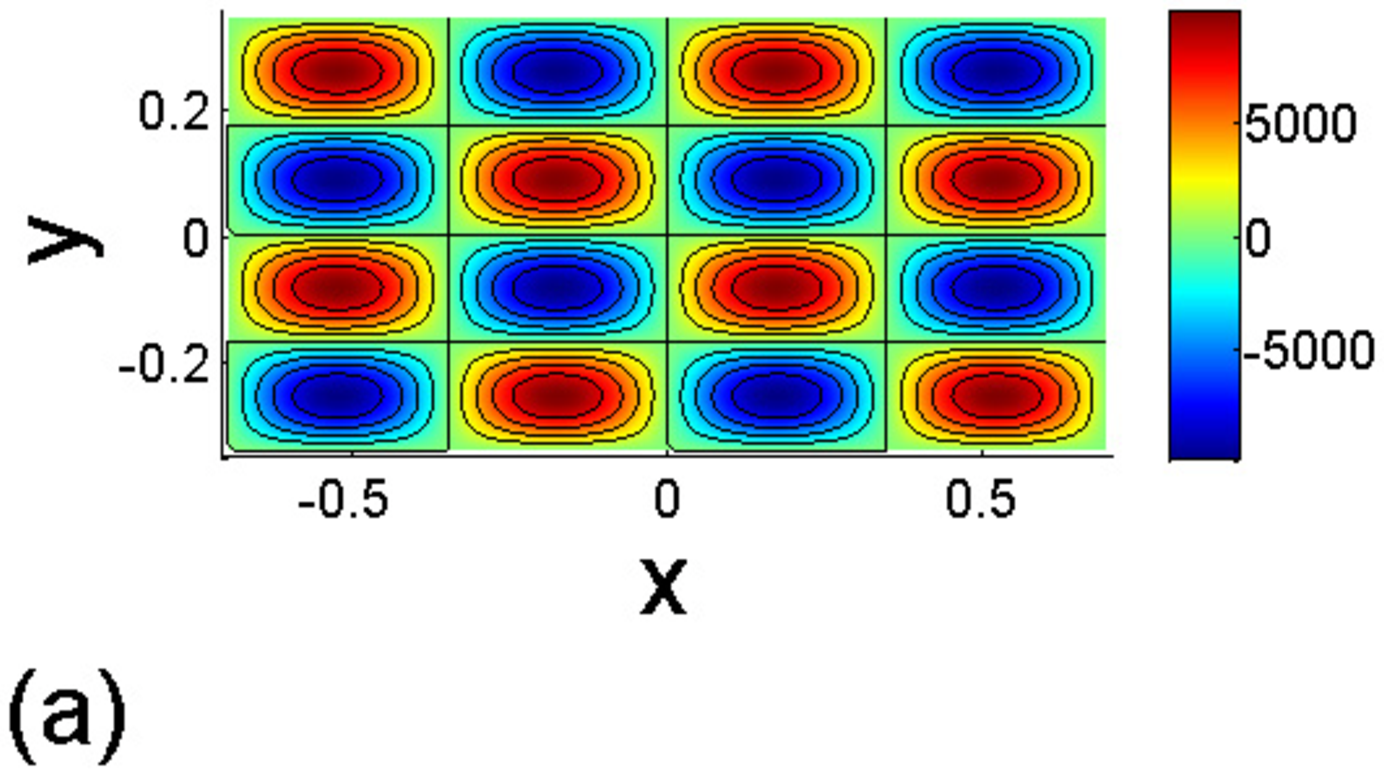}
\includegraphics[width=4cm,keepaspectratio]{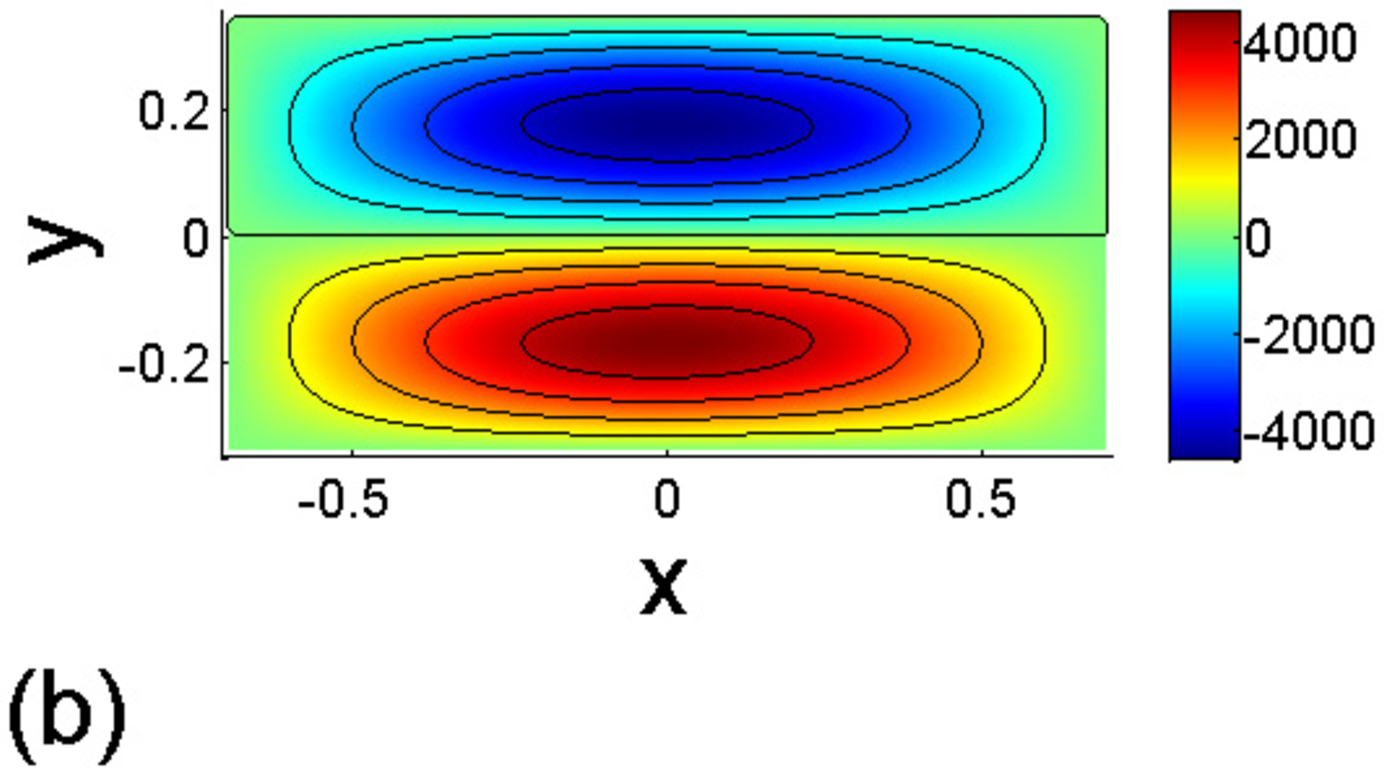}
\includegraphics[width=4cm,keepaspectratio]{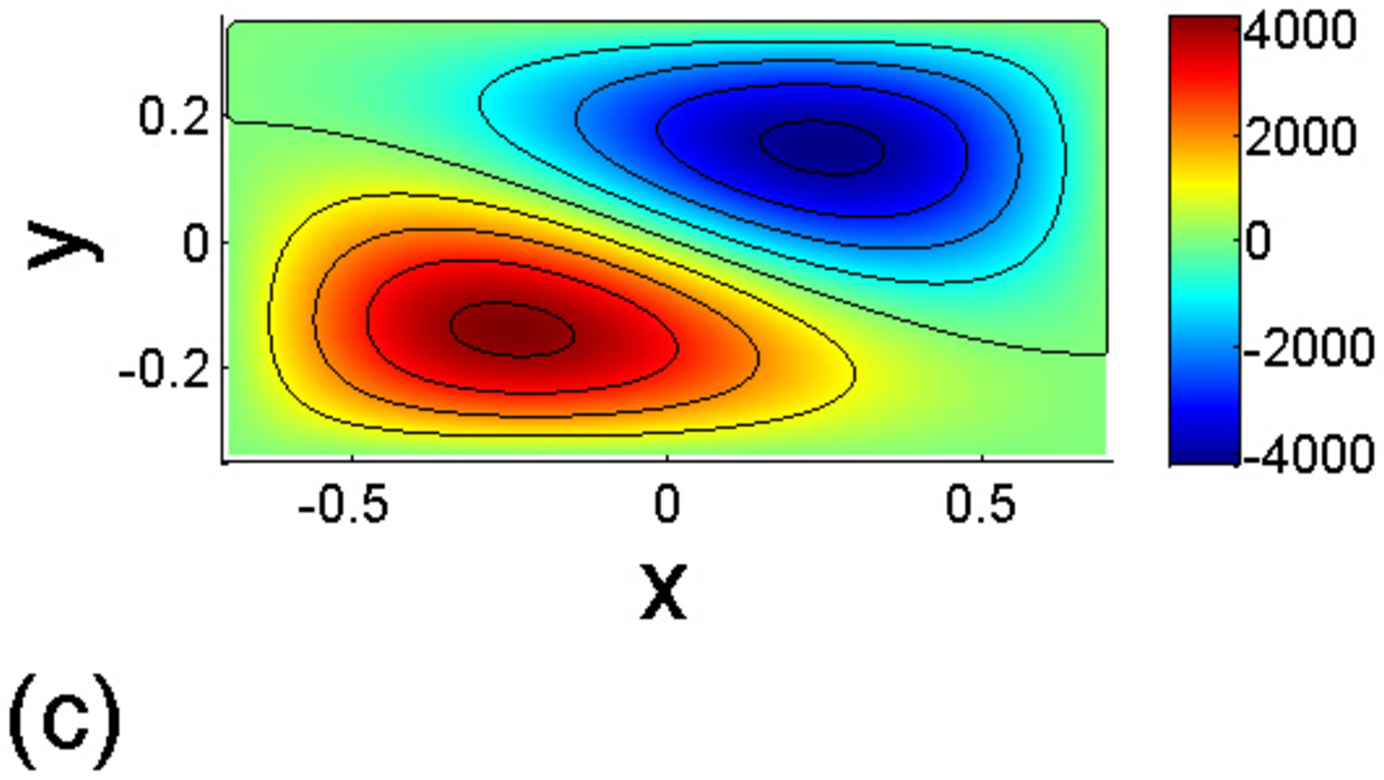}
\includegraphics[width=4cm,keepaspectratio]{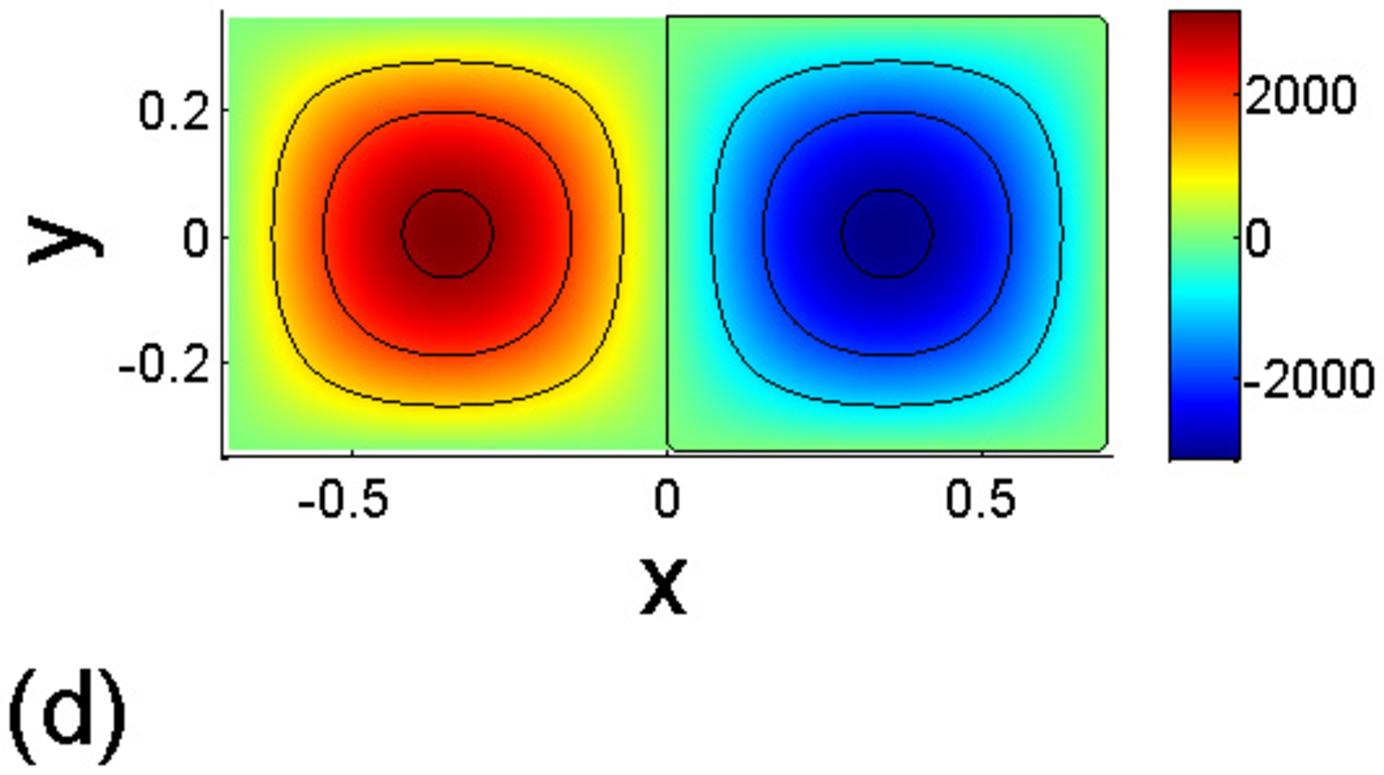}
\caption{\label{rel_f5} Potential vorticity at $t=0$, $10$, $21$, $60$, for
$\Gamma=0$ and $1/E\simeq 1.6.10^{-5}$ (high energy) in a rectangular domain of
aspect ratio $\tau=2$. The system first relaxes towards the vertical dipole
(saddle point). It then destabilizes spontaneously and converges towards the
horizontal dipole (stable steady state). In that case, the stable state is not
influenced by the topography.}
\end{figure}

\begin{figure}[h]
\center
\includegraphics[width=4cm,keepaspectratio]{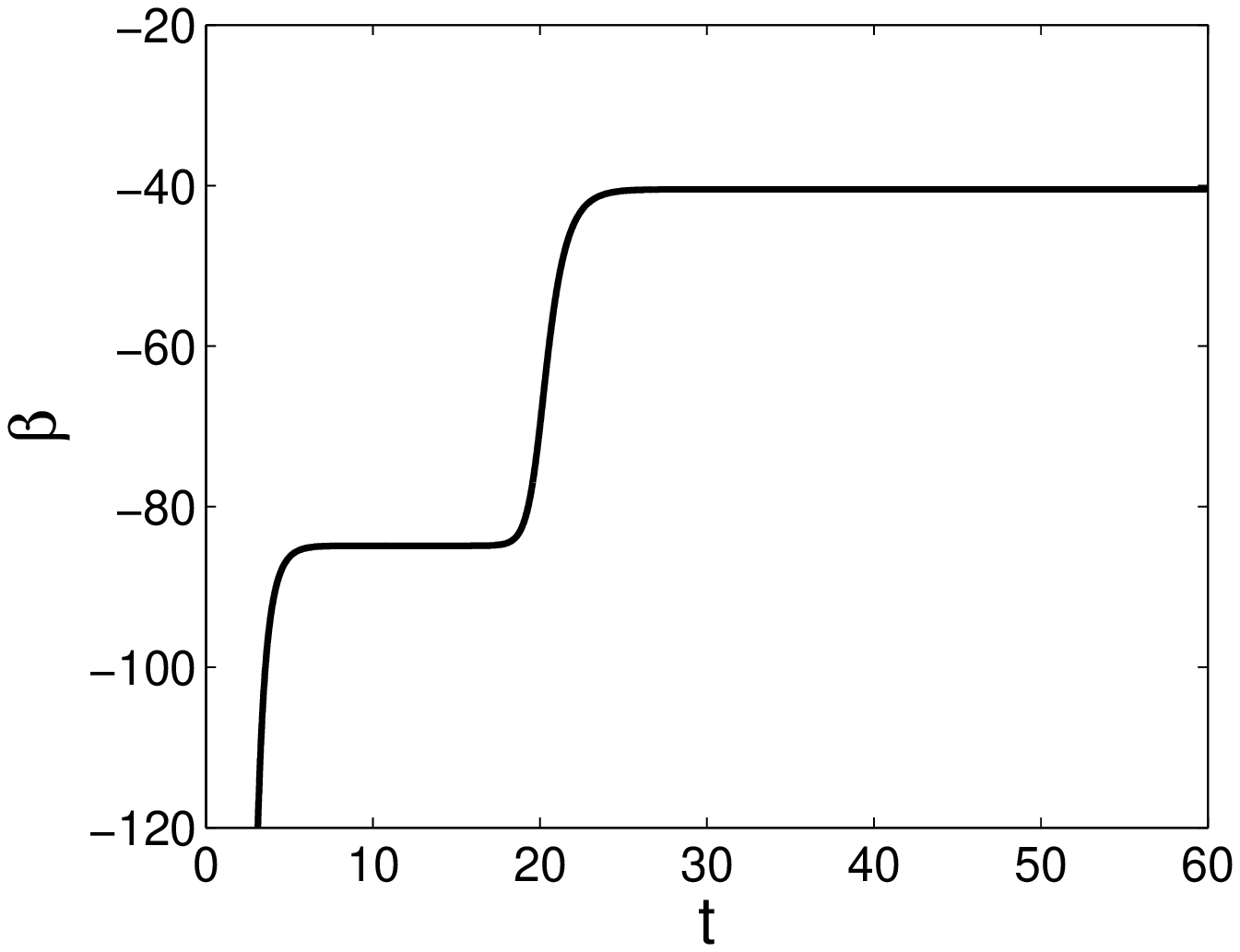}
\includegraphics[width=4cm,keepaspectratio]{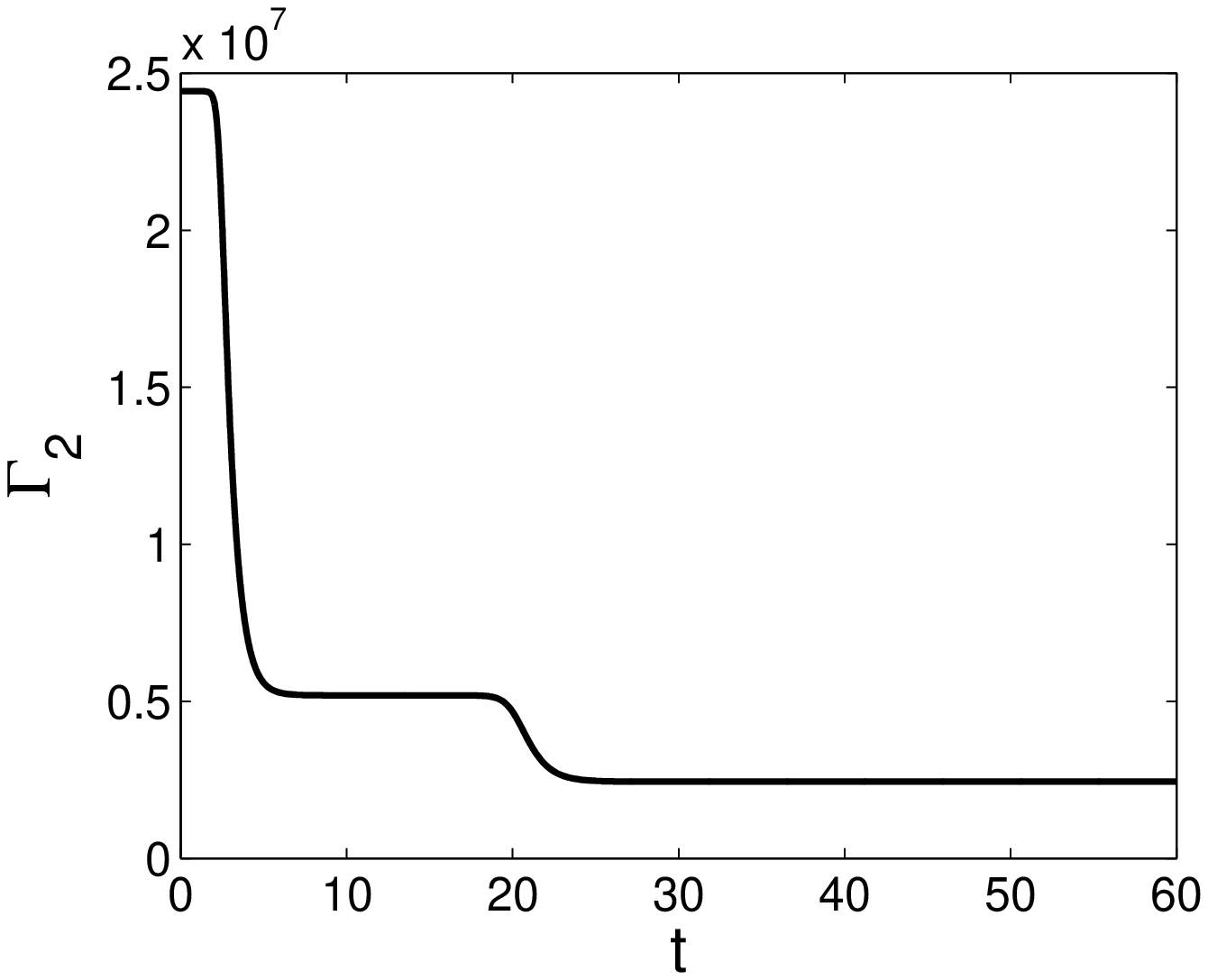}
\caption{\label{rel_f6} Time evolutions of the inverse temperature $\beta$ and
of the enstrophy $\Gamma_2$, for $\Gamma=0$ and $1/E\simeq 1.6.10^{-5}$ (high
energy) in a rectangular domain of aspect ratio $\tau=2$, corresponding to
Fig.~\ref{rel_f5}.}
\end{figure}

To summarize, we have illustrated the fact that in an
antisymmetric square
domain with linear topography, if $\Gamma=0$, the stable state is
strongly correlated to the topography at low energy (Figs.~\ref{rel_f1}
and \ref{rel_f2}), but is not influenced by it at high energy
(Figs.~\ref{rel_f3} and \ref{rel_f4}). At high energy, it is influenced
by the domain geometry: in a square domain, we get a monopole whereas
in a rectangle sufficiently elongated in the $x$ direction
($\tau>\tau_c=1.12$), we get a dipole. To illustrate this remark, we integrate
numerically Eq.~(\ref{rel_1}), with the constraints (\ref{rel_2},\ref{rel_3}),
in an antisymmetric rectangular domain of aspect ratio
$\tau=2$, starting from an initial condition of zero circulation and of high
energy ($1/E\simeq 1.6.10^{-5}$). The resulting potential vorticity density is
plotted at different times in Fig.~\ref{rel_f5}. The system first converges
towards the vertical dipole (saddle point), with $\beta\simeq\beta_{12}$ (see
Fig.~\ref{rel_f5}(b) and first plateaus of Fig.~\ref{rel_f6}). It then
destabilizes \textit{spontaneously}, and converges towards the horizontal dipole
(stable state) (see Fig.~\ref{rel_f5}(d) and second plateaus of
Fig.~\ref{rel_f6}), with $\beta\simeq\beta_{21}$. These results can be compared
to Fig.~\ref{E_beta_2}.

Other numerical
simulations of the relaxation equations illustrating phase transitions
in geophysical flows with nonlinear topography are reported in
\cite{cnd}. Interestingly, the initial conditions of high energy that we have
considered in the present paper are similar to those used in Fig. 7(c) (square
domain) and 8 (rectangular domain of aspect ratio $\tau=2$) of \cite{cnd}:
$q(t=0)$ is proportional to $\psi_{44}$ and the energies are identical. We find
that the final states are the same with both topographies. This is to be
expected since the topography should not influence the maximum entropy state in
the limit of high energy.   However, we observe that  the relaxation equations
with the nonlinear topography of \cite{cnd} directly converge towards the
equilibrium state whereas, with the linear topography, the system first
converges towards the vertical dipole (saddle point). Therefore, at very high
energy, while the topography does not influence the stable states, it seems to
play a role in the dynamics of the relaxation. Of course, this
conclusion is reached on the basis of our relaxation equations. It would be
interesting to know whether it remains valid for more realistic
parameterizations.

\section{Conclusion}
\label{sec_conc}

In this paper, we have studied the nature of phase transitions in
simple models of oceanic circulation described by the QG equations
with an arbitrary topography $h$. We have assumed a linear
relationship between potential vorticity and stream function
corresponding to minimum potential enstrophy states. We have given
several interpretations of this minimum potential enstrophy principle
in connection with statistical mechanics, phenomenological selective
decay principles, and nonlinear dynamical stability. We have
explicitly treated the case of a rectangular basin and an
antisymmetric
linear topography $h=by$ like in Fofonoff's classical paper. For small
energies, we recover Fofonoff's westward jet solution. In that case,
the flow is strongly influenced by the topography and only weakly by
the domain geometry. For large energies, we obtain geometry induced
phase transitions between monopoles and dipoles like in the study of
Chavanis \& Sommeria \cite{jfm}. In that case, the flow is strongly
influenced by the domain geometry and only weakly by the
topography. In rectangular domains elongated in the
$x$ direction ($\tau>\tau_c$), as we decrease the energy of the flow,
we describe symmetry breaking phase transitions between horizontal
dipoles and Fofonoff flows. Alternatively, in rectangular domains
elongated in the $y$ direction ($\tau<1/\tau_c$), the smooth
transitions between vertical dipoles and Fofonoff flows occur without
symmetry breaking. This phenomenology, illustrated in a rectangular
domain with a linear topography, has been generalized to arbitrary
domains and arbitrary topography.

Our study returns and confirms the results previously obtained by
Venaille \& Bouchet \cite{vb,vbf} by a different method. These authors
provide a very detailed statistical analysis of the problem,
emphasizing the notions of bicritical points, azeotropy and ensemble
inequivalence. Our approach, that is less abstract and illustrated by
several explicit calculations, provides a useful complement to their
study. It gives another way of describing the complicated and rich
bifurcations that occur in geophysical flows.  Our theoretical results
extend the work of Chavanis \& Sommeria \cite{jfm} to the case of
geophysical flows (with a topography) and allow to interpret the phase
transitions studied numerically by Chavanis {\it et al.} \cite{cnd} in
the case of complex topographies.

\appendix

\section{Minimum potential enstrophy principle}
\label{sec_visc}

We shall here discuss the difficulty to justify a minimum potential enstrophy
principle based on the viscosity. In the presence of viscosity, the QG equations
become
\begin{eqnarray}
\label{visc1}
\frac{\partial q}{\partial t}+{\bf u}\cdot\nabla q=\nu\Delta\omega,
\end{eqnarray}
where $q=\omega+h$. We emphasize that the quantity dissipated by viscosity is
the vorticity $\omega$, not the potential vorticity $q$. Therefore, the rate of
change of potential enstrophy $\Gamma_2=\int q^2\, d{\bf r}$ is
\begin{eqnarray}
\label{visc2}
\dot\Gamma_2=2\nu\int q\Delta\omega\, d{\bf r}=2\nu\int q\Delta (q-h)\, d{\bf
r}.
\end{eqnarray}
For the 2D Navier-Stokes equation ($h=0$), we obtain after an integration by
parts
\begin{eqnarray}
\label{visc3}
\dot\Gamma_2=-2\nu\int  (\nabla \omega)^2\, d{\bf r}\le 0,
\end{eqnarray}
so that the enstrophy decreases monotonically under the effect of viscosity.
However, for the viscous QG equations, we do not have a monotonic decay of
potential enstrophy. Following Bretherton \& Haidvogel \cite{bretherton}, we
must assume $h\ll q$ in Eq. (\ref{visc2}) in order to  have a monotonic decay of
potential enstrophy. In more general cases, the phenomenological minimum
potential enstrophy principle is not clearly justified.  These remarks also
apply to the functionals (\ref{dims}). Indeed, while these functionals increase
monotonically  under the effect of viscosity for the 2D Navier-Stokes equations
since $\dot S=\nu\int C''(\omega)(\nabla\omega)^2\, d{\bf r}\ge 0$, their
monotonic increase is not guaranted for the QG equations unless  $h\ll q$.

Therefore it is difficult to justify a principle of the form (\ref{dim}) for the
viscous  QG equations. By contrast, the interpretations of this principle that
we have  given in the framework of the {\it inviscid} QG equations are valid.

\section{Link with the Neptune effect}
\label{sec_nept}

In this Appendix, we  show that the relaxation equations of Sec. \ref{sec_relax}
are consistent with the Neptune effect of Holloway \cite{hollowaynept}. This
discussion is related to the one given by Kazantsev {\it et al.}
\cite{kazantsev} but the justification of the relaxation equations that we
consider is different.

Let us first consider the case $h=0$ and $R=\infty$ and derive the relaxation
equation for the velocity field. We start from the equation for the vorticity
field (\ref{two16}) that becomes
\begin{equation}
\label{nept1}
\frac{\partial \omega}{\partial t}+{\bf u}\cdot \nabla \omega=\nabla\cdot
\left\lbrack D\left (\nabla \omega+\frac{\beta(t)}{C''(\omega)}\nabla\psi\right
)\right\rbrack,
\end{equation}
where we recall that $D$ can depend on position and time. For a 2D field, we
have the identity  $\nabla\times ({\bf z}\times {\bf a})=(\nabla\cdot {\bf
a}){\bf z}$. Therefore, we can rewrite the foregoing equation as
\begin{eqnarray}
\label{nept2}
\left (\frac{\partial \omega}{\partial t}+{\bf u}\cdot \nabla \omega\right )
{\bf z}=\nabla\times  \left\lbrack {\bf z}\times  D\left (\nabla
\omega+\frac{\beta(t)}{C''(\omega)}\nabla\psi\right )\right\rbrack. \nonumber\\
\end{eqnarray}
The corresponding equation for the velocity field is
\begin{eqnarray}
\label{nept3}
\frac{\partial {\bf u}}{\partial t}+({\bf u}\cdot \nabla) {\bf
u}=-\frac{1}{\rho}\nabla p+ {\bf z}\times  D\left (\nabla
\omega+\frac{\beta(t)}{C''(\omega)}\nabla\psi\right ),\nonumber\\
\end{eqnarray}
where $p$ is the pressure and $\rho$ the density. To pass from Eq. (\ref{nept2})
to Eq. (\ref{nept3}), we have used the identity $({\bf u}\cdot\nabla){\bf
u}=\nabla({\bf u}^2/2)-{\bf u}\times\omega {\bf z}$ and the identity
$\nabla\times ({\bf u}\times\omega {\bf z})=-({\bf u}\cdot\nabla)\omega {\bf z}$
that is valid for a 2D incompressible flow, so that finally $\nabla\times\lbrack
({\bf u}\cdot\nabla){\bf u}\rbrack=({\bf u}\cdot \nabla)\omega {\bf z}$. Now,
using ${\bf u}=-{\bf z}\times \nabla\psi$ and the identity
\begin{eqnarray}
\label{nept4}
\Delta {\bf u}=\nabla (\nabla\cdot {\bf u})-\nabla\times(\nabla\times {\bf
u})\nonumber\\
=-\nabla\times(\omega {\bf z})={\bf z}\times \nabla\omega,
\end{eqnarray}
valid for a 2D incompressible flow, we finally obtain
\begin{eqnarray}
\label{nept5}
\frac{\partial {\bf u}}{\partial t}+({\bf u}\cdot \nabla) {\bf
u}=-\frac{1}{\rho}\nabla p+   D\left (\Delta {\bf
u}-\frac{\beta(t)}{C''(\omega)}{\bf u}\right ).\nonumber\\
\end{eqnarray}
We see that the drift term in the equation for the vorticity takes the form of a
friction in the equation for the velocity. Furthermore, the friction coefficient
is given by an Einstein-type formula $\xi=D\beta(t)$ involving the diffusion
coefficient and the inverse temperature. At equilibrium,
\begin{eqnarray}
\label{nept6}
\Delta {\bf u}=\frac{\beta}{C''(\omega)}{\bf u}.
\end{eqnarray}
This equation can be directly derived from the relation
$C'(\omega)=-\beta\psi-\alpha$ determining the steady states of Eq.
(\ref{nept1}). Combining $\Delta {\bf u}={\bf z}\times \nabla\omega$,
$\nabla\omega=\omega'(\psi)\nabla\psi$  and $C''(\omega)=-\beta/\omega'(\psi)$,
we recover Eq. (\ref{nept6}). In particular, if $S$ is the opposite of the
enstrophy $S=-(1/2)\int {\omega}^{2}\, d{\bf
r}$, the relaxation equation (\ref{nept5}) becomes 
\begin{eqnarray}
\label{nept7}
\frac{\partial {\bf u}}{\partial t}+({\bf u}\cdot \nabla) {\bf
u}=-\frac{1}{\rho}\nabla p+   D\left (\Delta {\bf u}-\beta(t){\bf u}\right ).
\end{eqnarray}
At equilibrium, $\Delta {\bf u}=\beta{\bf u}$.

Let us now consider the relaxation equation (\ref{two16}) including a topography
$h$. It can be rewritten
\begin{equation}
\label{nept8}
\frac{\partial \omega}{\partial t}+{\bf u}\cdot \nabla q=\nabla\cdot
\left\lbrack D\left (\nabla \omega+\nabla
h+\frac{\beta(t)}{C''(q)}\nabla\psi\right )\right\rbrack.
\end{equation}
Proceeding as before, the corresponding equation for the velocity field is
\begin{eqnarray}
\label{nept9}
\frac{\partial {\bf u}}{\partial t}&+&({\bf u}\cdot \nabla) {\bf u} + h{\bf
z}\times {\bf u} =-\frac{1}{\rho}\nabla p+ \nonumber\\  &D&\left (\Delta {\bf
u}+{\bf z}\times \nabla h-\frac{\beta(t)}{C''(q)}{\bf u}\right ),
\end{eqnarray}
where we have used $\nabla\times ({\bf z}\times h{\bf u})=(\nabla\cdot
(h{\bf u})){\bf z}=({\bf u}\cdot \nabla h){\bf z}$ for a 2D
incompressible velocity field.  In particular, if $S$ is the opposite
of the potential enstrophy $S=-(1/2)\int {q}^{2}\, d{\bf r}$,
the relaxation equations (\ref{nept8}) and (\ref{nept9})
become
\begin{equation}
\label{nept8b}
\frac{\partial \omega}{\partial t}+{\bf u}\cdot \nabla q=\nabla\cdot
\left\lbrack D\left (\nabla \omega+\nabla h+{\beta(t)}\nabla\psi\right
)\right\rbrack,
\end{equation}
and
\begin{eqnarray}
\label{nept10}
\frac{\partial {\bf u}}{\partial t}&+&({\bf u}\cdot \nabla) {\bf u}+ h{\bf
z}\times {\bf u}=-\frac{1}{\rho}\nabla p+ \nonumber\\  &D&\left (\Delta {\bf
u}+{\bf z}\times \nabla h-\beta(t){\bf u}\right ).
\end{eqnarray}
Inspired by the original idea of Holloway
\cite{hollowaynept}, we introduce a velocity field
based on the topography
\begin{eqnarray}
\label{nept11}
{\bf u}_*=\frac{1}{\beta(t)}{\bf z}\times \nabla h,\quad
\psi_*=-\frac{h}{\beta(t)},\quad \omega_*=\frac{\Delta h}{\beta(t)}.
\end{eqnarray}
With these notations, the  equation for the velocity field can
be rewritten
\begin{eqnarray}
\label{nept12}
\frac{\partial {\bf u}}{\partial t}&+&({\bf u}\cdot \nabla) {\bf u}+ h{\bf
z}\times {\bf u}=-\frac{1}{\rho}\nabla p+ \nonumber\\  &D& \Delta {\bf
u}-D\beta(t)({\bf u}-{\bf u}_*).
\end{eqnarray}
It involves a turbulent viscosity and a friction force proportional to the
difference between the velocity field ${\bf u}$ and the velocity field ${\bf
u}_*$ based on the topography. The friction coefficient is given by an Einstein
relation $\xi=D\beta$. This term ``pushes'' the flow towards the topographic
flow ${\bf u}_*$. This corresponds to the so-called ``Neptune effect'' of
Holloway \cite{hollowaynept}. On the other hand, the diffusion term allows some
deviation with respect to the topographic flow.  At equilibrium
\begin{eqnarray}
\label{nept13}
\Delta {\bf u}=\beta({\bf u}-{\bf u}_*).
\end{eqnarray}
If $D$ is constant, the equation  (\ref{nept8b})  for the vorticity
can be written 
\begin{equation}
\label{nept14}
\frac{\partial \omega}{\partial t}+{\bf u}\cdot \nabla q=D\Delta
\omega-D\beta(t)(\omega-\omega_*).
\end{equation}
At equilibrium,
\begin{equation}
\label{nept15}
\Delta \omega=\beta(\omega-\omega_*).
\end{equation}
If we neglect the Laplacian, we get $\omega=\omega_*$. This is valid in the
limit $\beta\rightarrow +\infty$. This is equivalent to neglecting
the Laplacian in the fundamental differential equation
$-\Delta\psi+h=-\beta\psi-\alpha$. This leads to
$\psi=-\frac{1}{\beta}(h+\alpha)$, equivalent to ${\bf u}={\bf u}_*$ and
$\omega=\omega_*$. As we have seen, this corresponds to the standard  Fofonoff
\cite{fofonoff} flows that are completely determined by the topography (far from
the boundaries). More generally, Eq. (\ref{nept15}) takes into account finite
temperature effects that can induce deviations to the standard Fofonoff flows.
These finite temperature effects (corresponding to
sufficiently high energies) are precisely those that have been described in this
paper.

In conclusion, the relaxation equation (\ref{two16}) derived from the Maximum
Entropy Production Principle (MEPP) \cite{proc} is relatively consistent with
the oceanographic parametrization of Holloway \cite{hollowaynept}, especially
when the (generalized) entropy is the neg-enstrophy. This is
interesting because the parametrization of Holloway \cite{hollowaynept} has been
used in realistic oceanic modeling where the Neptune effect was shown to play a
significant role. This suggests that our parametrization can be of relevance also
in the physics of the oceans.

\section{Modal decomposition}
\label{sec_md}

We define the eigenfunctions and eigenvalues of the Laplacian by
\begin{eqnarray}
\label{ei1}
\Delta\psi_n=\beta_n\psi_n,
\end{eqnarray}
with $\psi_n=0$ on the domain boundary. These eigenfunctions are orthogonal and
normalized such that $\langle\psi_n\psi_m\rangle=\delta_{nm}$. Since $-\int
(\nabla\psi_n)^2\, d{\bf r}=\beta_n\int \psi_n^2\, d{\bf r}$, we note that
$\beta_n<0$. Following Chavanis \& Sommeria \cite{jfm}, we distinguish two types
of eigenmodes: the odd eigenmodes $\psi_n'$ such that $\langle \psi_n'\rangle=0$
and the even eigenmodes $\psi_n''$ such that $\langle \psi_n''\rangle\neq 0$. We
note $\beta_n'$ and $\beta_n''$ the corresponding eigenvalues.

In a rectangular domain of unit area whose sides are denoted $a=\sqrt{\tau}$ and
$b=1/\sqrt{\tau}$ (where $\tau=a/b$ is the aspect ratio), the eigenmodes and
eigenvalues are
\begin{eqnarray}
\label{ei5}
\psi_{mn}=2\sin(m\pi (x/\sqrt{\tau}+1/2))\sin(n\pi(\sqrt{\tau}y+1/2)),\quad
\end{eqnarray}
\begin{eqnarray}
\label{ei6}
\beta_{mn}=-\pi^2\left (\frac{m^2}{\tau}+\tau n^2\right ),
\end{eqnarray}
where the origin of the Cartesian frame is taken at the center of the domain.
The integer $m\ge 1$ gives the number of vortices along the $x$-axis and $n\ge
1$ the number of vortices along the $y$-axis. We have $\langle \psi_{mn}\rangle
=0$ if $m$ or $n$ is even and  $\langle \psi_{mn}\rangle \neq 0$ if $m$ and $n$
are odd. The largest eigenvalue with non zero mean is $\beta_{11}$. The largest
eigenvalue with zero mean is $\beta_{21}$ for $\tau>1$ and $\beta_{12}$ for
$\tau<1$.

We want to solve the differential equations
\begin{eqnarray}
\label{f69}
-\Delta\phi_1+\beta\phi_1=1,
\end{eqnarray}
and
\begin{eqnarray}
\label{f70}
-\Delta\phi_2+\beta\phi_2=-H,
\end{eqnarray}
with $\phi_1=0$ and $\phi_2=0$ on the domain boundary.  This can be done by
decomposing the solutions on the eigenmodes of the Laplacian using
$f=\sum_{mn}\langle f\psi_{mn}\rangle\psi_{mn}$.

For $\beta\neq \beta_{mn}$, the solution of Eq. (\ref{f69}) is unique and given
by
\begin{eqnarray}
\label{f72}
\phi_1=\sum_{mn} \frac{\langle\psi_{mn}\rangle}{\beta-\beta_{mn}}\psi_{mn}.
\end{eqnarray}
From this relation, we obtain
\begin{eqnarray}
\label{f73}
\langle \phi_1\rangle =\sum_{mn}
\frac{\langle\psi_{mn}\rangle^2}{\beta-\beta_{mn}},
\end{eqnarray}
\begin{eqnarray}
\label{f74}
\langle \phi_1^2\rangle =\sum_{mn}
\frac{\langle\psi_{mn}\rangle^2}{(\beta-\beta_{mn})^2}=-\frac{d\langle
\phi_1\rangle}{d\beta},
\end{eqnarray}
\begin{eqnarray}
\label{f75}
\langle \phi_1 H\rangle =\sum_{mn} \frac{\langle\psi_{mn}\rangle\langle
H\psi_{mn}\rangle}{\beta-\beta_{mn}}.
\end{eqnarray}

For $\beta\neq \beta_{mn}$, the solution of Eq. (\ref{f70}) is unique and given
by
\begin{eqnarray}
\label{f76}
\phi_2=-\sum_{mn} \frac{\langle H\psi_{mn}\rangle}{\beta-\beta_{mn}}\psi_{mn}.
\end{eqnarray}
From this relation, we obtain
\begin{eqnarray}
\label{f77}
\langle \phi_2 H\rangle =-\sum_{mn} \frac{\langle
H\psi_{mn}\rangle^2}{\beta-\beta_{mn}},
\end{eqnarray}
\begin{eqnarray}
\label{f78}
\langle \phi_2^2\rangle =\sum_{mn} \frac{\langle
H\psi_{mn}\rangle^2}{(\beta-\beta_{mn})^2}= \frac{d\langle \phi_2
H\rangle}{d\beta},
\end{eqnarray}
\begin{eqnarray}
\label{f79}
\langle \phi_2\rangle =-\sum_{mn} \frac{\langle
H\psi_{mn}\rangle\langle\psi_{mn}\rangle }{\beta-\beta_{mn}}.
\end{eqnarray}
We also have
\begin{eqnarray}
\label{f80}
\langle \phi_1\phi_2\rangle =-\sum_{mn} \frac{\langle
H\psi_{mn}\rangle\langle\psi_{mn}\rangle }{{(\beta-\beta_{mn})^2}}.
\end{eqnarray}
Finally, we remark that
\begin{eqnarray}
\label{f81}
{\langle \phi_1 H\rangle =-\langle \phi_2\rangle}.
\end{eqnarray}
On the other hand, if the topography is antisymmetric with
respect to $y=0$ (as in the case of a linear topography
$H=y$), we have $\langle H\psi_{mn}\rangle\langle\psi_{mn}\rangle =0$
implying
\begin{eqnarray}
\label{f82}
\langle \phi_1 H\rangle =\langle \phi_2\rangle=\langle \phi_1\phi_2\rangle=0.
\end{eqnarray}
In that case, $\phi_1$ is even and $\phi_2$ is odd with respect to the variable 
$y$. This can be directly seen at the level of the
differential equations (\ref{f69}) and (\ref{f70}).

\section{The case $\beta=0$}
\label{sec_z}

It can be interesting to consider the case $\beta=0$ specifically. This regime
of infinite temperatures corresponds to uniform potential vorticity $q=-\alpha$.
The differential equation (\ref{f13}) reduces to
\begin{eqnarray}
\label{z1}
-\Delta\psi=\Gamma-b H.
\end{eqnarray}
The solution is $\psi=\Gamma\phi_1+b\phi_2$ where $\phi_1$ and $\phi_2$ are the
solutions of Eqs. (\ref{f18}) and (\ref{f19}) with $\beta=0$. The equation for
the energy (\ref{f26}) can be written
\begin{eqnarray}
\label{f65b}
\frac{2E}{b^2}=\langle\phi_1\rangle\left (\frac{\Gamma}{b}\right
)^2+(\langle\phi_2\rangle-\langle\phi_1 H\rangle)\frac{\Gamma}{b}-\langle\phi_2
H\rangle,
\end{eqnarray}
where we have used $c=b/\Gamma$ according to Eq. (\ref{f23}). For given
$\Gamma$, Eq. (\ref{f65b})  determines the energy $E(\Gamma)$ for which
$\beta=0$ (i.e. $q$ is uniform). It is given by a parabola.
If the topography is antisymmetric with respect to $y=0$, using
Eq. (\ref{f82}),  the foregoing equation reduces to
\begin{eqnarray}
\label{f66}
\frac{2E}{b^2}=\langle\phi_1\rangle\frac{\Gamma^2}{b^2}-\langle\phi_2 H\rangle.
\end{eqnarray}

\section{The low energy limit}
\label{sec_low}

In the limit $\beta\rightarrow +\infty$, a boundary layer approximation can be made. As a first approximation, the Laplacian can be neglected in the differential equations (40) and (41), everywhere in the domain except close to the boundary.  To leading order, the function $\phi_1$ is given by $\phi_1=1/\beta$. The correction, close to the boundary, behaves like $Ae^{-\sqrt{\beta}\zeta}$, where $\zeta$ is a coordinate perpendicular to the boundary pointing towards the inside of the domain. The constant $A$ is determined by the condition $\phi_1=0$ on the boundary. Similarly, to leading order, the function $\phi_2$ is given by $\phi_2=-y/\beta$ with a correction scaling like  $Ae^{-\sqrt{\beta}\zeta}$ close to the boundary.

In this limit, we obtain 
\begin{eqnarray}
\label{low1}
\langle\phi_1\rangle=\frac{1}{\beta}-\frac{2\left (\sqrt{\tau}+\frac{1}{\sqrt{\tau}}\right )}{\beta^{3/2}},
\end{eqnarray}
\begin{eqnarray}
\label{low2}
\langle\phi_1^2\rangle=-\frac{d\langle \phi_1\rangle}{d\beta}=\frac{1}{\beta^2}-\frac{3\left (\sqrt{\tau}+\frac{1}{\sqrt{\tau}}\right )}{\beta^{5/2}}.
\end{eqnarray}
Similarly,
\begin{eqnarray}
\label{low3}
\langle\phi_2 y\rangle=-\frac{1}{12\beta\tau}+\frac{1}{6\tau^{3/2}\beta^{3/2}}+\frac{1}{2\sqrt{\tau}\beta^{3/2}}. 
\end{eqnarray}
\begin{eqnarray}
\label{low4}
\langle\phi_2^2\rangle=\frac{d\langle \phi_2 y\rangle}{d\beta}=\frac{1}{12\beta^2\tau}-\frac{1}{4\tau^{3/2}\beta^{5/2}}-\frac{3}{4\sqrt{\tau}\beta^{5/2}}. 
\end{eqnarray}
Finally, substituting these results in the energy equation (64), we obtain for $\beta\rightarrow +\infty$:
\begin{eqnarray}
\label{low5}
\frac{2E}{b^2}=\left (\frac{\Gamma}{b}\right )^{2}\frac{1}{4\left (\sqrt{\tau}+\frac{1}{\sqrt{\tau}}\right )\sqrt{\beta}}+\frac{1}{12\beta^{3/2}}\left (\frac{1}{\tau^{3/2}}+\frac{3}{\sqrt{\tau}}\right ). \nonumber\\
\end{eqnarray}

\section{Technical construction of the caloric curves}
\label{sec_technical_constr}

\subsection{Antisymmetric linear topography
and $\Gamma=0$} \label{techn_S_eq}

Several cases must be considered in order to
construct the caloric curve. In the first one, $\Gamma=0$ and
$\langle\psi\rangle=0$ (i.e. $\alpha=0$, $c=\infty$). Since $\psi=b\phi_2$ with
$\langle\phi_2\rangle=0$, these solutions exist for any value of $\beta$. The
curve $\beta({E})$ relating their temperature to the energy is given by Eq.
(\ref{f26vfb}). This forms the main curve (see Figs. \ref{E_beta_1},
\ref{E_beta_1z}, \ref{E_beta_2} and \ref{E_beta_3}). The entropy of these
solutions is given  by Eq. (\ref{f27hjb}).

Another possible situation is $\Gamma=0$ and
$\langle\psi\rangle\neq 0$ (i.e. $\alpha\neq 0$, $c$ finite). According to Eq.
(\ref{f31}), we see that $c$ is finite iff  $1-\beta\langle\phi_1\rangle=0$,
i.e. $\beta=\beta_*^{(k)}$. Therefore, the temperatures of these solutions take
only discrete values. The value of $c$ is determined by the energy $E$ according
to Eq. (\ref{f32}) with $\beta$ replaced by $\beta_*^{(k)}$. This determines two
solutions $\pm c(E)$ (i.e $\pm \alpha(E)$) for each value of the energy (see for
instance Fig.~\ref{pm_a}).  The ensemble of these solutions form a {plateau}
$\beta=\beta_{*}^{(k)}$ (see Figs. \ref{E_beta_1}, \ref{E_beta_1z},
\ref{E_beta_2} and \ref{E_beta_3}) starting at  $1/E=0$ ($c=0$,
$\alpha=\pm\infty$) and connecting the main curve at  $1/E=1/E_*^{(k)}(\tau)$
($c=\pm\infty$, $\alpha=0$). When $1/E=0$, we find $S/E=\beta_*^{(k)}$.  On the
other hand, $E_*^{(k)}(\tau)$ is given by Eq. (\ref{f26vfb}) with $\beta$
replaced by $\beta_*^{(k)}$. The evolution of $E_*(\tau)$ with $\tau$ is shown
in Fig. \ref{phasediag1}.

Finally, it will be assumed that $\beta$ is
equal to an eigenvalue $\beta_{mn}$ of the Laplacian. In that case,
$\langle\psi\rangle$ is necessarily equal to $0$ (i.e.
$\alpha=0$, $c\rightarrow \infty$), which corresponds to $\psi=b\phi_2$. Four
cases must be distinguished:

$\bullet$ Case 1: $\beta=\beta_{mn}$ with $\langle
y\psi_{mn}\rangle\neq 0$ and $\langle \psi_{mn}\rangle\neq 0$. This is not
possible for an antisymmetric topography.

$\bullet$ Case 2: $\beta=\beta_{mn}$ with $\langle
y\psi_{mn}\rangle\neq 0$ and $\langle \psi_{mn}\rangle= 0$ ($m$ odd and $n$
even) like $\beta_{12}$. In that case, Eq. (\ref{f19}) has no solution for
$\beta=\beta_{mn}$ and $\phi_2$ diverges like
\begin{eqnarray}
\label{f45}
\phi_2\sim -\frac{\langle y\psi_{mn}\rangle}{\beta-\beta_{mn}}\psi_{mn},
\end{eqnarray}
for $\beta\rightarrow \beta_{mn}$. Then, we find from Eqs. (\ref{f26vfb}) and
(\ref{f27hjb}) that $1/E\rightarrow 0$ and $S/E\rightarrow \beta_{mn}$. This is
a limit case of the main curve. Note that the sign of $\phi_2$ (hence $q$)
changes as we pass from $\beta_{mn}^+$ to $\beta_{mn}^-$.

$\bullet$ Case 3: $\beta=\beta_{mn}$ with $\langle
y\psi_{mn}\rangle= 0$ and $\langle \psi_{mn}\rangle= 0$ ($m$ even and $n$
arbitrary) like $\beta_{21}$. In that case, the  solution of Eq. (\ref{f19})
with $\beta=\beta_{mn}$ is not unique. To the solution (\ref{f77}) of Eq. 
(\ref{f19}) with $\beta=\beta_{mn}$ we can always superpose the eigenmode
$\psi_{mn}$ with an amplitude $\chi$. This corresponds to a ``mixed solution''
\begin{eqnarray}
\label{f49}
\phi_2\rightarrow \phi_2+\chi \psi_{mn},
\end{eqnarray}
which has zero average as required. The amplitude $\chi$ is determined by the
energy constraint (\ref{f26vfb}) with $\beta$ replaced by $\beta_{mn}$. This
determines two solutions $\pm\chi(E)$ for each value of the energy (note that
these two distinct solutions have the same value of the chemical potential
$\alpha=0$). The ensemble of these mixed solutions form a plateau
$\beta=\beta_{mn}$ (see Figs. \ref{E_beta_1}, \ref{E_beta_1z}, \ref{E_beta_2}
and \ref{E_beta_3}) starting from  $1/E=0$ ($\chi=\pm\infty$) and connecting the
main curve at $1/E=1/E_{mn}(\Gamma=0,\tau)$ ($\chi=0$). For  $1/E=0$, we find
$S/E=\beta_{mn}$. On the other hand, $E_{mn}(\Gamma=0,\tau)$ is given by Eq.
(\ref{f26vfb}) with $\beta$ replaced by $\beta_{mn}$. The evolution of
$E_{21}(\Gamma=0,\tau)$ with $\tau$ is shown in Fig. \ref{phasediag1}. Note that
in a square domain ($\tau=1$), since $\beta_{mn}=\beta_{nm}$ by symmetry, we
find that $1/E=0$ for $m$ even and $n$ odd according to Case 2 (see Figs.
\ref{E_beta_1} and \ref{E_beta_1z}). In that case, the plateau reduces to a
point.

$\bullet$ Case 4: $\beta=\beta_{mn}$ with $\langle
y\psi_{mn}\rangle= 0$ and $\langle \psi_{mn}\rangle\neq 0$ ($m$ odd and $n$ odd)
like $\beta_{11}$. In that case, the  solution of Eq. (\ref{f19}) with
$\beta=\beta_{mn}$ is not unique and we expect a mixed solution of the form
$\phi_2\rightarrow \phi_2+\chi \psi_{mn}$. However, $\psi=b\phi_2$ has a zero
average value (as required) only for  $\chi=0$. Hence, we are just left with a
limit case of the main curve. It corresponds to the energy
$E_{mn}(\Gamma=0,\tau)$  given by Eq. (\ref{f26vfb}) with $\beta$ replaced by
$\beta_{mn}$.

\subsection{Antisymmetric linear topography
and $\Gamma\neq 0$} \label{techn_S_neq}

Once more, several cases must be considered. It
can be first assumed that $\Gamma\neq 0$ and $\Gamma+\beta \langle\psi\rangle=0$
(i.e. $\alpha=0$, $c\rightarrow \infty$), which corresponds to $\psi=b\phi_2$.
Since $\langle\phi_2\rangle=0$, there are no such solutions for $\Gamma\neq 0$
and $\beta\neq\beta_{mn}$.

Another possibility is $\Gamma\neq 0$ and
$\Gamma+\beta \langle\psi\rangle\neq 0$ (i.e. $\alpha\neq 0$, $c$ finite).
These solutions exist for any value of $\beta$. The   curve $\beta({E})$
relating their temperature to their energy is given by Eqs. (\ref{f31}) and
(\ref{f32}). This forms the main curve (see Figs.
\ref{E_beta_4}-\ref{E_beta_9}). Their entropy is given  by Eq. (\ref{f33}). For
$\beta=\beta_*^{(k)}$, we have $c=0$ leading to $1/E=0$. In that case,
$S/E=\beta_*^{(k)}$. The caloric curve $\beta(E)$ is unchanged when
$\Gamma\rightarrow -\Gamma$  since only $\Gamma^2$ appears in the equation for
the energy. However, the sign of $c$ (hence of $\alpha$) changes when
$\Gamma\rightarrow -\Gamma$ so that the flow structure is different. In the
figures, we consider $\Gamma>0$.

Finally, if $\beta$ is equal to an eigenvalue
$\beta_{mn}$ of the Laplacian, four cases must be distinguished.

$\bullet$ Case 1: $\beta=\beta_{mn}$ with $\langle
y\psi_{mn}\rangle\neq 0$ and $\langle \psi_{mn}\rangle\neq 0$. This is not
possible for an antisymmetric topography.

$\bullet$ Case 2: $\beta=\beta_{mn}$ with $\langle
y\psi_{mn}\rangle\neq 0$ and $\langle \psi_{mn}\rangle= 0$ ($m$ odd and $n$
even) like $\beta_{12}$. We are necessarily  in the case $\Gamma+\beta
\langle\psi\rangle\neq 0$ (i.e. $\alpha\neq 0$, $c$ finite). We see that Eq.
(\ref{f19}) has no solution for $\beta=\beta_{mn}$ and that
$\phi_2$ diverges like
\begin{eqnarray}
\label{f37}
\phi_2\sim -\frac{\langle y\psi_{mn}\rangle}{\beta-\beta_{mn}}\psi_{mn},
\end{eqnarray}
when $\beta\rightarrow \beta_{mn}$. In that case  $1/E\rightarrow 0$ and
$S/E\rightarrow \beta_{mn}$. This is just a limit case of the main curve. Note
that the sign of $\phi_2$ (hence $q$) changes as we pass from $\beta_{mn}^+$ to
$\beta_{mn}^-$.

$\bullet$ Case 3: $\beta=\beta_{mn}$ with $\langle
y\psi_{mn}\rangle= 0$ and $\langle \psi_{mn}\rangle= 0$ ($m$ even and $n$
arbitrary) like $\beta_{21}$. We are necessarily in the case $\Gamma+\beta
\langle\psi\rangle\neq 0$ (i.e. $\alpha\neq 0$, $c$ finite). The solutions of 
Eqs. (\ref{f18}) and (\ref{f19}) are not unique since we can always superpose to
Eqs. (\ref{f72}) and (\ref{f76}) an eigenfunction $\psi_{mn}$ of the Laplacian.
Thus
\begin{eqnarray}
\label{f40}
\phi_1\rightarrow \phi_1+\chi\psi_{mn},
\end{eqnarray}
\begin{eqnarray}
\label{f41}
\phi_2\rightarrow \phi_2+\chi\psi_{mn}.
\end{eqnarray}
The amplitude $\chi$ is determined by the energy constraint (\ref{f32}) with
$\beta_{mn}$ replacing $\beta$. This determines two solutions $\pm\chi(E)$ for
each value of $E$ (note that these two distinct solutions have the same value of
$c(E)$, hence of $\alpha(E)$, given by Eq. (\ref{f31})).  The ensemble of these
mixed solutions form a plateau $\beta=\beta_{mn}$ (see Figs.
\ref{E_beta_4}-\ref{E_beta_9}) starting from $1/E=0$ ($\chi\rightarrow +\infty$)
and reaching the main curve at $1/E=1/E_{mn}(\Gamma,\tau)$ ($\chi=0$). For
$1/E=0$, we find $S/E\rightarrow \beta_{mn}$.
On the other hand, $E_{mn}(\Gamma,\tau)$ is given by Eqs.  (\ref{f31}) and
(\ref{f32}) replacing $\beta$ by $\beta_{mn}$. It corresponds to a parabola of
the form $E_{mn}(\Gamma,\tau)=a(\tau)\Gamma^2+c(\tau)$ with $a>0$ (since
$\langle\phi_1\rangle-\beta\langle\phi_1^2\rangle\ge 0$ \cite{jfm}).
The evolution of $E_{21}(\Gamma,\tau)$ with $\tau$ is shown in Fig.
\ref{phasediag2} for different values of $\Gamma$. Note that in a square domain
($\tau=1$), since $\beta_{mn}=\beta_{nm}$ by symmetry, we find that $1/E=0$ for
$m$ even and $n$ odd according to Case 2 (see Figs.
\ref{E_beta_4}-\ref{E_beta_5}). In that case, the plateau is reduced to a
point.

$\bullet$ Case 4: $\beta=\beta_{mn}$ with $\langle
y\psi_{mn}\rangle= 0$ and $\langle \psi_{mn}\rangle\neq 0$ ($m$ odd and $n$ odd)
like $\beta_{11}$. If we consider $\Gamma+\beta \langle\psi\rangle\neq 0$ (i.e.
$\alpha\neq 0$, $c$ finite), we see that  Eq. (\ref{f18}) has no solution for
$\beta=\beta_{mn}$ and that $\phi_1$ diverges like
\begin{eqnarray}
\label{f38}
\phi_1\sim \frac{\langle \psi_{mn}\rangle}{\beta-\beta_{mn}}\psi_{mn},
\end{eqnarray}
for $\beta\rightarrow \beta_{mn}$. Then we find that $c\sim -(b/\Gamma)\beta
\langle\psi_{mn}\rangle^2/(\beta-\beta_{mn})\rightarrow \infty$ and that $E$
tends to a finite value
\begin{eqnarray}
\label{krl}
\frac{2E}{b^2}=-\frac{1}{\beta
\langle\psi_{mn}\rangle^2}\frac{\Gamma^2}{b^2}
-\beta\langle\phi_2^2\rangle-\langle\phi_2 y\rangle,
\end{eqnarray}
with $\beta=\beta_{mn}$. It corresponds to a parabola of the form
$E_{mn}(\Gamma,\tau)=a(\tau)\Gamma^2+c(\tau)$.  This is just a limit case of the
main curve. If we consider $\Gamma+\beta \langle\psi\rangle=0$ (i.e. $\alpha=0$,
$c$ infinite) so that $\psi=b\phi_2$, we see that the solution of Eq.
(\ref{f19}) is not unique for $\beta=\beta_{mn}$ since we can always superpose
to Eq.  (\ref{f76}) an eigenfunction $\psi_{mn}$ of the Laplacian. Thus
\begin{eqnarray}
\label{f39}
\phi_2\rightarrow \phi_2+\chi \psi_{mn}.
\end{eqnarray}
The condition  $\Gamma+\beta \langle\psi\rangle=0$, reducing to
$\Gamma/b+\beta\chi\langle\psi_{mn}\rangle=0$, determines the value of $\chi$.
Then, using  Eq. (\ref{f26vfb}), we find that this solution exists for a unique
value of the energy $E_{mn}(\Gamma,\tau)$ given by Eq. (\ref{krl}). This is to
be expected since $c\rightarrow +\infty$ when $\beta\rightarrow \beta_{mn}$. In
conclusion, the case $\beta\rightarrow \beta_{mn}$ or $\beta=\beta_{mn}$ with
$m$ and $n$ odd is just a limit case of the main curve.

\subsection{Non-symmetric linear topography and
$\Gamma\neq\Gamma_*$} \label{techn_NS_neq}

The first case that will be treated is the one
for which $\Gamma\neq \Gamma_*$ and $\Gamma+\beta \langle\psi\rangle\neq 0$
(i.e. $\alpha\neq 0$, $c$ finite). These solutions exist for any value of
$\beta$. The curve $\beta({E})$ relating their temperature to their energy is
given by Eqs. (\ref{f26}), (\ref{f28}) and (\ref{f23}). This forms the main
curve. Their entropy is given  by Eq. (\ref{f27}). For $\beta=\beta_*^{(k)}$ we
have $c=0$ leading to $1/E=0$. In that case, $S/E=\beta_*^{(k)}$.

Another possibility is $\Gamma\neq \Gamma_*$
and $\Gamma+\beta \langle\psi\rangle=0$ (i.e. $\alpha=0$, $c\rightarrow
\infty$). In that case $\psi=b\phi_2$ which implies $\Gamma+\beta
b\langle\phi_2\rangle=0$. Therefore, this solution exists for a very particular
value of the inverse temperature that will be  denoted $\beta_0$. The
corresponding energy is given by Eq. (\ref{f26vf}) where $\beta$ is replaced by
$\beta_0$. It turns out that this is just a particular point of the main branch
corresponding to $c\rightarrow +\infty$ so that it does not bring any new
solution.

In the case where $\beta$ is equal to an eigenvalue
$\beta_{mn}$ of the Laplacian, four cases must be
distinguished.

$\bullet$ Case 1: $\beta=\beta_{mn}$ with $\langle
H\psi_{mn}\rangle\neq 0$ and $\langle \psi_{mn}\rangle\neq 0$ ($m$ odd and $n$
odd) like $\beta_{11}$. In that case, Eqs. (\ref{f18}) and (\ref{f19}) do not
have any solution for $\beta=\beta_{mn}$ and the functions
$\phi_1$ and $\phi_2$ behave like
\begin{eqnarray}
\label{f52}
\phi_1\sim \frac{\langle \psi_{mn}\rangle}{\beta-\beta_{mn}}\psi_{mn},
\end{eqnarray}
\begin{eqnarray}
\label{f53}
\phi_2\sim -\frac{\langle H\psi_{mn}\rangle}{\beta-\beta_{mn}}\psi_{mn},
\end{eqnarray}
when $\beta\rightarrow \beta_{mn}$. We are in the case $\Gamma+\beta
\langle\psi\rangle\neq 0$ (i.e. $\alpha\neq 0$, $c$ finite). Substituting the
asymptotic expansions (\ref{f52}) and (\ref{f53}) in Eqs. (\ref{f28}) and
(\ref{f23}), we find that
\begin{eqnarray}
\label{soir1}
c&\simeq& \frac{\langle\psi_{mn}\rangle}{\langle H\psi_{mn}\rangle}
+\frac{1}{\beta_{mn}\langle H\psi_{mn}\rangle}\nonumber\\
&\times&\left (\frac{\Gamma}{b\langle
H\psi_{mn}\rangle}-\frac{1}{\langle\psi_{mn}\rangle}\right )(\beta-\beta_{mn}),
\end{eqnarray}
and
\begin{eqnarray}
\label{soir2}
\phi\simeq -\frac{1}{\beta_{mn}}\left (\frac{\Gamma}{b\langle
H\psi_{mn}\rangle}-\frac{1}{\langle\psi_{mn}\rangle}\right )\psi_{mn}.
\end{eqnarray}
Therefore $c$ and $\phi$ are finite as  $\beta\rightarrow \beta_{mn}$.
Substituting these results in Eq. (\ref{f26}), we obtain a finite energy
\begin{eqnarray}
\label{soir2b}
\frac{2E}{b^2}=-\frac{1}{\langle\psi_{mn}\rangle^2}\frac{1}{\beta_{mn}}\left
(\frac{\Gamma}{b}-\frac{\langle
H\psi_{mn}\rangle}{\langle\psi_{mn}\rangle}\right )^2.
\end{eqnarray}
This solution is a particular case of the main curve. We note that the energy
(\ref{soir2b}) is of the form
$E_{mn}(\Gamma,\tau)=a(\tau)\Gamma^2+b(\tau)\Gamma+c(\tau)$ with $a>0$ so that
it forms a parabola. The minimum value of this parabola is $E_{mn}^{min}=0$
reached for $\Gamma=\Gamma_{mn}^{min}(\tau)$ with
\begin{eqnarray}
\label{soir2c}
\frac{\Gamma_{mn}^{min}(\tau)}{b}=\frac{\langle H\psi_{mn}\rangle}{\langle
\psi_{mn}\rangle}.
\end{eqnarray}

$\bullet$ Case 2: $\beta=\beta_{mn}$ with $\langle
H\psi_{mn}\rangle\neq 0$ and $\langle \psi_{mn}\rangle= 0$ ($m$ odd and $n$
even) like $\beta_{12}$. In that case, Eq. (\ref{f19}) does not have any
solution for $\beta=\beta_{mn}$ and $\phi_2$ diverges like
\begin{eqnarray}
\label{f55}
\phi_2\sim -\frac{\langle H\psi_{mn}\rangle}{\beta-\beta_{mn}}\psi_{mn},
\end{eqnarray}
when $\beta\rightarrow \beta_{mn}$. We note that $\langle\phi_2\rangle=0$.
Therefore, if $\Gamma\neq 0$, we are in the case $\Gamma+\beta
\langle\psi\rangle\neq 0$ (i.e. $\alpha\neq 0$). Then, $c$ is finite and
$\langle\phi_2^2\rangle\rightarrow +\infty$ so that $1/E\rightarrow 0$ according
to Eq. (\ref{f26}). If $\Gamma=0$, we are in the case $\Gamma+\beta
\langle\psi\rangle=0$ (i.e. $\alpha=0$). Then, $1/E\rightarrow 0$ according to
Eq. (\ref{f26vf}). This is a limit case of the main curve.

$\bullet$ Case 3: $\beta=\beta_{mn}$ with $\langle
H\psi_{mn}\rangle= 0$ and $\langle \psi_{mn}\rangle= 0$ ($m$ even and $n$
arbitrary) like $\beta_{21}$.  The solutions of  Eqs. (\ref{f18}) and
(\ref{f19}) are not unique since we can always superpose to Eqs. (\ref{f72}) and
(\ref{f76}) an eigenfunction $\psi_{mn}$ of the Laplacian. Thus
\begin{eqnarray}
\label{f56}
\phi_1\rightarrow \phi_1+\chi\psi_{mn},
\end{eqnarray}
\begin{eqnarray}
\label{f57}
\phi_2\rightarrow \phi_2+\chi \psi_{mn}.
\end{eqnarray}
The amplitude $\chi$ is determined by the energy constraint (\ref{f26}),
replacing $\beta$ by $\beta_{mn}$ and introducing Eqs. (\ref{f56}) and
(\ref{f57}). This determines two types of solutions $\chi_1(E)$ and $\chi_2(E)$
for each value of the energy $E$ (note that these two distinct solutions have
the same value of $c(E)$, hence the same value of $\alpha(E)$, given by Eq.
(\ref{f23})).  The ensemble of these mixed solutions form a plateau
$\beta=\beta_{mn}$ starting from $1/E=0$ ($\chi\rightarrow +\infty$) and
reaching the main curve at $1/E=1/E_{mn}(\Gamma,\tau)$ ($\chi=0$). For $1/E=0$,
we find $S/E\rightarrow \beta_{mn}$.
On the other hand, $E_{mn}(\Gamma,\tau)$ is given by Eqs.  (\ref{f26}),
(\ref{f28}) and (\ref{f23}) replacing $\beta$ by $\beta_{mn}$ and taking
$\chi=0$. It corresponds to a parabola of the form
$E_{mn}(\Gamma,\tau)=a(\tau)\Gamma^2+b(\tau)\Gamma+c(\tau)$ with $a>0$. We shall
denote by $E_{mn}^{min}(\tau)$ the minimum value of this parabola reached for
$\Gamma=\Gamma_{mn}^{min}(\tau)$. The expressions of these quantities can easily
be obtained from Eqs.  (\ref{f26}), (\ref{f28}) and (\ref{f23}) but they are not
particularly simple so we shall not give them explicitly. Note that in a square
domain ($\tau=1$), since $\beta_{mn}=\beta_{nm}$ by symmetry, we find that
$1/E=0$ for $m$ even and $n$ odd according to Case 2. In that case, the plateau
reduces to a point.

$\bullet$ Case 4: $\beta=\beta_{mn}$ with $\langle
H\psi_{mn}\rangle= 0$ and $\langle \psi_{mn}\rangle\neq 0$. This is not possible
for a topography of the form $h=b(y-y_0)$ with $y_0\neq 0$.

\subsection{Non-symmetric linear topography and $\Gamma =
\Gamma_*$} \label{techn_NS_eq}

It will be first assumed that $\Gamma=\Gamma_*$
and $\Gamma+\beta \langle\psi\rangle\neq 0$ (i.e. $\alpha\neq 0$, $c$ finite).
These solutions exist for any value of $\beta$. For $\beta\neq \beta_*^{(k)}$,
the curve $\beta({E})$ relating their temperature to their energy is given by
Eqs. (\ref{f26}), (\ref{f28}) and (\ref{f23}). This forms the main curve. Their
entropy is given  by Eq. (\ref{f27}). For $\beta=\beta_*^{(k)}$, the situation
is different because Eq. (\ref{f23}) takes in indeterminate
form. In that case, $c$ is determined by the energy constraint (\ref{f26}) with
$\beta$ replaced by $\beta_*^{(k)}$. This yields two
solutions $c_{1}(E)$ and $c_2(E)$  (and, correspondingly, two values of the
chemical potential $\alpha_{1}(E)$ and $\alpha_2(E)$) for each value of the
energy. The ensemble of these solutions form a  plateau $\beta=\beta_*^{(k)}$
starting from  $1/E=0$ ($c=0$, $\alpha=\infty$) and reaching the main curve at
$1/E=1/E_*^{(k)}(\tau)$ ($c=\infty$, $\alpha=0$). When $1/E=0$, we find
$S/E=\beta_*^{(k)}$.  On the other hand, $E_*^{(k)}(\tau)$ is given by Eq.
(\ref{f26vf}) with $\beta$ replaced by $\beta_*^{(k)}$.

A second possibility is $\Gamma=\Gamma_*$ and
$\Gamma+\beta \langle\psi\rangle=0$ (i.e. $\alpha=0$, $c\rightarrow \infty$). 
In that case $\psi=b\phi_2$, which implies $\Gamma_*+\beta
b\langle\phi_2\rangle=0$. These solutions exist for $\beta$ such that
$\beta\langle\phi_2\rangle=\beta_*\langle(\phi_2)_*\rangle$. Except in
an antisymmetric domain where they exist for any $\beta$ (see
Sec. \ref{sec_stz}), in a non-symmetric domain they exist only for
$\beta=\beta_*$ and for the corresponding value of the energy $E_*(\tau)$. This
is just a particular point of the main curve corresponding to $c\rightarrow
+\infty$ so this case does not bring any new solution.

The study of the eigenvalues $\beta=\beta_{mn}$ leads to the
same results as in the previous section.\\

\section{Interest of the relaxation equations}
\label{sec_interest}

Relaxation equations for two-dimensional flows have been
introduced by Robert \& Sommeria \cite{rsmepp} in the context of the MRS 
statistical theory and extended by Chavanis \cite{proc} to
more general situations. These relaxation equations are relatively
{\it ad hoc} since they are solely based on general ``principles''
similar to the first and second principles of thermodynamics:
conservation of energy (and possibly other constraints) and increase
of an ``entropic'' functional. Contrary to the Boltzmann equation, the
relaxation equations are not derived from first principles but just
assumed\footnote{An attempt to derive these relaxation equations from
a quasilinear theory of the 2D Euler equation is made in
\cite{quasi}.}. This is the reason why they exist in different
forms. There is no general consensus about the
importance and interest of these relaxation equations. Some
researchers argue that they are useless and that they have no physical
relevance. Other workers, on the other hand, consider that they are
interesting for the following reasons:

(i) They can be used as {\it numerical algorithms} to solve
maximization problems of the form (\ref{dim}) with a general
``entropic functional'' of the form (\ref{dims}). This maximization
problem can have various interpretations (as explained in Section
\ref{sec_inter} and in \cite{proc}) related to thermodynamical and/or
nonlinear dynamical stability of 2D Euler flows. Although the case of
a quadratic functional (\ref{f2}) can be treated analytically (as done
in our paper), the general maximization problem
(\ref{dim})-(\ref{dims}) is more complex and the relaxation equations
can be a way (among others) to solve it. Indeed, it is generally
difficult to directly solve the Euler-Lagrange equations associated
with (\ref{dim})-(\ref{dims}) and be sure that the critical points are
entropy maxima. By contrast, the relaxation equations guarantee (by
construction) that the relaxed state is an entropy maximum with
appropriate constraints. Furthermore, when several entropy maxima
(metastable states) exist for the same values of the constraints, the
relaxation equations can tell which (local) maximum will be reached by
a given initial condition, thus unveiling its {\it basin of
attraction}. Finally, by introducing in the relaxation equations a
variable diffusion coefficient (related to the local fluctuations of
vorticity), we can take into account the important effects of {\it
incomplete relaxation} and lack of ergodicity
\cite{rr,csr}. This is not possible if we consider just equilibrium
statistical mechanics.

(ii) They provide non trivial {\it dynamical systems},
consistent with
the equilibrium states, whose mathematical study is interesting in its
own right.  For example, we have observed that the system 
can remain blocked for a long time in an unstable state
before finally reaching the maximum entropy state.  Although this
effect might be an artefact of the relaxation equations, the same
phenomenon could also occur in more realistic
parametrizations. Therefore, the study of the relaxation equations can
suggest interesting ideas and developments.

(iii) The discussion of Appendix \ref{sec_nept} 
shows that the relaxation equations are remarkably consistent with the
parametrization of Holloway \cite{hollowaynept} based on relatively
different arguments. This is a hint that our {\it parametrization} may
contain ingredients that capture some features of real oceanic
circulation. For example, Holloway has shown that the Neptune effect
plays some role in the dynamics of the oceans. Therefore, making 
contact between the two approaches is of interest. Furthermore, it is
possible to include other geophysical effects in the relaxation
equations such as wind forcing like in Kazantsev {\it et al.} 
\cite{kazantsev}.

\end{document}